\documentclass[a4paper,11pt]{article}
\pdfoutput=1
\usepackage{jheppub}
\usepackage[T1]{fontenc}
\usepackage{mathtools}
\usepackage{bm}
\usepackage{longtable}
\usepackage{float}
\usepackage{slashed}
\usepackage[table]{xcolor}
\usepackage{tablefootnote}
\usepackage{multirow}
\usepackage{booktabs}
\usepackage{graphicx}
\usepackage[labelformat=simple]{subcaption}

\usepackage{dsfont}
\usepackage{upgreek}
\usepackage{textgreek}
\usepackage{cancel}
\usepackage{tikz}
\usepackage{hyperref}

\colorlet{origcolor}{.}
\definecolor{gray65}{gray}{.65}
\definecolor{ctcolorblack}{gray}{0}
\definecolor{ctcolorgray}{gray}{.5}
\definecolor{ctcolorgraylight}{gray}{.8}
\definecolor{ctcolorgraylighter}{gray}{.95}

\newcommand{\hdashline}{\arrayrulecolor{gray65}\midrule\arrayrulecolor{origcolor}}

\usepackage{listings}	
\newcommand{\prompt}[1]{
\vskip 0.25truecm
\noindent
\quad{\tt MG5\_aMC>}\ {\tt ~#1}
\vskip 0.25truecm
\noindent}
\newcommand{\madgraph}{\texttt{MadGraph5\_aMC@NLO}}
\newcommand{\mgshort}{\texttt{MG5\_aMC}}
\newcommand{\madsons}{\texttt{MadSONS}}
\newcommand{\madonia}{\texttt{MadOnia}}
 
\newcommand{\helaconia}{\texttt{HELAC-Onia}}
\newcommand{\ufo}{\textsc{UFO}}

\newcommand{\pythia}{\texttt{Pythia}}

\newcommand{\herwig}{\texttt{Herwig}}
\newcommand\ident{{\cal I}}
\newcommand\Ione{\ident_1}
\newcommand\Itwo{\ident_2}
\newcommand\swave{\ensuremath{\mathrm{S}}}
\newcommand\pwave{\ensuremath{\mathrm{P}}}

\renewcommand{\d}{\mathrm{d}}
\newcommand{\e}{\mathrm{e}}
\newcommand\swavetxt{\swave-wave}
\newcommand\pwavetxt{\pwave-wave}

\title{
Automated NRQCD and NRQED simulations of quarkonium and leptonium production with P-wave states and physical-mass effects
}
\author{Luca Maxia,}
\emailAdd{maxia@lpthe.jussieu.fr}
\author{Hua-Sheng Shao,}
\emailAdd{huasheng.shao@lpthe.jussieu.fr}
\author{and Lukas Simon}
\emailAdd{lsimon@lpthe.jussieu.fr}

\affiliation{Laboratoire de Physique Th\'eorique et Hautes Energies (LPTHE), UMR 7589, Sorbonne Universit\'e et CNRS, 4 place Jussieu, 75252 Paris Cedex 05, France}

\abstract{
We present the {\madsons} module, which extends the \madgraph\ framework by providing a fully automated solution for the simulation of arbitrary tree-level production processes involving non-relativistic bound states. More specifically, we focus on the production of quarkonia and leptonia in non-relativistic QCD and QED, respectively. This work constitutes the next step in the programme initiated in ref.~\cite{ColpaniSerri:2025vdz}, where the colour and spin projectors required for \swavetxt\ states were first incorporated into \madgraph. In the present development, we implement the remaining projectors associated with orbital and total angular momentum, thereby enabling the event generation of processes involving \pwavetxt\ states. The derivatives required for the orbital-angular-momentum projection are evaluated using dual numbers. In addition, we implement a momentum-reshuffling procedure that accounts for physical bound-state mass effects in the phase space. The resulting framework applies to a wide range of collider environments while remaining compatible with standard multi-purpose Monte Carlo event generators for parton showering and hadronisation. As a first phenomenological application, we revisit $J/\psi + \psi(2\swave)$ production at the LHC and obtain an improved description of the LHCb data.
}

\begin{document} 
\maketitle
\flushbottom

\section{Introduction}

Quarkonium and leptonium systems constitute unique laboratories for investigating bound-state dynamics in quantum field theory.
Quarkonia probe both short- and long-distance effects of the strong interaction, making their production and decay mechanisms a long-standing testing ground for theoretical descriptions of hadronic bound states. The production and decay of quarkonia are studied in a variety of experimental environments, with measurements performed,
for instance, at electron-positron, lepton-hadron, and hadron-hadron colliders. For reviews, we refer the reader to refs.~\cite{Brambilla:2010cs,Lansberg:2019adr,Chapon:2020heu,Boer:2024ylx}. 
Leptonia, on the other hand, have not yet been observed at high-energy collider facilities. Positronium, the first predicted “onium” state in 1931~\cite{Mohorovicic1934}, was experimentally observed in 1951 through its formation in low-pressure gases~\cite{Deutsch:1951zza}.
Nowadays, these exotic systems serve, for instance, in low-energy experiments for precision studies of electromagnetic interactions~\cite{Karshenboim:2005iy} and for tests of the equivalence principle with antimatter~\cite{AEgIS:2023lpw}. Moreover, they can be exploited to investigate possible violations of discrete spacetime symmetries such as CPT invariance~\cite{Bernreuther:1988tt,Yamazaki:2009hp}. Here, C, P, and T denote charge conjugation, parity, and time reversal, respectively.

Non-relativistic QCD (NRQCD)~\cite{Bodwin:1994jh} is a rigorous low-energy effective field theory of QCD, which
enables systematic improvements in the calculation of quarkonium production and decay rates and remains, to date, the most widely adopted theoretical framework for quarkonium studies. Similarly, its QED counterpart, non-relativistic QED (NRQED)~\cite{Caswell:1985ui}, provides a framework for understanding leptonium states.
Within these frameworks, production cross sections are factorised into short-distance coefficients and long-distance matrix elements (LDMEs). Within these non-relativistic quantum field theories (NRQFTs), the spectroscopy, production rates, and decay rates of quarkonia and leptonia admit a controlled expansion in the relative velocity $v$ of their constituents.
This expansion organises the underlying interactions according to distinct power-counting schemes, commonly referred to as velocity-scaling rules.
While the so-called \swavetxt\ configurations, characterised by vanishing intrinsic orbital angular momentum, are among the most widely studied states both theoretically and experimentally, they are not the only ones of interest. States with non-vanishing orbital angular momentum provide complementary probes to test the factorisation of the employed NRQFTs and the validity of the velocity expansion. Among these, we focus in this work on \pwavetxt\ states, which possess one unit of orbital angular momentum.

In heavy-quarkonium physics, \pwavetxt\ states offer direct sensitivity to spin-orbit and tensor interactions in the heavy-quark potential and have long served as precision probes of quarkonium spectroscopy. Beyond this, they also play a central role in the theoretical description of quarkonium production within NRQCD. 
For instance, the transverse-momentum spectrum of inclusive $J/\psi$ hadroproduction receives significant contributions from colour-octet (CO) \pwavetxt\ channels, despite their LDMEs being suppressed relative to the colour-singlet (CS) \swavetxt\ state.
Moreover, the production of \pwavetxt\ states highlights important subtleties of NRQCD factorisation, such as the cancellation of infrared (IR) divergences between short-distance coefficients and LDMEs~\cite{Bodwin:1992ye,Bodwin:1994jh,Petrelli:1997ge,AH:2024ueu}.
Experimentally, \swavetxt\ quarkonium yields may receive significant feed-down contributions from decays of \pwavetxt\ states. A precise description of these yields therefore requires accurate modelling of \pwavetxt\ production and decay mechanisms.
{\pwavetxt}s are thus indispensable for understanding inclusive quarkonium production in high-energy particle collisions and remain a key ingredient in global fits of CO LDMEs~\cite{Ma:2010vd,Ma:2010yw,Butenschoen:2010rq,Ma:2010jj,Butenschoen:2011yh,Butenschoen:2012px,Chao:2012iv,Wang:2012is,Gong:2012ug,Shao:2012fs,Butenschoen:2012qr,Gong:2013qka,Shao:2014fca,Bodwin:2014gia,Shao:2014yta,Han:2014kxa,Feng:2015wka,Bodwin:2015iua,Feng:2018ukp,Feng:2020cvm,Brambilla:2021abf,Brambilla:2022rjd,Butenschoen:2022qka,Brambilla:2022ayc,Brambilla:2024iqg}.

In the QED sector, leptonium systems, such as positronium, muonium, and their heavier siblings, provide a complementary and theoretically cleaner arena in which analogous \pwavetxt\ states can be studied. Governed solely by QED, these systems allow for high-precision theoretical predictions without the complications of confinement or non-perturbative hadronic effects. Experimentally, \pwavetxt\ leptonic bound states have already been observed in low-energy experiments~\cite{PhysRevLett.34.177,PhysRevLett.102.133202}. Theoretically, their properties are well understood within NRQED and can be calculated with high precision~\cite{dEnterria:2022alo}, making them promising probes of higher-order QED radiative corrections, fundamental Standard Model (SM) constants such as the fine-structure constant and lepton masses~\cite{dEnterria:2023yao}, as well as potential physics beyond the SM (BSM).

Given their relevance, public tools that support \pwavetxt\ generation to some extent are available. However, owing to the increased theoretical and computational complexity of \pwavetxt\ calculations, such tools are fewer in number than those publicly available for \swavetxt\ production. In general-purpose Monte Carlo (MC) event generators, \pwavetxt\ quarkonia can be generated using parton showers in \pythia~{\tt v8.3}~\cite{Cooke:2023ldt} and \herwig~{\tt v7.4}~\cite{Masouminia:2025kec}. In addition, \pythia\ provides matrix elements for a limited set of $2\to 2$ single \pwavetxt\ quarkonium production processes~\cite{Sjostrand:2000wi,Sjostrand:2006za,Sjostrand:2007gs,Sjostrand:2014zea,Bierlich:2022pfr}. 
More specialised implementations include exclusive double-diffractive $\chi_{Q}$ production in the \texttt{SuperChic} generator~\cite{Khoze:2000jm,Harland-Lang:2009kjv}, as well as simulations based on the phenomenological Wigner density matrix formalism in \texttt{EPOS4}~\cite{Song:2017phm,Villar:2022sbv,Zhao:2025cnp}, which do not rely on NRQCD factorisation. A dedicated parton-level generator, \texttt{BCVEGPY}~{\tt v2.0}~\cite{Chang:2005hq}, provides leading-order (LO) simulations of inclusive hadroproduction of excited \pwavetxt\ $B_c$ states, while \texttt{FDCHQHP}~\cite{Wan:2014vka} enables next-to-leading order (NLO) calculations of differential cross sections for inclusive \swave- and \pwavetxt\ charmonium and bottomonium production at hadron colliders.
Besides these largely process-specific tools, two process-independent implementations are available, \madonia~\cite{Artoisenet:2007qm} and \helaconia~\cite{Shao:2012iz,Shao:2015vga}. 
They provide parton-level event generation for processes involving \swave- and \pwavetxt\ quarkonia at tree level within NRQCD. 
\madonia, based on \texttt{MadGraph/MadEvent}~{\tt v4}~\cite{Alwall:2007st} and no longer actively maintained, is restricted to final states containing at most one \swave- or \pwavetxt\ quarkonium. By contrast, \helaconia, built upon \texttt{HELAC-PHEGAS}~\cite{Kanaki:2000ey,Papadopoulos:2000tt,Kanaki:2000ms,Papadopoulos:2006mh}, is more flexible and imposes no such restriction on the number of final-state quarkonia.\footnote{In the current public versions of \helaconia, the number of \pwavetxt\ quarkonia cannot exceed two, while no restriction applies to \swavetxt\ states. This limitation is expected to be lifted in future versions.}
Both programs generate unweighted LO events that can be interfaced with general-purpose MC generators via the Les Houches Event (LHE) file format~\cite{Alwall:2006yp}. Finally, there are also public software packages, such as {\tt FeynOnium}~\cite{Brambilla:2020fla} and {\tt AmpRed}~\cite{Chen:2024xwt}, for semi-automatic symbolic or analytical calculations of matrix elements  within NRQFTs.

Despite these considerable efforts, the aforementioned tools still present important limitations, the most common of which is their restriction to LO calculations. This is a notable shortcoming, since LO predictions are often insufficient to meet the precision requirements of modern experiments, and higher-order perturbative corrections can be sizeable. While process-by-process calculations remain possible, they lack the high degree of automation achieved for processes involving only elementary particles. Automated tools are therefore highly desirable to simplify calculations involving bound states.
Within the \madgraph\ framework~\cite{Alwall:2014hca,Frederix:2018nkq} (hereafter referred to as \mgshort), a programme is underway to overcome such limitations, aiming to develop an all-encompassing tool for bound-state production studies. The first necessary step is to extend \mgshort\ to simulate arbitrary \swave- and \pwavetxt\ quarkonium and leptonium production processes at tree-level within NRQCD and NRQED.
The programme was initiated in an earlier work~\cite{ColpaniSerri:2025vdz}, where the \swavetxt\ implementation was presented. 
Here, we upgrade this framework by automating the generation of events with an arbitrary number of \pwavetxt\ states, thereby completing the LO automation of \swave- and \pwavetxt\ quarkonium and leptonium production processes.
Our implementation of the \pwavetxt\ states and the associated projectors relies on dual numbers~\cite{Clifford:1871aaa}; this choice differs from those adopted in the literature and is, a priori, well suited for further extensions towards NLO calculations.
The \swave- and \pwavetxt\ implementations are combined within a unified automated framework, denoted as the \textcolor{teal}{\madsons} (\texttt{\textcolor{teal}{Mad}Graph} \textcolor{teal}{S}imulations \textcolor{teal}{O}f \textcolor{teal}{N}RQFT \textcolor{teal}{S}tates) module.

As a second novel development in this module, we introduce an improved treatment of quarkonium masses, which are conventionally approximated within NRQCD by the sum of the masses of the heavy quark-antiquark pair. 
This approximation is required to preserve fundamental properties of the $S$-matrix in the calculation of short-distance amplitudes.
To account for the difference between the physical quarkonium mass and the sum of the constituent masses, we propose to retain the physical quarkonium masses during phase-space generation. The resulting phase space is then mapped onto parton-level four-momenta used for matrix-element evaluation through a momentum-reshuffling procedure, a technique widely used in high-energy MC tools, for instance in parton-shower algorithms~\cite{Bengtsson:1987kr,Marchesini:1987cf}. This strategy offers several advantages. 
First, strictly speaking, a consistent treatment of physical mass effects would require the inclusion of higher-order relativistic corrections in both the matrix element and the phase space. While their inclusion in the former is highly non-trivial, the present method provides a simple way to incorporate them in the latter, thereby improving MC simulations near threshold with minimal effort. Second, without momentum reshuffling, generated events assign a common mass to all quarkonia composed of the same heavy quarks.\footnote{For instance, in $J/\psi + \psi(2\swave)$ production, the two physical masses differ by approximately $590\,{\rm MeV}$, yet both quarkonium masses are taken to be twice the chosen charm-quark mass.} As a result, variations in the constituent-quark mass can introduce sizeable uncertainties in the predicted cross sections. Fixing the phase space through momentum reshuffling is, a priori, expected to reduce this dependence, particularly in threshold regions. Finally, the approach automatically accounts for mass differences between the produced quarkonia, feed-down contributions from heavier states, and CO intermediate states, which can be taken heavier than the corresponding quarkonium state, thereby matching, for instance, the hadronisation models used in \pythia{\tt 8}. This, in turn, enhances the automation of interfaces to general-purpose MC generators.

The remainder of the paper is organised as follows. In section~\ref{sec:nrqft}, we review the NRQFT factorisation framework underlying NRQCD and NRQED, which provides the foundation for our implementation. The technical details of the implementation are presented in section~\ref{sec:implementation}. Validation tests and benchmark studies are discussed in section~\ref{sec:benchmark}, followed by phenomenological applications to quarkonium and leptonium production in sections~\ref{sec:quarkonium} and~\ref{sec:leptionium}, respectively. Our conclusions are given in section~\ref{sec:conclusions}. Unless explicitly stated otherwise, all numerical computations presented in this work are carried out using the default setup described in appendix~\ref{app:setup}.

\section{Theoretical framework}\label{sec:nrqft}

We base our calculation of quarkonium and leptonium yields on the NRQCD~\cite{Bodwin:1994jh} and NRQED~\cite{Caswell:1985ui} approaches. 
We follow a notation similar to that used in refs.~\cite{AH:2024ueu, ColpaniSerri:2025vdz}, based on the covariant projection method to single out each contributing Fock state $n$ \cite{Kuhn:1979bb,Guberina:1980dc,Berger:1980ni,Petrelli:1997ge,Maltoni:1997pt}. Fock states are denoted in spectroscopic notation as $n = {}^{2 S + 1} L_J^{[C]}$, where $S$ is the spin of the pair, $L$ its orbital angular momentum, $J$ the total angular momentum, and $C$ the colour configuration, with $C = 1$ indicating a CS state and $C = 8$ a CO state. While ref.~\cite{ColpaniSerri:2025vdz} implemented bound states $\rm B$ for quarkonia (${\rm B} = {\cal Q}$) and leptonia (${\rm B} = {\cal L}$) resctricted to $L=0$, in this paper we extend the framework to states with $L=1$. This upgrade constitutes the first iteration of the {\madsons} module, enabling fully differential cross-section computations and parton-level event generation within {\mgshort} for tree-level processes involving the production of an arbitrary number of NRQFT bound states. 
In particular, all phenomenologically relevant states up to order $v^7$ in the NRQCD expansion (see table~\ref{tab:NRQCD_scaling}) can be computed, whereas for leptonia only the LO contribution in $v$ is included.\footnote{As QED states are purely CS, the expansion in $v$ involves relativistic corrections, whose implementation goes beyond the scope of this work.}
In the remainder of the manuscript, for a bound state $\rm B$ formed by the pair ${\cal C} {\cal C}^\prime$, the notation ${\cal B}[n]$ denotes that the projected state ${\cal C C}^\prime [n]$ has selected the transition channel to $\rm B$, whose probability is encoded in the LDME $\langle {\cal O}_n^{\rm B} \rangle$. 

\begin{table}[t!]
\centering
\renewcommand*{\arraystretch}{1.3}
\begin{tabular}[t]{ccccc}
\toprule
\parbox[c]{2cm}{\centering \textbf{power}\\\textbf{counting}}  & \multirow{1}{*}{$\eta_Q,B_c^{\pm}$} & \multirow{1}{*}{$\psi,\Upsilon,B_c^{\ast \pm}$} & \multirow{1}{*}{$h_Q,B_{c1}(L)^{\pm}$} & \multirow{1}{*}{$\chi_{QJ},B_{cJ}^{\ast\pm},B_{c1}(H)^{\pm}$} \\
\midrule
$v^3$ & $^1\swave_0^{[1]}$ & $^3\swave_1^{[1]}$ & --- & --- \\
$v^5$ & --- & --- & $^1\pwave_1^{[1]},{}^1\swave_0^{[8]}$ & $^3\pwave_J^{[1]},{}^3\swave_1^{[8]}$ \\
$v^7$ & $^1\swave_0^{[8]},{}^3\swave_1^{[8]},{}^1\pwave_1^{[8]}$ & $^1\swave_0^{[8]},{}^3\swave_1^{[8]},{}^3\pwave_J^{[8]}$ & --- & --- \\
\bottomrule
\end{tabular}
\caption{\it\small Hierarchy of Fock-state contributions to different quarkonium states, organised according to the leading powers in the NRQCD velocity scaling~\cite{Bodwin:1994jh}. Note that $J=0,1,2$ for $\chi_{QJ}$ mesons, whereas only $J=0,2$ occur for $B_{cJ}^{\ast \pm}$, in agreement with the PDG notation~\cite{ParticleDataGroup:2024cfk}.\textsuperscript{\ref{fnt:bc_mesons}}}
\label{tab:NRQCD_scaling}
\end{table}
\addtocounter{footnote}{1}\footnotetext{For the \pwavetxt{} CS states of the charmed bottom mesons, we adopt the same naming convention used for the physical mesons, although we refer to the corresponding eigenstates in the Fock basis. Accordingly, the $B_{c1}(L)^\pm$ and $B_{c1}(H)^\pm$ states correspond to pure ${}^1\pwave_1^{[1]}$ and ${}^3\pwave_1^{[1]}$ configurations, respectively, without mixing, unlike the corresponding physical mass eigenstates~\cite{Davies:1996gi,LHCb:2025uce}.
\label{fnt:bc_mesons}}

For completeness, we summarise the key steps in the computation of a generic $2 \to {\rm N}$ cross section. 
To simplify the discussion, we consider the case in which only the last particle, with index “${\rm N}+2$”, is a bound state of elementary particles ${\cal C}$ and ${\cal C}^\prime$. The derivation can be straightforwardly generalised to an arbitrary number of bound states and to arbitrary positions in the final state.
We denote by $\d \widetilde \sigma$ the differential cross section at the event-generation level
\begin{equation}
    \d \widetilde \sigma_{{\rm N},{\cal B}[n]} \equiv \d\widetilde{\sigma}( {\rm A}_1 + {\rm A}_2 \rightarrow \ident_3 + \dots + \ident_{{\rm N}+1} + {\cal B}[n] + \widetilde{X}_p)\,.
\label{eq:mg5_xsec_output}
\end{equation}
Note that eq.~\eqref{eq:mg5_xsec_output} corresponds to the {\mgshort} output. 
The tilde notation ($\d \widetilde \sigma$) distinguishes this quantity from the fully hadronic cross section $\d \sigma$, which includes additional (non-)perturbative contributions in the final state, such as soft radiation (showering) and the hadronisation of elementary particles into composite ones.
For clarity, if the eventual (soft) gluon emission associated with a state $n$ serves only to neutralise colour and match the quantum numbers of ${\cal B}[n]$ to those of ${\rm B}$, then summing eq.~\eqref{eq:mg5_xsec_output} over $n$ yields the cross section for producing ${\rm B}$ at the event-generation level.
In eq.~\eqref{eq:mg5_xsec_output}, the initial state is defined at the “hadronic” level, with ${\rm A}_i$ denoting a nucleus/ion, a QCD meson or baryon, or an electroweak elementary particle.
The final state, instead, consists of $\ident_i$, which are elementary particles in the theory, and ${\cal B}[n]$, which is defined above.
$\widetilde{X}_p$ represents beam remnants and additional parton-level initial-state radiation. 
The purely partonic counterpart of eq.~\eqref{eq:mg5_xsec_output} is given by
\begin{equation}
    \d \hat \sigma_{{\rm N},{{\cal C C}^\prime}[n]} \equiv \d\hat{\sigma}\big( \Ione + \Itwo \rightarrow \ident_3 + \dots + \ident_{{\rm N}+1} + {{\cal C C}^\prime}[n]\big)\,.
\label{eq:partonic_xsec}
\end{equation}
In the initial state, each ${\rm A}_i$ is replaced by its daughter $\ident_i$, which are elementary particles carrying a light-cone fraction $x_i$ of the momentum of their mother particle.
In the final state, ${\cal B}[n]$ is replaced by the projection of the meson constituent pair onto the state $n$, which has not yet been multiplied by the LDME. The transition from the pure partonic cross section to the event-generation-level cross section is achieved through the relation
\begin{equation}
    \d \widetilde \sigma_{{\rm N},{\cal B}[n]} = \sum_{\Ione, \Itwo}\int \d x_1\, \d x_2 f_{\Ione/\mathrm{A}_1}(x_1)\, f_{\Itwo/\mathrm{A}_2}(x_2)\, \d\hat\sigma_{{\rm N},{{\cal CC}^\prime}[n]}\, \langle \mathcal O_{n}^{\rm B} \rangle \,.
\label{eq:from_part-xsec_to_mg5-xsec}
\end{equation}
The sum runs over all partons $\ident_j$ that can be found inside the initial particle ${\rm A}_j$. The partonic cross section is therefore convoluted with the parton distribution functions (PDFs) $f_{\ident_j/\mathrm{A}_j}$ and weighted by the LDME $\langle \mathcal O_{n}^{\rm B}\rangle$.
The former measures the probability of finding $\ident_j$ inside ${\rm A}_j$, carrying a fraction $x_j$ of the mother momentum. It is understood that $f_{\ident_j/\mathrm{A}_j}(x_j) = \delta(1 - x_j)$ if $\ident_j = {\rm A}_j$ and no additional radiation is present. Eq.~\eqref{eq:partonic_xsec} is related to the squared amplitude of the $2{\to}{\rm N}$ process in which the elementary-particle pair ${\cal C C}^\prime$ is projected onto the state $n$. To show this, we first define the partonic $2{\to}{\rm N}+1$ process describing the open production of ${\cal C}$ and ${\cal C}^\prime$ as
\begin{equation}
    r \equiv \Ione (k_1)+\Itwo(k_2) \to \ident_3(k_3) + \dots + \ident_{{\rm N}+2}(k_{{\rm N}+2}) + \ident_{{\rm N}+3}(k_{{\rm N}+3})\,,
\end{equation}
where the four-momentum of each particle is given in parentheses, and we identify ${\cal C} = \ident_{{\rm N}+2}$ and ${\cal C}^\prime = \ident_{{\rm N}+3}$. When ${\cal C}$ and ${\cal C}^\prime$ are projected onto the state $n$, the pair is treated as a single particle with momentum $K = k_{{\rm N}+2} + k_{{\rm N}+3}$, and the partonic process becomes
\begin{equation}
    \dot r = r^{\ident_{{\rm N}+2} \bigoplus \ident_{{\rm N}+3},\cancel\ident_{{\rm N}+3}}\,.
\label{eq:mg5_process_definition}
\end{equation}
This notation follows that adopted in \mgshort~\cite{Frederix:2009yq}, where the “$\bigoplus$” symbol indicates that two elementary particles are bound together, while the slash marks the corresponding elementary particle as removed from both the phase space and the matrix element. The LO amplitudes associated with the processes $r$ and $\dot r$ are, respectively, ${\cal A}^{({\rm N}+1,0)}(r)$ and ${\mathds A}^{({\rm N},0)}(\dot{r})$. The latter is obtained from the former by applying the colour (${\mathds P}_C$),\footnote{Relevant only for quarkonia. We explicitly retain the colour projector in the following to keep the notation unified, although it is understood to reduce to the identity for leptonia.} spin (${\mathds P}_S$), orbital (${\mathds P}_L$), and total angular-momentum (${\mathds P}_J$) projectors
\begin{equation}
    {\mathds A}^{({\rm N},0)}(\dot{r}) = {\cal A}^{({\rm N}+1,0)}_{\left\{[C],S,L,J\right\}}(r) = \sum_{\lambda_J} \left(\sum_{\lambda_L,\lambda_S} {\mathds P}_J \left[ {\mathds P}_L {\mathds P}_S {\mathds P}_{[C]} {\cal A}^{({\rm N}+1,0)}(r)\right]_{q = 0} \right),
    \label{eq:projected_amplitude}
\end{equation}
where $q = |k_{{\rm N}+2} - k_{{\rm N}+3}|/2$ is the relative velocity of the pair, and $\lambda_J$, $\lambda_L$, and $\lambda_S$ are the magnetic quantum number associated with $J$, $L$, and $S$, respectively.
The explicit forms of the projectors introduced in eq.~\eqref{eq:projected_amplitude} are in order.
\begin{itemize}
    \item
    The colour projector for CS ($C=1$) and CO ($C=8$) states is
        \begin{equation}
            {\mathds P}_{[C]} =
                \begin{cases}
                    \delta_{c_1 c_2}/\sqrt{N_c} & \quad {\rm for}\ C=1\,, \\
                    \sqrt{2}\, t_{c_1 c_2}^{c_{12}} & \quad {\rm for}\ C=8\,, \\
                \end{cases}
        \label{eq:proj_colour2}
        \end{equation}
        where $N_c = 3$ in QCD, $c_1$ and $c_2$ are the colour indices of ${\cal C}$ and ${\cal C}^\prime$, and $t_{ij}^a$ denotes the $\mathrm{SU}(3)$ generators in the fundamental representation.
    \item
    The spin projector for spin singlet ($S=0$) and triplet ($S=1$) states is
        \begin{equation}
            {\mathds P}_{S} = \dfrac{\bar{v}_{\lambda_{{\cal C}^\prime}}\, \Gamma_S\, u_{\lambda_{{\cal C}}}}{2 \sqrt{2\, m_{{\cal C}} m_{{\cal C}^\prime}}}\,,\quad {\rm where}\ \Gamma_S =
            \begin{cases}
                \gamma_5 & \quad {\rm for}\ S=0\,, \\
                \slashed{\varepsilon}_{\lambda_S}^*(K) & \quad {\rm for}\ S=1\,, \\
            \end{cases}
        \label{eq:proj_spin2}
        \end{equation}
        the constituent helicities satisfy $\lambda_{{\cal C^{(\prime)}}} = \pm 1/2$, the spin magnetic quantum number takes the values $\lambda_S=0,\pm1$, and $\varepsilon$ is the polarisation vector associated with the outgoing bound state. The quantities $u$ and $v$ denote the spinors of the constituent particle and antiparticle, respectively.
    \item
    The orbital angular-momentum projector is
        \begin{equation}
            {\mathds P}_{L} = \bigg( \varepsilon_{\lambda_L}^{*\mu}(K) \frac{{\rm d}}{{\rm d}q^\mu}\bigg)^L\,,
        \label{eq:proj_oam}
        \end{equation}
        with $\lambda_L = 0, \pm 1$ for $L=1$. The derivative acts on all terms to its right, including both ${\cal A}^{({\rm N}+1,0)}(r)$ and ${\mathds P}_{S}$, as indicated in eq.~\eqref{eq:projected_amplitude}.
    \item 
    The total angular-momentum projector is given by the Clebsch-Gordan coefficients
        \begin{equation}
            {\mathds P}_{J} = \langle L, \lambda_L; S, \lambda_S | J, \lambda_J \rangle\,,
        \label{eq:proj_tam}
        \end{equation}
        with $J$ taking values in the range $\vert L-S\vert\le J \le L+S$ in integer steps, and $\lambda_J = \lambda_S + \lambda_L$.\\
\end{itemize}
Finally, the explicit LO relation between $\d\hat\sigma_{{\rm N},{\cal C}{\cal C}^\prime[n]}$ and $\mathds{A}^{({\rm N},0)}(\dot{r})$ is 
\begin{align}
   \d\hat\sigma_{{\rm N},{\cal C}{\cal C}^\prime[n]} & =
   \frac{\d {\phi_{\rm N}}(\dot{r})}{{\cal N}(\dot{r})}\!\left(\frac{1}{4k_1\cdot k_2} \frac{1}{\omega(\Ione)\omega(\Itwo)} \right)\! \frac{1}{(2J+1)N_{[C]}} {\frac{1}{2\mu_{\cal B}}}
   \sum_{\substack{\rm colour\\ \rm spin}} |  {\mathds A}^{({\rm N},0)}(\dot{r})|^2\,,
\label{eq:partonic_xsec_explicit}
\end{align}
where $\mu_{{\cal B}} = (m_{{\cal C}}m_{{\cal C}^\prime})/(m_{{\cal C}}+m_{{\cal C}^\prime})$ is the reduced mass of the ${\cal C C}^\prime$ system. 
The $\d{\phi_{\rm N}}(\dot r)$ factor denotes the ${\rm N}$-body phase-space measure of the process $\dot r$, while ${\cal N}(\dot{r})$ is the final-state symmetry factor. The quantity $N_{[C]}$ is a colour factor, with $N_{[1]} = 2 N_c$ and ${N_{[8]} = N_c^2-1}$ for quarkonia, and $N_{[1]} = 1$ for leptonia. Moreover, $\omega(\ident_i)$ accounts for the initial-state colour and spin averaging at the $\dot{r}$ level, while $4k_1\cdot k_2 \approx 2\hat s$ corresponds to the flux factor.

\section{Implementation}
\label{sec:implementation}

\subsection[{The {\tt sm\_onia} {\ufo} model}]{The {\ttfamily{\fontseries{b}\selectfont sm\_onia}} {\ufo} model}

\begin{lstlisting}[caption={\it\small Definitions of the $\chi_{c2}$ meson Fock state $^3\pwave_2^{[1]}$ in {\tt boundstates.py} of the updated \ufo\ model {\tt sm\_onia}.},captionpos=b,language=Python,label={lst:boundstates.py},float=tp,floatplacement=tbp]
# chi meson
chic2_13p21 = Boundstate(
    pdg_code = 445,
    name = 'chic2(1|3P21)',
    particles = ['c','c~'],
    mass = 3.56,
    principal = 1,
    spin = 3,
    orbital = 1,
    J = 2,
    color = 1,
    charge = 0,
    ldme = 0.5371,
    texname = 'chic213P21'
)
\end{lstlisting}

Alongside this article, an updated version of the Universal Feynman Output (\ufo)~\cite{Degrande:2011ua,Darme:2023jdn} model \texttt{sm\_onia} is provided. 
The model is part of the \mgshort\ suit and is automatically downloaded if it is not already present in the local installation when the command
\prompt{import~model~sm\_onia}\noindent
is called.
The elementary particle content of this model, together with all associated default parameter values, is identical to that of the {\tt sm} \ufo\ model, which is loaded automatically when \mgshort\ is launched. Consequently, by default, \texttt{sm\_onia} features zero lepton masses and adopts a four-flavour scheme ($4$FS) with a massless charm quark. To generate charmonia, charmed bottom mesons, and leptonia, additional model restrictions are needed. Specifically, 
\prompt{import~model~sm\_onia-c\_mass}\noindent
sets a non-zero mass for the charm quark, while
\prompt{import~model~sm\_onia-lepton\_masses}\noindent
is used for leptons. The default masses are reported in table~\ref{tab:sm_input} of appendix~\ref{app:setup}.

In \texttt{sm\_onia}, the Fock states associated with a quarkonium or leptonium state are defined in the \texttt{Boundstate} class and provided in the dedicated file \texttt{boundstates.py}. An example of an implemented Fock-state configuration for $\chi_{c2}$ is given in listing~\ref{lst:boundstates.py}.
In this way, the generation of processes involving bound states follows the same logic as for ordinary elementary particles, namely by specifying either the Fock-state \texttt{pdg\_code} or its \texttt{name}. In both cases, we adopt the same convention as that used in ref.~\cite{ColpaniSerri:2025vdz}.
The Fock-state PDG numbering scheme follows the convention established in \pythia{\tt 8}~\cite{Bierlich:2022pfr} and is consistent with the PDG MC guidelines~\cite{ParticleDataGroup:2024cfk} (cf.\@ section~$45$ therein). 
The only additional feature introduced here concerns the updated number $n_s$ used for CO states, which identifies their spin and orbital quantum numbers: \texttt{0} for $^3\swave_1^{[8]}$, \texttt{1} for $^1\swave_0^{[8]}$, \texttt{2} for $^3\pwave_J^{[8]}$, \texttt{3} for $^3\pwave_0^{[8]}$, \texttt{4} for $^3\pwave_1^{[8]}$, \texttt{5} for $^3\pwave_2^{[8]}$, and \texttt{6} for $^1\pwave_1^{[8]}$.
If a particle is specified by its {\tt name} without selecting a particular Fock state, the complete collection of states up to order $v^7$ is included automatically. Furthermore, for Fock states with $L=1$ and $S=1$, we allow users to collectively denote ${}^{3}\pwave_J^{[8]}$ contribution using the notation \texttt{name(N|3PJ8)}, where \texttt{N} is the energy level of the particle (the principal number number for leptonia and the radial quantum number for quarkonia). For example, \texttt{jpsi(1|3PJ8) = jpsi(1|3P08) jpsi(1|3P18) jpsi(1|3P28)} for the $J/\psi$.
In the {\tt sm\_onia} model, the same symbol is used for {\swave}- and {\pwavetxt} leptonia; e.g., all {\pwavetxt} positronia adopt the symbol {\tt PS} rather than {\tt he}, {\tt chie0}, {\tt chie1}, and {\tt chie2}.  
Different states are then uniquely distinguished by their quantum-number content, which must always be specified when generating leptonia. Moreover, we point out that, for \pwavetxt\ leptonia, a different convention for the principal number $N$ is adopted, since the lowest-energy state corresponds to $N=2$ rather than $N=1$, in the case of quarkonia. More details on the different conventions will be provided in the dedicated discussion of section~\ref{sec:leptionium}.
The list of bound-state names implemented in the local \mgshort\ installation can be displayed in the command-line interface using the command
\prompt{display~boundstates}\noindent
after loading the \ufo\ model.

\begin{lstlisting}[caption={\it\small Example of the {\tt onia\_card.dat} in case of the production of the $^3\pwave_2^{[1]}$ $\chi_{c2}$ state.},captionpos=b,language=Python,label={lst:onia_card.dat},float=tp,floatplacement=tbp]
#*********************************************************************
# Long distance matrix elements (LDME)                               *
#*********************************************************************
Block ldme
   445  0.5371  # LDME for chic2(1|3P21)
   
#*********************************************************************
# Mass                                                               *
#*********************************************************************
Block onium_mass
   445  3.56    # mass for chic2(1|3P21)
\end{lstlisting}

Compared with the first version of the model introduced in ref.~\cite{ColpaniSerri:2025vdz}, we have added two properties to the \texttt{Boundstate} class: \texttt{mass} and \texttt{ldme}.
Their corresponding default values are reported in tables~\ref{tab:onia_masses},~\ref{tab:ldme_charmonia},~\ref{tab:ldme_bottomonia}, and~\ref{tab:ldme_Bc} of appendix~\ref{app:setup}.
When a process involving at least one Fock state is generated and the {\tt output} command has been executed, each LDME and mass parameter is copied to the \texttt{onia\_card.dat} file, as shown in the example in listing~\ref{lst:onia_card.dat}. The user can modify each parameter to the desired value at runtime following the same procedures used for accessing the {\tt run\_card.dat} and {\tt param\_card.dat}. Explicitly, the commands for setting the LDME and mass parameters are
\prompt{set~onia\_card~ldme~<pdg\_code>~<value>}\noindent
and
\prompt{set~onia\_card~onium\_mass~<pdg\_code>~<value>}\noindent
respectively.

\subsection{Automatic differentiation with dual numbers}

{\pwavetxt} states, in contrast to {\swavetxt} states, are characterised by a vanishing wavefunction at the origin. Hence, their production is controlled by the first derivative of the wavefunction at the origin. In momentum space, the same derivative is encoded in the projector introduced in eq.~\eqref{eq:proj_oam}.
Consequently, the implementation of \pwavetxt{} states in \mgshort{} requires the automated evaluation of derivatives within the amplitude-generation framework.
To this end, we employ dual numbers ${\cal D}$~\cite{Clifford:1871aaa}, which provide an exact and numerically stable solution. These properties are essential in view of a future extension of \madsons\ to NLO accuracy, where the local cancellation between IR-singular real-emission contributions and their corresponding FKS subtraction terms demands high numerical precision. 
This solution is also scalable and can be generalised to higher derivative orders with minimal intervention, allowing for possible future extensions of the package to incorporate states with $L>1$, such as ${\rm D}$-waves, or to provide a systematic treatment of relativistic corrections in the matrix element.
Beyond our specific application, automatic differentiation based on dual numbers has previously been used in a variety of contexts, including machine learning~\cite{Baydin:2015tfa},\footnote{While the dual numbers used here correspond to a dedicated forward-mode automatic differentiation implementation, existing machine-learning-based approaches in the \mgshort{} ecosystem, such as {\tt MadJax}~\cite{Heinrich:2022xfa} and {\tt MadNIS}~\cite{Heimel:2022wyj,Heimel:2023ngj}, rely on reverse-mode automatic differentiation for backpropagation, as provided by frameworks such as {\tt JAX}~\cite{jax2018github} and {\tt PyTorch}~\cite{NEURIPS2019_9015pytorch}.} the computer algebra system {\tt Symbolica}~\cite{Symbolica}, and high-energy physics within the Local Unitarity formalism~\cite{Capatti:2020xjc,Capatti:2022tit} for a constructive, integrand-level realisation of the Kinoshita-Lee-Nauenberg (KLN) theorem~\cite{Kinoshita:1962ur,Lee:1964is}.
We finally emphasise that the techniques used to support {\pwavetxt} quarkonia in \madonia\ and \helaconia\ differ significantly from the dual-number strategy employed in {\tt MadSONS}. Indeed, \madonia, which supports only single-quarkonium production, relies on numerical differentiation, whereas \helaconia\ introduces dedicated {\pwavetxt} currents into the recursion relations. Although the latter approach is exact and numerically stable, its scalability remains limited. For completeness, we briefly describe the rationale behind our implementation of dual numbers in \mgshort{} in this section.

The dual number representation of a complex-valued variable $x(q)$ is defined as
\begin{equation}
    {\cal D}(x) = x + \epsilon\, \frac{\d x}{\d q}\,,
\end{equation}
which consists of a primal component and a dual (or tangent) component. The latter is proportional to the nilpotent quantity $\epsilon$, which satisfies $\epsilon^2 = 0$ while $\epsilon \neq 0$. For our purposes, we employ the multivariate generalisation
\begin{align}
    {\cal D}_n(x) = \sum_{k=0}^n \frac{1}{k!} \left( \sum_{j=1}^n \epsilon_j \frac{\d}{\d q_j} \right)^{\!k} x\,,
\label{eq:mulvar_dual_numb}
\end{align}
in which $n$ directions in the $q_k$ parameter space are considered, each associated with a distinct nilpotent element $\epsilon_j$, satisfying $\epsilon_j^2 = 0$ and ${\epsilon_i\, \epsilon_j \neq 0}$ for $i \neq j$.
For practical implementations, the primal and dual components of eq.~\eqref{eq:mulvar_dual_numb} can be represented by the array
\begin{align}
    {\cal D}_n^{\scriptscriptstyle{\rm arr}}(x) & = \bigg({\cal D}_{n,0}^{\scriptscriptstyle{\rm arr}}(x),{\cal D}_{n,1}^{\scriptscriptstyle{\rm arr}}(x),{\cal D}_{n,2}^{\scriptscriptstyle{\rm arr}}(x),\ldots,{\cal D}_{n,2^{n}-1}^{\scriptscriptstyle{\rm arr}}(x)\bigg) \nonumber\\
    & = \bigg(
        x,\,
        \frac{\d x}{\d q_1},\,
        \frac{\d x}{\d q_2},\,
        \frac{\d^2 x}{\d q_1\d q_2},\,
        \frac{\d x}{\d q_3},\,
        \frac{\d^2 x}{\d q_1\d q_3},\,
        \frac{\d^2 x}{\d q_2\d q_3}, \frac{\d^3 x}{\d q_1\d q_2\d q_3},
        \dots,
        \frac{\d^n x}{\d q_1\dots\d q_n}
    \bigg)\,.
\label{eq:dual_array}
\end{align}
The derivative component associated with the $i^{\mathrm{th}}$ entry, with $i\in\left\{0,1,\ldots,2^{n}-1\right\}$, is determined by the binary decomposition
\begin{equation}
    i = \sum_{k=0}^{n-1} 2^k b_{i,k}\, , 
    \qquad  
    b_{i,k}\in\{0,1\}\, ,
\label{eq:binary_decomposition}
\end{equation}
such that
\begin{equation}
    {\cal D}_{n,i}(x)  = \prod_{k=1}^{n}\left({\epsilon_k^{b_{i,k-1}}} \right){\cal D}_{n,i}^{\scriptscriptstyle{\rm arr}}(x)= \prod_{k=1}^{n} \left(\epsilon_k \frac{\d}{\d q_k} \right)^{b_{i,k-1}} x\, .
\label{eq:dual_component}
\end{equation}
Here, each bit $b_{i,k-1}$ specifies whether the derivative with respect to $q_{k}$ is present in the corresponding mixed derivative.
The full multivariate dual representation is then obtained by summing over all components,
\begin{equation}
    {\cal D}_n(x) = \sum_{i=0}^{2^n-1} {\cal D}_{n,i}(x)\, .
\end{equation}

One advantage of dual numbers is the automatic propagation of derivatives through the chain rule using operator overloading. For elementary operations such as addition and multiplication, the propagation follows directly from the nilpotency of the dual unit:
\begin{align}
    {\cal D}(x + y) & = x + y + \epsilon \left( \frac{\d x}{\d q} + \frac{\d y}{\d q} \right)={\cal D}(x)+{\cal D}(y)\, , \\
    {\cal D}(x{\cdot}y) & = x{\cdot}y + \epsilon \left( \frac{\d x}{\d q}{\cdot}y + x{\cdot}\frac{\d y}{\d q} \right)={\cal D}(x)\cdot{\cal D}(y)\,,
\end{align}
where the $\epsilon^2$ term vanishes due to the nilpotency condition. For non-linear functions, the corresponding derivatives must be implemented explicitly. For a generic function $f(x)$,\footnote{We leave the generalisation of the formula for a multivariable function $f(\vec{x})$ to the reader.} the chain rule gives 
\begin{equation}
    {\cal D}\big( f(x) \big) = f(x) + \epsilon\,  \frac{\d f(x)}{\d q} = f(x) + \epsilon\, \left( \frac{\partial f(x)}{\partial x}\right)\frac{\d x}{\d q}\,.
\label{eq:dual_chain_rule}
\end{equation}
Hence, once the analytic form of the derivative of $f(x)$ is implemented, the derivative propagation proceeds automatically.
Examples of eq.~\eqref{eq:dual_chain_rule} for a general power law, the logarithm, and the cosine are given below:
\begin{align}
    {\cal D}(x^p) & = x^p + \epsilon \left( p\, x^{p-1}\frac{\d x}{\d q} \right) && \Rightarrow && \frac{\partial f(x)}{\partial x} = p\, x^{p-1}\,, \nonumber\\
    {\cal D}(\log{(x)}) & = \log{(x)} + \epsilon \left( \frac{1}{x}\,\frac{\d x}{\d q} \right) && \Rightarrow && \frac{\partial f(x)}{\partial x} = \frac{1}{x}\,, \nonumber\\
    {\cal D}(\cos{(x)}) & = \cos{(x)} + \epsilon \left( -\sin(x)\,\frac{\d x}{\d q} \right) && \Rightarrow && \frac{\partial f(x)}{\partial x} =  -\sin{(x)}\,.\nonumber
\end{align}
The multivariate extension of the dual-number representation in eq.~\eqref{eq:dual_chain_rule} is given by
\begin{align}
    {\cal D}_n\big(f(x)\big) = \sum_{k=0}^n \frac{1}{k!} \left( \sum_{j=1}^n \epsilon_j \frac{\d}{\d q_j} \right)^{\!k} f(x) \,,
\label{eq:mulvar_dual_chain}
\end{align}
which can be represented in array form as
\begin{align}
    {\cal D}_n^{\rm arr}\big(f(x)\big) & = 
    \bigg(
        f(x),\,
        \frac{\d x}{\d q_1}\,\frac{\partial f(x)}{\partial x},\,
        \frac{\d x}{\d q_2}\,\frac{\partial f(x)}{\partial x},\,\nonumber\\ & \phantom{=\bigg(}\quad
        \frac{\d^2 x}{\d q_1\d q_2}\,\frac{\partial f(x)}{\partial x} + \frac{\d x}{\d q_1}\,\frac{\d x}{\d q_2}\,\frac{\partial^2 f(x)}{\partial x^2},
        \frac{\d x}{\d q_3}\,\frac{\partial f(x)}{\partial x},
        \dots
    \bigg),
\label{eq:dual_chain_array}
\end{align}
in analogy with eqs.~\eqref{eq:mulvar_dual_numb} and~\eqref{eq:dual_array}. Here, the array entries represent derivatives of the composite function $f(x(q))$, obtained through repeated application of the multivariate chain rule. Despite these similarities, the binary decomposition alone is not sufficient to directly determine the $i^{\mathrm{th}}$ contribution $\mathcal{D}_{n,i}(f(x))$ of eq.~\eqref{eq:mulvar_dual_chain}. Instead, these contributions can be systematically organised through partitions of the binary-support set associated with a given array index $i$.
Using the binary decomposition introduced in eq.~\eqref{eq:binary_decomposition}, we define the binary-support set
\begin{equation}
    B_i = \left\{2^k \,\middle|\, k \in {\mathbb N}_0\,,\, b_{i,k}=1 \right\}.
\end{equation}
The set of all non-empty subsets of $B_i$ is denoted by
\begin{equation}
    P^\ast\!(B_i) = \left\{ g \,\middle|\, g \subseteq B_i\,, g \neq \varnothing \right\},
\end{equation}
while the corresponding partition set is then defined as
\begin{equation}
    P(B_i) = \left\{ \{g_\alpha\} \,\middle|\, g_\alpha \in P^\ast\!(B_i)\,, \bigcup g_\alpha = B_i\,, \forall\,g_\alpha\neq g_\beta \in P^\ast\!(B_i): g_\alpha\cap g_\beta=\varnothing \right\}.
\end{equation}
For each partition, we further define the partition-sum map
\begin{equation}
    S(P) = \left\{ \left\{ \sum_{\beta\in g}\beta ~\bigg\vert~ g\in p
    \right\}~\Bigg\vert~ p\in P\right\}.
\end{equation}
In other words, for each partition $p\in P$, the map $S$ assigns the set of sums of the elements in each block $g\in p$.
As examples, we explicitly show the derivation of the partition-sum map for the numbers $6$ and $7$, whose binary decompositions are $b_6 = 110$ and $b_7 = 111$, respectively.
For the former, we have
$$B_6 = \{2,4\} \Rightarrow P(B_6) = \{\{\{2,4\}\},\ \{\{2\},\{4\}\}\} \Rightarrow S(P(B_6)) = \{\{6\},\ \{2,4\}\}\,,$$
while for the latter, we have instead
\begin{align*}
    B_7 = \{1,2,4\} 
    & \Rightarrow P(B_7) =\{\{\{1,2,4\}\},\ \{\{1\},\{2,4\}\},\ \{\{2\},\{1,4\}\},\\
    & \phantom{\Rightarrow P(B_7) =\{\ } \{\{4\},\{1,2\}\},\ \{\{1\},\{2\},\{4\}\}\} \\ 
    & \Rightarrow S(P(B_7)) = \{\{7\},\ \{1,6\},\ \{2,5\},\ \{4,3\},\ \{1,2,4\}\}\,.
\end{align*}
Using these definitions, the $i^{\mathrm{th}}$ component of eq.~\eqref{eq:mulvar_dual_chain} can be written as
\begin{equation}
    {\cal D}_{n,i}\big( f(x)\big) = \sum_{s\in S(P(B_i))} \left\{ \left[   
    \left(\frac{\partial}{\partial x}\right)^{|s|}f(x)\right] \left[\prod_{k \in s}  \left( \prod_{j=1}^{n} \left( \epsilon_j \frac{\d}{\d q_{j}}\right)^{\!b_{k,j-1}} x \right) \right] \right\},
\label{eq:dual_chain_component}
\end{equation}
where $|s|$ denotes the number of elements in the set $s$. The full multivariate dual representation is then obtained by summing over all components,
\begin{equation}
     {\cal D}_{n}\big( f(x)\big) = \sum_{i=0}^{2^n-1} {\cal D}_{n,i}\big( f(x)\big)\, .
\end{equation}

For the implementation of the projector in eq.~\eqref{eq:proj_oam} at tree level, only the elementary operations ($+$, $-$, $\times$, $\div$) and general power functions with arbitrary exponents are required. Dual numbers and their overloaded operations are implemented in \mgshort{} as an independent {\tt Fortran} module contained in the file {\tt dual\_variables.f}.
Whenever at least one \pwavetxt{} state is detected during the process-generation stage, \mgshort{} automatically activates the {\tt dual\_mode}. In eq.~\eqref{eq:mulvar_dual_numb}, the parameter $n$ then corresponds to the number of independent dual directions, which in this application is equal to the number of \pwavetxt{} states in the process. Each variable $q_k$ is promoted to a four-vector $q_k^\mu$, associated with the relative momentum of the ${\cal C}_k{\cal C}_k^\prime$ pair forming the $k^{\mathrm{th}}$ \pwavetxt{} state. The derivative propagation is performed component-wise in Lorentz space, while the same dual variable $\epsilon_k$ is shared among all Lorentz components. Within the {\tt MATRIX} routines of the generated {\tt matrix.f} files, the {\tt DUAL} type is introduced wherever required, and the code automatically performs the sums over the orbital- and spin-angular-momentum projections $\lambda_{L}$ and $\lambda_{S}$.
The derivative propagation originates from the initialisation of the partonic momenta. For a process involving $n$ \pwavetxt{} states, the momenta associated with the constituents of the $k^{\mathrm{th}}$ \pwavetxt{} state are initialised as
\begin{equation}
\begin{aligned}
    {\cal D}_{n,0}(k_{{\cal C}_k}^\mu) & = \frac{m_{{\cal C}_k}}{m_{{\cal C}_k}+m_{{\cal C}^\prime_k}} K_k^\mu\,,
    \qquad
    {\cal D}_{n,2^{k-1}}(k_{{\cal C}_k}^\mu) = +\epsilon_{k}\varepsilon_{\lambda_{L_k}}^{*\mu}\!(K_k)\,,
    \\
    {\cal D}_{n,0}(k_{{\cal C}_k^\prime}^\mu) & = \frac{m_{{\cal C}^\prime_k}}{m_{{\cal C}_k}+m_{{\cal C}^\prime_k}}K_k^\mu\,,
    \qquad 
    {\cal D}_{n,2^{k-1}}(k_{{\cal C}_k^\prime}^\mu) = -\epsilon_{k}\varepsilon_{\lambda_{L_k}}^{*\mu}\!(K_k)\,,
\end{aligned}
\label{eq:dual_mom_init}
\end{equation}
while all remaining dual components vanish. This construction directly reproduces the projector definition in eq.~\eqref{eq:proj_oam}.

\subsection{Momentum reshuffling for improved mass treatment}\label{sec:resh}

In conventional perturbative NRQFT calculations, the mass of a bound state $\mathcal{B}$ is approximated by the sum of the masses of its constituents $\mathcal{C}$ and $\mathcal{C}^\prime$, $M_{\mathcal{B}}=m_{\mathcal{C}}+m_{\mathcal{C}^\prime}$. However, this approximation can differ significantly from the physical bound-state mass and thereby introduce a systematic uncertainty into theoretical predictions. The effect is particularly pronounced in kinematic threshold regions and in processes involving multiple bound states with identical constituents but different physical masses, such as the associated production of $J/\psi + \psi(2\swave)$ discussed in section~\ref{sec:quarkonium}, where the conventional treatment assigns the same constituent-mass approximation to both states.
This approximation originates from the requirement that the matrix element in eq.~\eqref{eq:projected_amplitude} must be evaluated with on-shell constituent particles. Nevertheless, a more realistic treatment of the physical bound-state masses is expected to improve the description of observables sensitive to kinematic thresholds and phase-space boundaries. To this end, \mgshort\ allows the user to specify physical bound-state masses through the \texttt{onia\_card.dat} input file, as described previously, cf.\@ listing~\ref{lst:onia_card.dat}. The phase-space integration is then performed using these physical masses, ensuring that each bound state contributes with its measured mass to the available phase-space volume. Meanwhile, through the momentum-reshuffling procedure described in this section, the constituent masses entering the matrix-element level remain fixed to their associated parameter values.
 
We consider a generic $2\to {\rm N}$ process
\begin{equation}
    \dot{r} \equiv \Ione (k_1)+\Itwo(k_2) \to \ident_3(k_3) + \dots + \ident_{{\rm N}+2}(k_{{\rm N}+2})\,.
\end{equation}
The phase-space generator samples ``physical'' momenta $k_i$ satisfying the on-shell conditions $k_i^2=m_i^2$, where $m_i$ denotes the physical mass of the particle $\ident_i$. Since the matrix element must be evaluated at the constituent level, bound-state external particles must instead satisfy the on-shell conditions corresponding to the sum of their constituent masses. We therefore introduce a set of ``reshuffled'' momenta $\overline{k}_i$, satisfying $\overline{k}_i^2=\overline{m}_i^2$, where $\overline{m}_i=m_i$ for elementary particles, while $\overline{m}_i=m_{{\cal C}_i}+m_{{\cal C}^\prime_i}$ for a bound state composed of the constituents ${\cal C}_i$ and ${\cal C}_i^\prime$. Since $m_i \simeq \overline{m}_i$ and the amplitude $\mathds{A}^{({\rm N},0)}(\dot r)$ depends only weakly on small variations in the external momenta, we approximate
\begin{equation}
    \mathds{A}^{({\rm N},0)}(\dot{r}) \approx \mathds{A}^{({\rm N},0)}\left(\Ione (\overline{k}_1)+\Itwo(\overline{k}_2) \to \ident_3(\overline{k}_3) + \dots + \ident_{{\rm N}+2}(\overline{k}_{{\rm N}+2})\right)\,.
\label{eq:resh_matrix_element_approx}
\end{equation}
Whenever this approximation holds, the phase space can be generated using the exact masses $m_i$, while the matrix element is evaluated using the on-shell condition of the elementary particles. 
This translates to an approximated inclusion of higher-order relativistic corrections in the cross section, due to the improved mass treatment in the phase-space measure.

The mapping $\{{k_i}\}\rightarrow\{{\overline{k}_i}\}$ is not unique. We therefore introduce different momentum-reshuffling prescriptions based on either initial-state or final-state recoil schemes.
These constructions follow the same general philosophy as those employed in \mgshort\ for resonance-aware event generation~\cite{Artoisenet:2012st,Frixione:2019fxg}, suitably adapted to non-relativistic bound-state production. Table~\ref{tab:momreshuff} summarises the available momentum-reshuffling options implemented in \mgshort{} and the corresponding values of the \texttt{mom\_resh\_type} parameter to activate them introduced with this update.

To describe the algorithms, it is convenient to distinguish between the momenta of elementary particles and those of bound states. We therefore define
\begin{align}
    &\{p_{i^\prime}\} = \{k_i\, \vert\, \ident_i\in \mathfrak{E}\}\,,\quad \{q_{j^\prime}\} = \{k_j\, \vert\, \ident_j\in\mathfrak{B}\}\,,\\
    &\{\overline{p}_{i^\prime}\} = \{\overline{k}_i\, \vert\, \ident_i\in\mathfrak{E}\}\,,\quad \{\overline{q}_{j^\prime}\} = \{\overline{k}_j\, \vert\, \ident_j\in\mathfrak{B}\}\,,
\end{align}
where $\mathfrak{E}$ ($\mathfrak{B}$) corresponds to the collection of elementary particles (bound states) in the theory, while $i^\prime$ and $j^\prime$ denote relabellings of the indices $i$ and $j$, running from $1$ to ${\rm N}_e$ (${\rm N}_e\geq 2$) for elementary particles and from $1$ to ${\rm N}_b$ (${\rm N}_b\geq 1$) for bound states, respectively.
The relabelling preserves the original relative ordering within each subset, thereby ensuring a unique inverse mapping.
The separation between momenta $p$ and $q$ ensures that all the momenta $k_i$ for which $\overline{m}_i \neq m_i$ belong to the second subset. By construction, if all constituent masses are chosen such that $\overline{m}_i = m_i$ for all $i$, then $\overline{k}_i = k_i$ for all $i$ (and consequently also for all $q_i$).
The reshuffled momenta are required to preserve overall four-momentum conservation, from which we obtain
\begin{align}
    &k_1+k_2 = \sum_{i=3}^{{\rm N}+2} k_i = \sum_{i^\prime=3}^{{\rm N}_e} p_{i^\prime} + \sum_{j^\prime=1}^{{\rm N}_b} q_{j^\prime}=p_1+p_2\,,\\
    &\overline{k}_1+\overline{k}_2 = \sum_{i=3}^{{\rm N}+2} \overline{k}_i = \sum_{i^\prime=3}^{{\rm N}_e} \overline{p}_{i^\prime} + \sum_{j^\prime=1}^{{\rm N}_b} \overline{q}_{j^\prime}=\overline{p}_1+\overline{p}_2\,. 
    \label{eq:mom_tilde}
\end{align}

\begin{table}[t]
    \centering
    \begin{tabular}{|c|l|}
    \hline
        {\tt mom\_resh\_type} & momentum-reshuffling option \\
        \hline
        0 & initial-state reshuffling \\
        1 & final-state reshuffling with smooth $\varphi$ [cf. eq.~\eqref{eq:varphiformomreshtype1}] \\
        2 & final-state reshuffling with step-function $\varphi$ [cf. eq.~\eqref{eq:varphiformomreshtype2}] \\\hline
    \end{tabular}
    \caption{\label{tab:momreshuff}\it\small Possible values of the {\tt mom\_resh\_type} 
parameter and the corresponding momentum-reshuffling options.}
\end{table}

\paragraph{Initial-state momentum reshuffling:}
Working in the hadronic centre-of-mass (c.m.) frame, the physical initial-state momenta of the elementary particles $\Ione$ and $\Itwo$ can be parametrised as
\begin{equation}
    p_1 = x_1\dfrac{\sqrt{s}}{2}(1,0,0,1)^\mathrm{T},\quad p_2 = x_2\dfrac{\sqrt{s}}{2}(1,0,0,-1)^\mathrm{T},
\end{equation}
where $\sqrt{s}$ denotes the hadronic collision energy. The corresponding reshuffled momenta are given by
\begin{equation}
    \overline{p}_1 = \overline{x}_1\dfrac{\sqrt{s}}{2}(1,0,0,1)^\mathrm{T},\quad \overline{p}_2 = \overline{x}_2\dfrac{\sqrt{s}}{2}(1,0,0,-1)^\mathrm{T}.
\end{equation}
In the initial-state reshuffling procedure, only the initial-state momenta $p_1$, $p_2$, and the bound-state momenta $q_{j^\prime}$ are modified, while all other final-state momenta remain unchanged,
\begin{equation}
    \overline{p}_{i^\prime} = p_{i^\prime}\,\ \ \forall\ \ i^\prime \ge 3\,.
\end{equation}
The bound-state momenta are adjusted to satisfy the corresponding on-shell conditions while keeping their spatial components fixed,
\begin{equation}
    \overline{q}_{j^\prime}^{0} = \sqrt{\overline{m}_{j^\prime}^2+\vert \bm{q}_{j^\prime} \vert^2}\,,\quad \overline{\bm{q}}_{j^\prime}^{\phantom{0}} = \bm{q}_{j^\prime}^{\phantom{0}}\,.
\end{equation}
Using eq.~\eqref{eq:mom_tilde}, we then obtain
\begin{align}
    \overline{x}_1 & = \dfrac{1}{\sqrt{s}}\left(\sum_{i^\prime=3}^{{\rm N}_e}p_{i^\prime}^0+\sum_{j^\prime=1}^{{\rm N}_b}q_{j^\prime}^0\right)+\dfrac{x_1-x_2}{2}\,,\\
    \overline{x}_2 & = \dfrac{1}{\sqrt{s}}\left(\sum_{i^\prime=3}^{{\rm N}_e}p_{i^\prime}^0+\sum_{j^\prime=1}^{{\rm N}_b}q_{j^\prime}^0\right)-\dfrac{x_1-x_2}{2}\,.
\end{align}
This momentum reshuffling option is activated by setting \texttt{mom\_resh\_type=0} in the input file \texttt{run\_card.dat}.

\paragraph{Final-state momentum reshuffling:}
The final-state reshuffling is based on a recoil scheme in which the desired on-shell conditions are enforced primarily through modifications of the final-state momenta, while the initial-state kinematics are modified only through the corresponding change of the partonic c.m.\@ energy when required by kinematic threshold constraints. The construction is performed in the c.m.\@ frame of the $\Ione$-$\Itwo$ system with physical momenta. If the events are generated in the hadronic frame, as is the case in \mgshort, the reshuffling is performed by first boosting the momenta into this frame, applying the reshuffling procedure, and finally boosting the reshuffled momenta back to the hadronic frame. The initial-state momenta are given by
\begin{align}
    p_1 & = \dfrac{\sqrt{\hat s}}{2}(1,0,0,1)^\mathrm{T}\,,\quad p_2 = \dfrac{\sqrt{\hat s}}{2}(1,0,0,-1)^\mathrm{T}\,,\\
    \overline{p}_1 & = \dfrac{\sqrt{\hat{\overline{s}}}}{2}(1,0,0,1)^\mathrm{T}\,,\quad \overline{p}_2 = \dfrac{\sqrt{\hat{\overline{s}}}}{2}(1,0,0,-1)^\mathrm{T},
\end{align}
where $\sqrt{\hat s}$ ($\sqrt{\hat{\overline{s}}}$) denotes the partonic c.m.\@ energy before (after) reshuffling. The total momentum of the non-bound-state recoil system is defined as
\begin{equation}
    P_X = \sum_{i^\prime=3}^{{\rm N}_e} p_{i^\prime}\,,\quad \overline{P}_X = \sum_{i^\prime=3}^{{\rm N}_e} \overline{p}_{i^\prime}\,,
\end{equation}
before and after reshuffling, respectively. Its invariant mass is preserved under the reshuffling,
\begin{equation}
    P_X^2=\overline{P}_X^2\equiv M_\mathrm{rec}^2\,.
\end{equation}
To construct the reshuffled bound-state kinematics, we introduce the intermediate systems
\begin{align}
    &Q_{j^\prime} = \sum_{k=j^\prime}^{{\rm N}_b} q_k\,,\quad M_{j^\prime} = \sqrt{Q_{j^\prime}^2}\,,\label{eq:invM}\\
    &\overline{Q}_{j^\prime} = \sum_{k=j^\prime}^{{\rm N}_b} \overline{q}_{k}\,,\quad \overline{M}_{j^\prime} = \sqrt{\overline{Q}_{j^\prime}^2}\,.\label{eq:invOverlineM}
\end{align}
The reshuffled invariant masses are determined recursively, starting from $j^\prime={\rm N}_b$ and decreasing the index successively according to
\begin{equation}
    \overline{M}_{j^\prime} = \varphi(M_{j^\prime},m_{j^\prime},M_{j^\prime+1},\overline{m}_{j^\prime},\overline{M}_{j^\prime+1})\,,\quad 1\leq j^\prime<{\rm N}_b\,,
\end{equation}
with the boundary conditions
\begin{equation}
    M_{{\rm N}_b} = m_{{\rm N}_b}\,,\quad \overline{M}_{{\rm N}_b} = \overline{m}_{{\rm N}_b}\,.
\end{equation}
The function $\varphi(m,m_1,m_2,\overline{m}_1,\overline{m}_2)$, defined for $m\geq m_1+m_2$, ensures that the generated invariant mass satisfies the condition
\begin{equation}
\varphi(m,m_1,m_2,\overline{m}_1,\overline{m}_2)
\geq
\overline{m}_1+\overline{m}_2\,.
\end{equation}
Its explicit form is not unique. In the default setup (\texttt{mom\_resh\_type=1}), we employ the smooth choice
\begin{align}
    & \varphi(m,m_1,m_2,\overline{m}_1,\overline{m}_2)
    \nonumber\\
    & \quad =\dfrac{1}{2}\left[ m+\overline{m}_1+\overline{m}_2+\sqrt{\left(m-\overline{m}_1-\overline{m}_2\right)^2+\dfrac{\left(m_1+m_2-\overline{m}_1-\overline{m}_2\right)^2}{\tilde{n}}} \right]
\label{eq:varphiformomreshtype1}
\end{align}
with $\tilde{n}=100$. Alternatively, a step-function prescription is available (\texttt{mom\_resh\_type=2}),
\begin{equation}
    \varphi(m,m_1,m_2,\overline{m}_1,\overline{m}_2) = 
    \begin{cases}
        \hfill m\,,& \text{if}\ m\ge \overline{m}_1+\overline{m}_2\,, \\
        \dfrac{\overline{m}_1+\overline{m}_2}{m_1+m_2}m\,,& \text{otherwise}\,.
    \end{cases}
\label{eq:varphiformomreshtype2}
\end{equation}
Once $\overline{M}_{j^\prime}$ is known for all $j^\prime=1,\dots,{\rm N}_b$, the reshuffled partonic invariant mass is fixed by
\begin{equation}
    \sqrt{\hat{\overline{s}}} = \varphi(\sqrt{\hat{s}},M_{1},M_\mathrm{rec},\overline{M}_{1},M_\mathrm{rec})\,.
\label{eq:overlineshat1}
\end{equation}
The recoil system momentum is then reconstructed as
\begin{align}
    &\overline{P}_X^0 = \varepsilon\!\left(\sqrt{\hat{\overline{s}}},M_\mathrm{rec},\overline{M}_{1}\right)\,,\\
    &\overline{\bm P}_X = \sqrt{\left(\overline{P}_X^0\right)^2-M_{\mathrm{rec}}^2}\frac{\bm{P}_X}{\vert \bm{P}_X \vert}=\pi\!\left(\sqrt{\hat{\overline{s}}},\overline{M}_{1},M_\mathrm{rec}\right)\frac{\bm{P}_X}{\vert \bm{P}_X \vert}\,,
\end{align}
where the auxiliary functions are defined as
\begin{align}
    \varepsilon(m,m_1,m_2) & = \sqrt{m_1^2 + \pi^2(m,m_1,m_2)}\,,\\
    \pi(m,m_1,m_2) & = \dfrac{m}{2}\lambda(m,m_1,m_2)\,,
\end{align}
with the dimensionless K\"all\'en factor
\begin{equation}
    \lambda(m,m_1,m_2) = \sqrt{1-\dfrac{(m_1+m_2)^2}{m^2}}\sqrt{1-\dfrac{(m_1 - m_2)^2}{m^2}}\,.
\end{equation}
The reshuffled recoil momenta are obtained through a double boost,
\begin{equation}
    \overline{p}_{i^\prime} = \mathds{B}_\mathrm{R}^{-1}(\overline{P}_X)\,\mathds{B}_\mathrm{R}(P_X)\,p_{i^\prime}\,,\quad \ \forall\ \ i^\prime\ge3\,,
\end{equation}
where the boost operator $\mathds{B}_\mathrm{R}(k)$ maps a four-vector $k$ into its rest frame,
\begin{equation}
    \mathds{B}_\mathrm{R}(k)\,k = (\sqrt{k^2},0,0,0)^\mathrm{T},
\end{equation}
and $\mathds{B}^{-1}_\mathrm{R}(k)$ denotes the corresponding inverse boost. Starting from the total reshuffled bound-state system momentum,
\begin{align}
    & \overline{Q}_{1}^0 = \varepsilon\!\left(\sqrt{\hat{\overline{s}}}, \overline{M}_{1}, M_\mathrm{rec}\right),
\label{eq:overlineQ101}\\
    & \overline{\bm Q}_{1} = \sqrt{\left( \overline{Q}_{1}^0 \right)^2 - \overline{M}_1^2} \frac{{\bm Q}_1}{\vert {\bm Q}_1\vert} = \pi\!\left(\sqrt{\hat{\overline{s}}}, \overline{M}_1, M_{\mathrm{rec}}\right) \frac{{\bm Q}_1}{\vert {\bm Q}_1\vert} = -\overline{\bm P}_X \, ,
\label{eq:overlineQ1vec1}
\end{align}
the individual bound-state momenta are constructed recursively according to
\begin{align}
    & \overline{q}_{j^\prime} = \mathds{B}_\mathrm{R}^{-1}(\overline{Q}_{j^\prime})\left(\varepsilon(\overline{M}_{j^\prime},\overline{m}_{j^\prime},\overline{M}_{j^\prime+1}),\pi(\overline{M}_{j^\prime},\overline{m}_{j^\prime},\overline{M}_{j^\prime+1})\,{\bm e}_{j^\prime}\right)^\mathrm{T},\\
    & \overline{Q}_{j^\prime+1} = \mathds{B}_\mathrm{R}^{-1}(\overline{Q}_{j^\prime})\left(\varepsilon(\overline{M}_{j^\prime},\overline{M}_{j^\prime+1},\overline{m}_{j^\prime}),-\pi(\overline{M}_{j^\prime},\overline{m}_{j^\prime},\overline{M}_{j^\prime+1})\,{\bm e}_{j^\prime}\right)^\mathrm{T}\,.
\end{align}
The direction unit vector ${\bm e}_{j^\prime}$ is determined from the original bound-state kinematics through
\begin{equation}
    \mathds{B}_\mathrm{R}(Q_{j^\prime})\,q_{j^\prime} = \left(\varepsilon(M_{j^\prime},m_{j^\prime},M_{j^\prime+1}),\pi(M_{j^\prime},m_{j^\prime},M_{j^\prime+1})\,{\bm e}_{j^\prime}\right)^\mathrm{T}.
\end{equation}
This procedure is iterated until $j^\prime=\mathrm{N}_b-1$. Finally, we assign $\overline{q}_{\mathrm{N}_b}=\overline{Q}_{\mathrm{N}_b}$. In the special case of $\mathrm{N}_e=2$, corresponding to the absence of final-state elementary particles to absorb recoil, eq.~\eqref{eq:overlineshat1} simplifies to
\begin{equation}
    \sqrt{\hat{\overline{s}}} = \overline{M}_1\,,
\label{eq:overlineshat2}
\end{equation}
while eqs.~\eqref{eq:overlineQ101} and \eqref{eq:overlineQ1vec1} reduce to
\begin{align}
    &\overline{Q}_{1}^0 = \overline{M}_{1}\,,\label{eq:overlineQ102}\\
    &\overline{\bm Q}_{1} ={\bm 0}\,.\label{eq:overlineQ1vec2}
\end{align}

A possible advantage of the final-state momentum-reshuffling options over the initial-state momentum-reshuffling procedure is that the invariant masses of final-state subsystems are preserved to the largest possible extent. This can be particularly relevant when final-state configurations contain sharply peaked resonant structures and the matrix elements exhibit a strong dependence on the corresponding invariant masses.

\section{Benchmark processes and validation}\label{sec:benchmark}

We dedicate this section to validating our LO implementation of the orbital and total angular-momentum projectors (eqs.~\eqref{eq:proj_oam} and \eqref{eq:proj_tam}), as well as the reshuffling routines {applied to eq.~\eqref{eq:partonic_xsec_explicit}.} For the projector implementation, in section~\ref{sec:Pwavetest} we compare our results for a single phase-space point (using the \textit{standalone mode}) and for the phase-space-integrated cross section against \helaconia~\cite{Shao:2012iz,Shao:2015vga}. Our tests involve processes up to three {\pwavetxt} quarkonia with both QCD and electroweak interactions included.
For the momentum reshuffling, we compare our algorithms against the non-reshuffling case in section~\ref{sec:momreshufftest}, obtained through an appropriate reparametrisation of the quark and quarkonium masses.
These benchmarks are essential to verify the correctness of our implementation and ensure the reliability of the applications discussed in sections~\ref{sec:quarkonium} and \ref{sec:leptionium}.

\subsection{Validation of the P-wave implementation}
\label{sec:Pwavetest}
With a slightly modified setup relative to the baseline configuration described in appendix~\ref{app:setup}, we benchmark the implementation of the new orbital and total angular momentum projectors. To match the parameter setup used in \helaconia{}, which treats the decay widths in internal propagators differently from \mgshort, we set the decay widths of the weak and Higgs bosons to zero, ${\Gamma_W = \Gamma_Z = \Gamma_H = 0}$. Moreover, we set all quarkonium masses to $M_{\cal Q}^{\phantom{2}}=m_{Q}^{\phantom{2}}+m_{\bar{Q}^\prime}$, where $m_{Q}^{\phantom{2}}$ and $m_{\bar{Q}^\prime}$ denote the masses of the heavy quark and antiquark constituents, respectively. This choice avoids momentum reshuffling, which is not implemented in \helaconia{}. Additionally, in the \textit{standalone mode}, we fix $\alpha_s=0.118$, while for integrated cross sections the value of $\alpha_s$ is obtained from {\tt LHAPDF6}~\cite{Buckley:2014ana} and evaluated at the renormalisation and factorisation scales defined in eq.~\eqref{eq:setup_mu}.

Our first comparison is performed at the level of the squared helicity amplitude, $|{\cal M}|^2$, averaged over the colours and helicities of the initial-state particles, summed over those of the final-state particles, and weighted by LDMEs. More precisely, these squared amplitudes correspond to the quantity obtained by multiplying the factor in front of the phase-space measure in eq.~\eqref{eq:partonic_xsec_explicit} by the corresponding LDMEs. For a $2 \to {\rm N}$ partonic process producing a set $\mathfrak{B}$ of bound states ${\cal B}[n]$, we obtain 
\begin{equation}
    \left|\mathcal{M}_{\mathrm{N},\mathfrak{B}}\right|^2 = \frac{1}{{\cal N}(\dot{r})}\!\left(\frac{1}{4k_1\cdot k_2} \frac{1}{\omega(\Ione)\omega(\Itwo)} \right)\! \prod_{{\cal B}[n] \in \mathfrak{B}} \!\left(\frac{1/(2\mu_{\cal B})}{(2J+1)N_{[C]}}\langle {\cal O}_n^{\rm B}\rangle\right)\!\sum_{\substack{\rm colour\\ \rm spin}} |  {\mathds A}^{(\mathrm{N},0)}(\dot{r})|^2\,.
\label{eq:me_square}
\end{equation}
Each squared amplitude is evaluated at a single phase-space point. While the {\it standalone mode} of \mgshort\ employs {\tt Rambo}~\cite{Kleiss1986RAMBO} by default, the generated phase-space points have limited numerical accuracy, leading to violations of the on-shell conditions in double-precision arithmetic. Therefore, we use an alternative implementation based on ref.~\cite{Byckling:1971vca}, which generates momenta with analytic precision. These momenta are then passed to both \helaconia\ and \mgshort\ in double precision. To avoid excessively boosted phase-space configurations, we set the partonic c.m.\@ energy for all processes to $\sqrt{\hat s} = \sum_i M_i + 20\,{\rm GeV}$, namely $20\,{\rm GeV}$ above threshold. 
The agreement between the two tools is quantified through the relative deviation, defined as
\begin{equation}
    \Delta_\mathrm{rel.} = 2\left|\dfrac{\left|\mathcal{M}_{N,\mathfrak{B}}\right|^2_{\mgshort} - \left|\mathcal{M}_{N,\mathfrak{B}}\right|^2_\helaconia}{\left|\mathcal{M}_{N,\mathfrak{B}}\right|^2_{\mgshort} + \left|\mathcal{M}_{N,\mathfrak{B}}\right|^2_\helaconia}\right|,
\label{eq:relative_ratio}
\end{equation}
where $|{\cal M}_{N,\mathfrak{B}}|^2_{\mgshort}$ and $|{\cal M}_{N,\mathfrak{B}}|^2_{\helaconia}$ denote the squared amplitudes as defined in eq.~\eqref{eq:me_square}, evaluated with \mgshort\ and \helaconia, respectively.
\begin{table}[H]
\centering
\renewcommand*{\arraystretch}{1.1}
\begin{tabular}[t]{lcc}
\toprule
\multirow{2}{*}{\textbf{process}}& \madgraph & \multirow{2}{*}{$\Delta_\mathrm{rel.}$}  \\
 & \helaconia  \\
\midrule
\multirow{2}{*}{$g g \to h_c\big[{}^1\pwave_{1}^{[1]}\big] + g u \bar{u}$} & $9.0256873979004725 \cdot 10^{-5}\,\mathrm{GeV}^{-4}$ & \multirow{2}{*}{$1.46 \cdot 10^{-14}$} \\
 & $9.0256873979006040 \cdot 10^{-5}\,\mathrm{GeV}^{-4}$ & \\ 
\hdashline

\multirow{2}{*}{$u \bar{u} \to \Upsilon\big[{}^3\pwave_{0}^{[8]}\big] + g g g$} & $1.1548805300001495 \cdot 10^{-5}\,\mathrm{GeV}^{-4}$ & \multirow{2}{*}{$1.38 \cdot 10^{-14}$} \\
 & $1.1548805300001654 \cdot 10^{-5}\,\mathrm{GeV}^{-4}$ & \\
\hdashline

{$g g \to \eta_c\big[{}^1\swave_{0}^{[8]}\big] + h_c\big[{}^1\pwave_{1}^{[1]}\big]$} & 
$4.7791765795384806 \cdot 10^{-16}\,\mathrm{GeV}^{-2}$ & \multirow{2}{*}{$9.06 \cdot 10^{-14}$} \\
\qquad\qquad $+ \Upsilon\big[{}^3\swave_{1}^{[1]}\big]$ & $4.7791765795389135 \cdot 10^{-16}\,\mathrm{GeV}^{-2}$ & \\
\hdashline

{$g \gamma \to \eta_b\big[{}^1\swave_{0}^{[1]}\big] + \eta_c\big[{}^1\swave_{0}^{[8]}\big]\ \textsuperscript{\ref{fnt:QED_flag}}$} & 
$1.0744719179339013 \cdot 10^{-13}\,\mathrm{GeV}^{-2}$ & \multirow{2}{*}{$2.94 \cdot 10^{-15}$} \\
\multirow{1}{*}{\qquad\qquad$ + \chi_c\big[{}^3\pwave_{2}^{[1]}\big]$} & 
$1.0744719179339045 \cdot 10^{-13}\,\mathrm{GeV}^{-2}$ & \\ 
\hdashline

{$u \bar{u} \to \chi_{c0}\!\big[{}^3\pwave_{0}^{[1]}\big] + \Upsilon\!\big[{}^3\pwave_{1}^{[8]}\big]$} & $2.9379230723426576 \cdot 10^{-16}\,\mathrm{GeV}^{-2}$ & \multirow{2}{*}{$5.71 \cdot 10^{-15}$} \\*
\qquad\qquad$+\eta_c\!\big[{}^1\pwave_{1}^{[8]}\big]$ & $2.9379230723426408 \cdot 10^{-16}\,\mathrm{GeV}^{-2}$ & \\
 \hdashline

{$g u \to J/\psi\big[{}^3\pwave_{1}^{[8]}\big] + \eta_b\big[{}^1\pwave_{1}^{[8]}\big]\ \textsuperscript{\ref{fnt:QED_flag}}$} & 
$1.0252237742951235 \cdot 10^{-18}\,\mathrm{GeV}^{-4}$ & \multirow{2}{*}{$1.10 \cdot 10^{-13}$} \\
\qquad\qquad$ + W^{+} d$ & $1.0252237742950112 \cdot 10^{-18}\,\mathrm{GeV}^{-4}$ & \\ 
\hdashline

\multirow{2}{*}{$g g \to \chi_b\big[{}^3\pwave_{1}^{[1]}\big] + Z H H \ \textsuperscript{\ref{fnt:QED_flag}}$} & $7.1410224887725591 \cdot 10^{-19}\,\mathrm{GeV}^{-4}$ & \multirow{2}{*}{$4.05 \cdot 10^{-16}$} \\
& $7.1410224887725562 \cdot 10^{-19}\,\mathrm{GeV}^{-4}$ & \\ 
\hdashline

\multirow{1}{*}{$ g g \to \chi_c\big[{}^3\pwave_{0}^{[1]}\big] + \chi_c\big[{}^3\pwave_{1}^{[1]}\big]$} & $9.8531189707367133 \cdot 10^{-13}\,\mathrm{GeV}^{-2}$ & \multirow{2}{*}{$6.76 \cdot 10^{-15}$} \\
\qquad\qquad$+ J/\psi\big[{}^3\swave_{1}^{[8]}\big]$ & $9.8531189707366466 \cdot 10^{-13}\,\mathrm{GeV}^{-2}$ & \\ 
\hdashline

\multirow{2}{*}{$g g \to J/\psi\big[{}^3\pwave_{2}^{[8]}\big] + \eta_c\big[{}^1\pwave_{1}^{[8]}\big] + c \bar{c}$} & $3.2515779884666702 \cdot 10^{-9}\,\mathrm{GeV}^{-4}$ & \multirow{2}{*}{$2.56 \cdot 10^{-14}$} \\
& $3.2515779884665871 \cdot 10^{-9}\,\mathrm{GeV}^{-4}$ & \\ 
\hdashline
 
{$u \bar{u} \to \chi_c\big[{}^3\pwave_{2}^{[1]}\big] + J/\psi\big[{}^3\pwave_{0}^{[8]}\big]$} & 
$1.9155132456571052 \cdot 10^{-14}\,\mathrm{GeV}^{-6}$ & \multirow{2}{*}{$3.29 \cdot 10^{-15}$} \\ 
{\qquad\qquad$ + \eta_c\big[{}^1\swave_{0}^{[1]}\big] + c \bar{c}$}& $1.9155132456570989 \cdot 10^{-14}\,\mathrm{GeV}^{-6}$ & \\ 
\hdashline

{$u \bar{u} \to \chi_c\big[{}^3\pwave_{0}^{[1]}\big] + \chi_c\big[{}^3\pwave_{0}^{[1]}\big]$} & 
$1.8708646257279440 \cdot 10^{-14}\,\mathrm{GeV}^{-4}$ & \multirow{2}{*}{$6.24 \cdot 10^{-15}$} \\ 
{\qquad\qquad$+ \eta_c\big[{}^1\swave_{0}^{[1]}\big] + \eta_c\big[{}^1\swave_{0}^{[1]}\big]$} & 
$1.8708646257279324 \cdot 10^{-14}\,\mathrm{GeV}^{-4}$ & \\ 
\hdashline 

{$g g \to \eta_c\big[{}^1\swave_{0}^{[1]}\big] + J/\psi\big[{}^3\swave_{1}^{[8]}\big] $} & 
$3.3023033137867921 \cdot 10^{-18}\,\mathrm{GeV}^{-4}$ & \multirow{2}{*}{$1.67 \cdot 10^{-14}$} \\ 
{\qquad\qquad$+\Upsilon\big[{}^3\swave_{1}^{[1]}\big] + \eta_c\big[{}^1\pwave_{1}^{[8]}\big]$} & 
$3.3023033137867371 \cdot 10^{-18}\,\mathrm{GeV}^{-4}$ & \\
\hdashline

{$u \bar{u} \to \chi_{c0}\!\big[{}^3\pwave_{0}^{[1]}\big]+ \chi_{b0}\!\big[{}^3\pwave_{0}^{[1]}\big]$} & $1.6923441255992509 \cdot 10^{-25}\,\mathrm{GeV}^{-4}$ & \multirow{2}{*}{$1.74 \cdot 10^{-14}$} \\*
{\qquad\qquad$+\chi_{b0}\!\big[{}^3\pwave_{0}^{[1]}\big]+\eta_c\!\big[{}^1\swave_{0}^{[8]}\big]$} & $1.6923441255992215 \cdot 10^{-25}\,\mathrm{GeV}^{-4}$ & \\

\bottomrule
\end{tabular}
\caption{\it\small Squared-amplitude benchmarks at individual phase-space points. The first column indicates the process. The second column reports the code outputs, with {\tt \mgshort} ({\tt \helaconia}) shown in the top (bottom) row. The last column gives the relative deviation defined in eq.~\eqref{eq:relative_ratio}. 
}
\label{tab:sa_check}
\end{table}\vspace*{-2\baselineskip}
\makeatletter
\renewcommand{\thefootnote}{QED}
\begingroup
  \renewcommand{\@makefnmark}{}%
  \footnotemark
\endgroup
\footnotetext{
This process is generated with the additional {\tt QCD=99 QED=99} flag, which incorporates subdominant contributions by including all diagrams of ${\cal O}(\alpha_s^m\alpha^n)$ for a fixed value of $m + n$. Without this option, \mgshort{} retains only the contributions with the largest non-vanishing power of $\alpha_s$.
\label{fnt:QED_flag}}
\renewcommand*{\thefootnote}{\arabic{footnote}}
\makeatother
\addtocounter{footnote}{-1}

We have performed this test for almost $1000$ processes. The complete list of benchmarked processes can be found in the supplemental material, while table~\ref{tab:sa_check} presents a selection of representative cases. 
Overall, we observe good agreement between the two codes, with relative deviations typically in the $10^{-14}\textrm{--}10^{-16}$ range, corresponding to double-precision accuracy. Only a small number of processes fail to reach this level of agreement. For these cases, we observe a clear correlation between the magnitude of the matrix element (typically below $10^{-25}$) and the number of coinciding digits between the two codes, with the latter gradually decreasing as the matrix element becomes smaller. This behaviour suggests that the observed discrepancies originate from numerical precision limitations, likely due to cancellations and the interplay between extremely small contributions and finite floating-point precision. 
Importantly, all phenomenologically relevant processes achieve double-precision accuracy.

\begin{table}[t]
\centering
\renewcommand*{\arraystretch}{1.}
\begin{tabular}[t]{lcc}
\toprule
\multirow{2}{*}{\textbf{process}}& \madgraph & \multirow{2}{*}{\textbf{pull}}  \\
 & \helaconia  \\
\midrule
\multirow{2}{*}{$p p \to g g \to \chi_{b2}\big[{}^3\pwave_{2}^{[1]}\big]$} & $\,{615.100(19)\,\rm nb} $ & \multirow{2}{*}{$0.34$} \\
 & $615.109(19)\,{\rm nb}$ & \\ \hdashline
 \multirow{2}{*}{$p p \to g g \to \eta_{b}\big[{}^1\pwave_{1}^{[8]}\big]$} & $\,{176.4070(54)\,\rm nb} $ & \multirow{2}{*}{$0.26$} \\
 & $176.4048(64)\,{\rm nb}$ & \\ \hdashline
\multirow{2}{*}{$p p \to g g \to \chi_{c0}\big[{}^3\pwave_{0}^{[1]}\big] + H  \ \textsuperscript{\ref{fnt:QED_flag}}$} & $474.64(12)\,{\rm zb} $ & \multirow{2}{*}{$0.40$} \\
 & $474.583(83)\,{\rm zb}$ & \\ \hdashline
\multirow{2}{*}{$p p \to g g \to \chi_{c0}\big[{}^3\pwave_{0}^{[1]}\big] + \eta_b\big[{}^1\swave_{0}^{[1]}\big]$} & $392.63(64)\,{\rm pb}$ & \multirow{2}{*}{$1.21$} \\
 & $393.48(29)\,{\rm pb}$ & \\ \hdashline
 \multirow{2}{*}{$pp \to gg \to \eta_c\big[{}^1\pwave_1^{[8]}\big] + \Upsilon\big[{}^3\pwave_2^{[8]}\big]$} & $18.213(36)\,{\rm pb}$ & \multirow{2}{*}{$0.12$} \\
 & $18.220(49)\,{\rm pb}$ & \\ \hdashline
 \multirow{2}{*}{$pp \to gg \to B^{\ast +}_{c2}\big[{}^3\pwave_2^{[1]}\big] + B^{\ast -}_c\big[{}^3\pwave_1^{[8]}\big]$} & $404.03(16)\,{\rm fb}$ & \multirow{2}{*}{$1.18$} \\
 & $402.92(93)\,{\rm fb}$ & \\ \hdashline
  \multirow{2}{*}{$pp \to u\bar{u} \to h_c\big[{}^1\pwave_1^{[1]}\big] + h_c\big[{}^1\pwave_1^{[1]}\big]$} & $563.32(26)\,{\rm fb}$ & \multirow{2}{*}{$0.81$} \\
 & $564.07(88)\,{\rm fb}$ & \\ \hdashline
 \multirow{2}{*}{$pp \to u\bar{u} \to B_{c1}(H)^+\big[{}^3\pwave_1^{[1]}\big] + b\bar{c}$} & $3.55150(89)\,{\rm pb}$ & \multirow{2}{*}{$1.70$} \\
 & $3.5577(35)\,{\rm pb}$ & \\ \hdashline
  \multirow{2}{*}{$e^+ e^- \to h_b\big[{}^1\pwave_1^{[1]}\big] + Z \ \textsuperscript{\ref{fnt:QED_flag}}$} & $48.2830(80)\,{\rm yb}$ & \multirow{2}{*}{$1.06$} \\
 & $48.2947(76)\,{\rm yb}$ & \\
 \hdashline
  \multirow{2}{*}{$e^+ e^- \to h_b\big[{}^1\pwave_1^{[1]}\big] + ZH \ \textsuperscript{\ref{fnt:QED_flag}}$} & $70.260(21)\,{\rm rb}$ & \multirow{2}{*}{$1.69$} \\
 & $70.314(24)\,{\rm rb}$ & \\ \hdashline
  \multirow{2}{*}{$e^+ e^- \to h_b\big[{}^1\pwave_1^{[1]}\big] + ZHH \ \textsuperscript{\ref{fnt:QED_flag}}$} & $449.23(46)\,{\rm qb}$ & \multirow{2}{*}{$0.53$} \\
 & $448.98(10)\,{\rm qb}$ & \\
\bottomrule
\end{tabular}
\caption{\it\small Cross-section benchmarks for selected $pp$ ($\sqrt{s}=13\,{\rm TeV}$) and $e^+ e^-$ ($\sqrt{s}=1\,{\rm TeV}$) collisions. The first column indicates the process; for $pp$ collisions, only one partonic channel is considered. The second column reports the code outputs, with {\tt \mgshort}~({\tt \helaconia}) shown in the top (bottom). MC uncertainties are given in parentheses. The third column gives the pull defined in eq.~\eqref{eq:pull}.
The meaning of the {\tt QED} flag is given in the footnote \ref{fnt:QED_flag}.}
\label{tab:xsec_check}
\end{table}

Our next benchmarks test LO integrated cross sections, where we compare the cross sections obtained from eq.~\eqref{eq:mg5_xsec_output} as evaluated in \mgshort\ and \helaconia\ for eleven different processes. The results of this comparison are summarised in table~\ref{tab:xsec_check}.
In this study, we focus on single and double {\pwavetxt} production and employ SM elementary particles to explore more complex phase-space configurations. For proton-proton ($pp$) collisions, we take $\sqrt s = 13\,{\rm TeV}$, while for electron-positron ($e^-e^+$) collisions we use $\sqrt s = 1\,{\rm TeV}$.
To quantify the agreement between the two codes, we define the pull as
\begin{equation}
    \textbf{pull}=\left|\dfrac{\sigma_\mgshort-\sigma_\helaconia}{\sqrt{\Delta_{\mgshort}^2+\Delta_{\helaconia}^2}}\right|,
\label{eq:pull}
\end{equation}
where $\sigma_\mgshort$ and $\sigma_\helaconia$ denote the total cross sections computed with \mgshort\ and \helaconia, respectively, while $\Delta_{\mgshort}$ and $\Delta_{\helaconia}$ denote their corresponding MC uncertainties.
Overall, the results exhibit pulls consistent with statistical fluctuations, demonstrating the stable and reliable performance of the newly implemented \mgshort\ features.

We conclude that the outcomes of our benchmarks provide a robust validation of the extension. This validation includes both the comparison of squared matrix elements at individual phase-space points in \textit{standalone mode} and integrated cross-section comparisons across a representative set of processes. In all cases, we find agreement with the corresponding results obtained with \helaconia, within the expected numerical precision. These tests confirm the reliability of the newly implemented \mgshort\ extension for the automated event generation of processes involving non-relativistic \swave- and \pwavetxt\ bound states.

\subsection{Momentum reshuffling benchmarks}\label{sec:momreshufftest}
To investigate the impact of momentum reshuffling, we study di-bottomonium production with the initial-state and final-state recoil schemes introduced in section~\ref{sec:resh}. In particular, we consider the $2\to2$ process ${gg \to \eta_b[^1 \swave_0^{[1]}]+ \Upsilon(3\swave)[^3 \swave_1^{[8]}]}$ and the $2\to3$ process ${gg \to \eta_b[^1 \swave_0^{[1]}]+\Upsilon(3\swave)[^3 \swave_1^{[8]}]+ Z}$. The large mass difference between the two quarkonium states (${M_{\Upsilon(3\swave)} - M_{\eta_b} = 960\,{\rm MeV}}$) provides an ideal playground for resolving reshuffling effects in an extreme scenario. 
For both processes we adopt two setups. The first configuration is our baseline (see appendix~\ref{app:setup}), with $m_b = 4.7\,\mathrm{GeV}$ corresponding to half of the $\eta_b$ mass ($M_{\eta_b} = 9.4\,\rm GeV$), the mass assigned to the CO state  being $M^{[8]}_{\Upsilon(3\swave)}=M_{\Upsilon(3\swave)}+200\,{\rm MeV}=10.56\,{\rm GeV}$ following the convention used in \pythia{\tt 8}, and the LDMEs taken at their default values as listed in table~\ref{tab:ldme_bottomonia}. In the second configuration, we consider an alternative bottom-mass choice by setting $m_b = M^{[8]}_{\Upsilon(3\swave)}/2 = 5.28\,{\rm GeV}$ and rescaling each LDME according to the mass-dependent factor $\left(5.28/4.7\right)^{3}$ with respect to the first setup, while keeping all other inputs unchanged.

\begin{figure}[t]
  \centering    
  \input{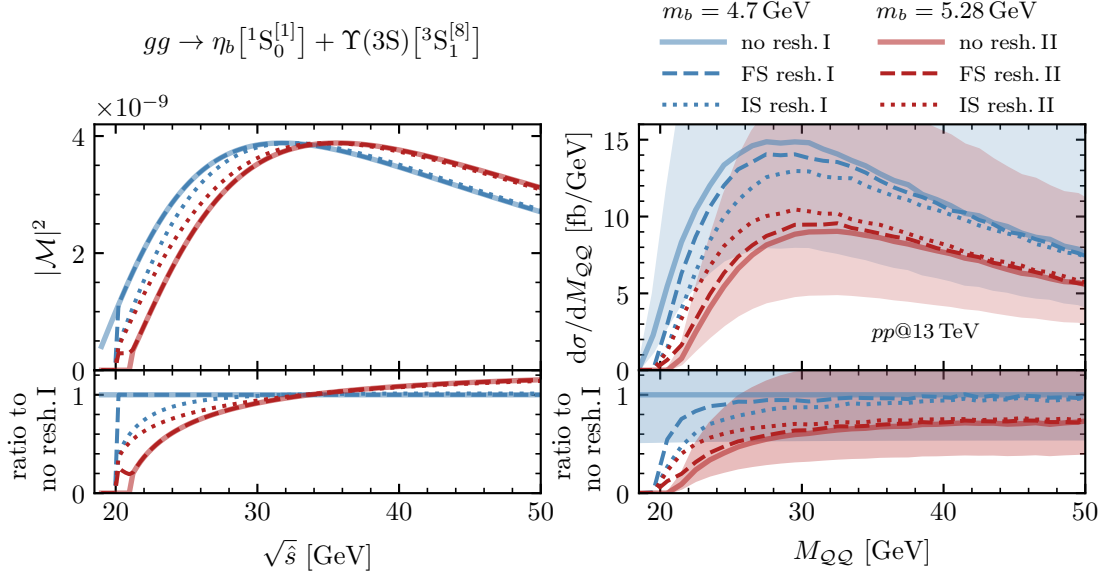}
  \caption{\it\small Effect of momentum reshuffling on the squared matrix element (left panel) and the cross section (right panel) as functions of the partonic c.m.\@ energy and the final-state invariant mass, respectively, for the process $pp\to gg \to \eta_b[^1 \swave_0^{[1]}]\, \Upsilon(3\swave)[^3 \swave_1^{[8]}]$ at $\sqrt{s}=13\,{\rm TeV}$. Blue (red) lines and bands correspond to the choice $m_b = 4.7\,{\rm GeV}$ ($m_b = 5.28\,{\rm GeV}$). The thick solid lines represent the non-reshuffling case with $M_{\cal Q} = 2 m_b$, while dashed and dotted lines correspond to the final-state and initial-state reshuffling schemes, respectively, with $M_{\eta_b} = 9.4\,{\rm GeV}$ and $M^{[8]}_{\Upsilon(3\swave)} = 10.56\,{\rm GeV}$. The lower panels show ratios with respect to the non-reshuffling case with $m_b=4.7\,\mathrm{GeV}$. The shaded bands indicate the $7$-point scale-variation uncertainties for the scenarios without momentum reshuffling.}
  \label{fig:etabUps_reshuffle}
\end{figure}

\begin{figure}[t]
    \centering
    \input{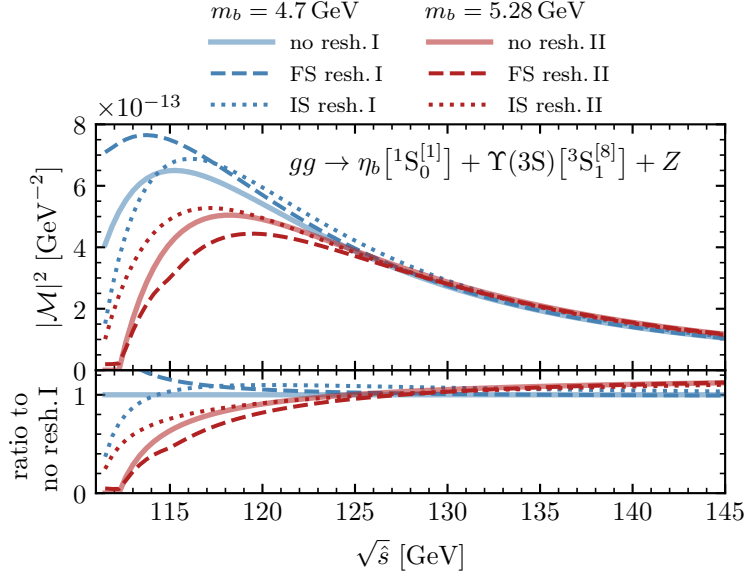}
    \caption{\it\small Same as the left panel of figure~\ref{fig:etabUps_reshuffle}, but with an additional $Z$ boson in the final state.}
    \label{fig:etabUpsZ_reshuffle}
\end{figure}

First, we discuss the effect of momentum reshuffling on the squared matrix element as a function of the partonic c.m.\@ energy. The results, shown in the left panel of figure~\ref{fig:etabUps_reshuffle} and in figure~\ref{fig:etabUpsZ_reshuffle}, are obtained by evaluating the squared matrix element at a single phase-space point for each energy value. The phase-space configuration is chosen such that it varies smoothly across the energy range, corresponding to a fixed point in the multidimensional phase-space hypercube. For example, the scattering angles in the $2\to2$ process are kept fixed, while the momenta are rescaled according to the available c.m.\@ energy.

The blue lines are obtained with the default setup ($m_b=4.7\,{\rm GeV}$), employing initial-state reshuffling (dotted), final-state reshuffling with {\tt mom\_resh\_type=1} (dashed), and without reshuffling with $M^{[8]}_{\Upsilon(3\swave)}=2m_b=9.4\,{\rm GeV}$ (solid).
The red lines are instead obtained using $m_b = 5.28\,{\rm GeV}$, combined with initial-state reshuffling (dotted), final-state reshuffling with {\tt mom\_resh\_type=1} (dashed), and without reshuffling with $M_{\eta_b}=2m_b=10.56\,{\rm GeV}$ (solid).

In the left panel of figure~\ref{fig:etabUps_reshuffle}, we observe that reshuffling effects are relevant only in the low-$\sqrt{\hat s}$ region. The no-reshuffling scenario predicts thresholds corresponding to $\sqrt{\hat s}=4m_b$, which are shifted relative to the physical threshold depending on the chosen value of $m_b$. In contrast, the reshuffled kinematics correctly reproduce the threshold at $\sqrt{\hat s}=M_{\eta_b}+M^{[8]}_{\Upsilon(3\swave)}$, independently of the constituent bottom-quark mass. At higher c.m.\@ energies, the results obtained with and without reshuffling converge, as mass effects become increasingly negligible. This behaviour is expected, since the relative difference between the two approaches scales as $(M_{\cal Q} - 2m_b)/\sqrt{\hat s}$. Thus, the impact of the mass mismatch decreases with increasing energy and vanishes in the asymptotic limit $\sqrt{\hat s}\to\infty$. At the same time, the ratio of the squared matrix elements obtained with the two bottom-quark mass choices approaches the corresponding LDME ratio, namely $(5.28/4.7)^6$.

The final-state reshuffling algorithm reproduces the no-reshuffling results across almost the entire kinematic range, modifying only the threshold behaviour discussed above. This is because in the present $2\to2$ process the reshuffling changes only the magnitudes of the outgoing quarkonium momenta to satisfy the on-shell conditions $M_{\cal Q}=2m_b$ for both $\eta_b$ and $\Upsilon(3\swave)$, while leaving the scattering angles and the partonic c.m.\@ energy unchanged. As a result, all Mandelstam invariants are preserved, and the Lorentz-invariant matrix element remains identical to that obtained without reshuffling.
By contrast, the initial-state recoil algorithm enforces the on-shell condition by modifying the momenta of the incoming partons. 
For $m_b=4.7\,\mathrm{GeV}$, the partonic c.m. energy compensates for the replacement of the (larger) $\Upsilon(3\swave)$ physical mass by $2m_b$.  
As a result, the matrix element corresponding to a physical collision energy $\sqrt{\hat s}$ is evaluated at a lower effective partonic c.m.\@ energy, shifting the spectrum towards larger $\sqrt{\hat{s}}$ values. For $m_b=5.28\,\mathrm{GeV}$, the opposite effect occurs, as the replacement of the (lower) $\eta_b$ physical mass by $2m_b$ requires an increase in the collision energy to compensate for the excess. The matrix element is therefore evaluated at a higher effective c.m.\@ energy, shifting the spectrum towards smaller $\sqrt{\hat{s}}$ values.

The $2\to3$ process shown in figure~\ref{fig:etabUpsZ_reshuffle} exhibits a more intricate dependence of the squared matrix element on the treatment of the bound-state masses. As in the previous case, the reshuffling methods reproduce the physical threshold, whereas the approaches without reshuffling exhibit a systematic shift depending on the bottom-quark mass. While this behaviour is expected, the structure of the matrix element in the threshold region becomes much richer and less predictable. This originates from the non-trivial dependence of the Mandelstam variables on the chosen reshuffling algorithm, leading to the complex shapes observed. These effects are particularly pronounced for the final-state reshuffling method, which deviates significantly from the non-reshuffling case.
In the high-energy limit $\sqrt{\hat{s}}\to\infty$, all methods again converge to the same asymptotic behaviour, as mass effects become increasingly negligible.

To extend the analysis beyond the matrix-element level and assess the phenomenological impact of reshuffling effects, we also study the invariant-mass spectrum of the process $gg \to \eta_b[^1 \swave_0^{[1]}] + \Upsilon(3\swave)[^3 \swave_1^{[8]}]$, shown in the right panel of figure~\ref{fig:etabUps_reshuffle}. At LO, the invariant mass $M_{\mathcal{QQ}}$ of the final-state $\eta_b$--$\Upsilon(3\swave)$ system coincides with the partonic c.m.\@ energy, $M_{\mathcal{QQ}}=\sqrt{\hat s}$. The spectrum therefore provides a direct connection between the squared matrix elements shown in the left panel and their corresponding contribution to the differential cross section.
The cross section, however, is not only affected by modifications of the matrix element induced by the reshuffling procedure, but also by changes in the available phase-space volume. For $m_b=4.7\,\mathrm{GeV}$, the physical meson masses exceed the constituent-mass values, leaving less kinetic energy available to populate the phase space. This effect suppresses the cross section in the threshold region and shifts the reshuffled spectra towards larger $M_{\mathcal{QQ}}$ values compared to the non-reshuffling case. For $m_b=5.28\,\mathrm{GeV}$, the opposite behaviour is observed. In this case, the physical threshold lies below the corresponding threshold at $4m_b$, such that the excess energy increases the accessible phase-space volume and enhances the cross section near threshold. The resulting spectra are therefore shifted towards smaller values of $M_{\mathcal{QQ}}$.
Since the final-state reshuffling algorithm leaves the matrix element essentially unchanged, the observed deviations in the corresponding cross sections originate almost entirely from the modified phase-space treatment. By contrast, the initial-state recoil prescription affects both the matrix element and the phase-space measure. Since these effects act in the same direction, they lead to a significantly larger deviation from the non-reshuffling prediction than that observed at the matrix-element level alone.

The more realistic phase-space treatment, in which the measured bound-state masses determine the available phase-space volume, suggests that applying a momentum-reshuffling prescription can significantly improve the physical description of the predictions obtained with \mgshort\ compared to calculations performed without reshuffling. However, we observe sizeable differences between the various reshuffling prescriptions. These differences reflect an intrinsic ambiguity in the mapping between the physical bound-state kinematics and the constituent-level kinematics required for the matrix-element evaluation. Thus, they should be regarded as an additional source of theoretical uncertainty.

\section{Quarkonium production}\label{sec:quarkonium}

In this section, we present the cross sections for selected processes, demonstrating the broad applicability of our code. Both $pp$ and $e^+ e^-$ colliders are considered. We use a c.m.\@ energy of $\sqrt{s} = 13\,{\rm TeV}$ for the former, while $\sqrt{s} = 10.58\,{\rm GeV}$ is adopted for the latter, corresponding to the asymmetric-energy configuration of $B$ factories such as SuperKEKB and Belle II ($E_{e^-} = 7\,{\rm GeV}$ and $E_{e^+}=4\,{\rm GeV}$). For each process, we provide the total (fiducial) cross section obtained with different setups. 
The first employs our default inputs (see appendix~\ref{app:setup}) and serves as the baseline.
The remaining setups are obtained by varying quark masses and LDME values. 
The non-default inputs are chosen to reproduce the physical masses of one of the produced quarkonia $\cal Q$, $m_{Q} + m_{\bar{Q}'} = M_{\cal Q}$, where $Q$ and $\bar{Q}'$ are charm or bottom (anti)quarks.
Each LDME is rescaled by the mass-dependent factors {$\left({m_c}/{1.55\,{\rm GeV}}\right)^n$} or {$\left({m_b}/{4.7\,{\rm GeV}}\right)^n$}, where $m_c$ and $m_b$ denote the charm and bottom masses used in the setup, and $n=3$ ($n=5$) for {\swavetxt}s ({\pwavetxt}s).
Note that this rescaling method is not the only viable choice for estimating the parametric dependence of the LDME, and for certain processes it may prove overly simplistic. For instance, a valid alternative, albeit more complex to implement, especially for the broad study presented here, would be to refit and evaluate all LDMEs directly at the quark mass used in the modified setup.
For each setup, we present both initial- and final-state reshuffled cross sections (the latter obtained with the option \texttt{mom\_resh\_type=1}) and compare them with the non-reshuffling scenario, in which the CS and CO quarkonium masses are set to the sums of their constituent-quark masses.
To quantify the impact of the mass treatment, we consider two quantities. The first, denoted by $\Delta\sigma^{\rm resh}$, measures the deviation of the averaged reshuffled cross sections from the non-reshuffled case,
\begin{equation}
    \Delta\sigma^{\rm resh} = \left\vert 1 - \frac{\sigma^{\rm FS\ resh.} + \sigma^{\rm IS\ resh.}}{2\,\sigma^{\rm no\ resh.}} \right\vert,
\end{equation}
capturing overall impact of using physical bound-state masses in the phase space generation. The second, denoted by $\Delta \sigma^{\rm algo}$, is the relative difference between the two reshuffling algorithms,
\begin{equation}\label{eq:unc_resh}
    \Delta\sigma^{\rm algo} = \left\vert \frac{\sigma^{\rm FS\ resh.} - \sigma^{\rm IS\ resh.}}{\sigma^{\rm FS\ resh.} + \sigma^{\rm IS\ resh.}} \right\vert,
\end{equation}
quantifying the uncertainty associated with the reshuffling procedure.
In phenomenological applications, neither algorithm is expected to be superior to the other. We therefore recommend performing computations with both prescriptions and using their difference as an estimate of the associated uncertainty. In this sense, eq.~\eqref{eq:unc_resh} provides an estimate of an additional theoretical uncertainty that should be included in phenomenological studies, as it is not negligible in general. For the LO predictions considered in this work, this uncertainty is expected to be subdominant compared to the scale variations for processes exhibiting a non-trivial scale dependence.

The inclusion of physical masses allows one to estimate the importance of missing relativistic corrections. Our implementation relies on the assumption that a phase-space treatment alone is sufficient to capture the effect of including physical masses, while the matrix element can still be evaluated at its original level of accuracy, as discussed in eq.~\eqref{eq:resh_matrix_element_approx}.
However, this assumption might not hold in general, as the neglected corrections to the matrix element could be larger than anticipated and might partially compensate for those already incorporated in the phase space. In such cases, the reshuffling procedure does not necessarily improve the parametric quark-mass dependence of the cross-section prediction and may even increase the associated uncertainty.
The size of the missing contributions to total cross sections can be estimated by comparing the $\Delta\sigma^{\rm resh}$ deviation with the expected scaling of ${\cal O} \left( \sum_{\cal Q}\Delta M_{\cal Q}/ M_{\cal Q}\right)$, where $\Delta M_{\cal Q} = \vert M_{\cal Q} - (m_Q + m_{\bar{Q}'}) \vert$.

The modular architecture of \mgshort\ records each partonic and Fock-state contribution individually in the {\tt HTML} output, such that the cross section associated with each Fock state combination can always be retrieved after process generation.
Moreover, an LHE file is saved at the end of the event-generation pipeline, enabling the decomposition of differential distributions into their individual contributions. This functionality will be demonstrated in the next sections.

Before moving to the description of some exemplary cross sections, we stress that these results are intended solely as a showcase of our code rather than as phenomenology studies.
In this sense, comparisons of the presented numbers with experimental data, where available, should be performed with care, as our numerical results do not include contributions beyond LO in QCD, do not account for possible feed-down effects, nor include possible double-parton-scattering (DPS) contributions~\cite{Kom:2011bd,Lansberg:2014swa,Lansberg:2015lva,Shao:2016wor,Lansberg:2016rcx,Lansberg:2016muq,Borschensky:2016nkv,Lansberg:2017chq,Shao:2019qob,Shao:2020kgj,Prokhorov:2020owf}. In addition, some processes may exhibit a strong dependence
on the chosen values of the CO LDMEs. The only exception is the discussion of $J/\psi\, + \psi(2\swave)$ in section \ref{sec:jpsipsi2sresult}, which is studied in greater depth to illustrate the full potential of our tool for phenomenological applications. In particular, although we restrict ourselves to single-parton-scattering (SPS) contributions, we consider not only direct \swave- and {\pwavetxt} contributions, but also $J/\psi$ feed-down contributions from $\psi(2\swave)$ and $\chi_{cJ}$ states. However, we refrain from presenting complete phenomenological studies of the other processes considered in this section, as this goes beyond the scope of this work.

\subsection{Single-quarkonium production in proton-proton collisions}

\begin{table}[t!]
\centering
\renewcommand*{\arraystretch}{1.4}
\textbf{$\boldsymbol{pp}$ collision: associated jet production}
\begin{tabular}[t]{lcccc}
\toprule
\multirow{2}{*}{\textbf{final state}}  & \multirow{2}{*}{$\left(\dfrac{m_c}{\rm GeV}, \dfrac{m_b}{\rm GeV}\right)$}& \multicolumn{3}{c}{$\boldsymbol{\sigma}~[\textbf{nb}]$} \\
& & \textbf{no resh.} & \textbf{FS resh.} & \textbf{IS resh.} \\
\midrule
\multirow{2}{*}{$\psi(2\swave) + j + X$} 
& $(1.55,4.7)$  & $137.7(2)$ & ${137.7(2)}$ & ${137.8(2)}$ \\ 
& $(1.845,4.7)$ & $138.5(2)$ & $142.1(2)$ & $1(2)$ \\ 
\hdashline
\multirow{2}{*}{$ \eta_c + j + X$} 
& $(1.55,4.7)$ & $828.6(13)$ & $847.3(12)$ & $847.2(12)$ \\ 
& $(1.49,4.7)$ & $809.4(12)$ & $820.6(12)$ & $820.1(12)$ \\ 
\hdashline
\multirow{2}{*}{$ \Upsilon + j + X$} 
& $(1.55,4.7)$ & $44.42(6)$ & $45.18(11)$ & $45.21(11)$ \\ 
& $(1.55,4.73)$ & $44.29(7)$ & $45.36(11)$ & $45.39(11)$ \\
\hdashline
\multirow{2}{*}{$\eta_b(2\swave) + j + X$} 
& $(1.55,4.7)$ & $168.4(3)$ & $164.8(2)$ & $165.8(2)$ \\
& $(1.55,5)$ & $171.6(2)$ & $171.8(3)$ & $171.8(3)$ \\
\hdashline
\multirow{2}{*}{$\chi_{c0} + j + X$} 
& $(1.55,4.7)$ & $86.03(14)$ & $87.19(15)$ & $87.22(15)$ \\  
& $(1.705,4.7)$ & $86.75(14)$ & $88.43(15)$ & $88.41(14)$ \\  
\hdashline
\multirow{2}{*}{$\chi_{c1} + j + X$} 
& $(1.55,4.7)$ & $283.4(5)$ & $285.2(4)$ & $285.3(5)$ \\ 
& $(1.755,4.7)$ & $291.0(5)$ & $294.5(5)$ & $294.5(5)$ \\ 
\hdashline
\multirow{2}{*}{$\chi_{c2} + j + X$} 
& $(1.55,4.7)$ & $382.6(7)$ & $387.3(6)$ & $386.3(6)$ \\ 
& $(1.78,4.7)$ & $377.0(7)$ & $384.9(6)$ & $384.8(6)$ \\ 
\hdashline
\multirow{2}{*}{$h_b + j + X$}
& $(1.55,4.7)$ & $3.680(6)$ & $3.624(5)$ & $3.651(5)$ \\ 
& $(1.55,4.95)$ & $3.941(6)$ & $3.930(6)$ & $3.938(6)$ \\ 
\bottomrule
\end{tabular}
\caption{\it\small
Total fiducial cross sections for associated inclusive quarkonium-jet production in $p p$ collisions at ${\sqrt{s} = 13\,{\rm GeV}}$. Kinematic cuts on the jet transverse momentum (${k_{Tj} > 10\,{\rm GeV}}$) and rapidity (${|\eta_j|< 5}$) are applied to remove the IR-sensitive regions. The no-reshuffling scenarios are obtained by fixing all quarkonium masses to the sum of the constituent-quark masses (given in column $2$). The first row for each final state corresponds to our default setup. For the other setups, the LDMEs are rescaled by a mass-dependent factor (see text). The MC uncertainties are given in parentheses.}
\label{tab:sigma_pp2Qj}
\end{table}

\begin{table}[t!]
\centering
\renewcommand*{\arraystretch}{1.4}
\textbf{$\boldsymbol{pp}$ collision: associated open heavy-quark production}
\begin{tabular}[t]{lcccc}
\toprule
\multirow{2}{*}{\textbf{final state}}  & \multirow{2}{*}{$\left(\dfrac{m_c}{\rm GeV}, \dfrac{m_b}{\rm GeV}\right)$}& \multicolumn{3}{c}{$\boldsymbol{\sigma}~[\textbf{nb}]$} \\
& & \textbf{no resh.} & \textbf{FS resh.} & \textbf{IS resh.} \\
\midrule
$ \psi(2\swave) + c \bar{c} + X$ 
& $(1.55,4.7)$ & {$410.6(7)$} & {$313.1(7)$} & {$342.9(9)$} \\ \hdashline
\multirow{2}{*}{$\eta_b(2\swave) + c \bar{c} + X$} 
& $(1.55,4.7)$ & {$37.06(6)$} & {$35.50(6)$} & {$34.78(6)$} \\
& $(1.55,5)$ & {$33.26(6)$} & {$33.24(6)$} & {$33.23(6)$} \\
\hdashline
$ B_c^+ + b \bar{c} + X$ 
& $(1.55,4.7)$ & {$15.404(15)$} & {$15.194(14)$} & {$15.272(14)$} \\ 
\hdashline
$ B_{c1}(L)^+ + b \bar{c} + X$ 
& $(1.55,4.7)$ & {$3.348(3)$} & {$2.867(3)$} & {$2.988(3)$} \\ 
\bottomrule
\end{tabular}
\caption{\it\small  Same as table~\ref{tab:sigma_pp2Qj}, but for the total cross sections of quarkonium production in association with an open heavy-quark pair. For quark masses different from the default values, the LDMEs are rescaled by a mass-dependent factor (see text).}
\label{tab:sigma_pp2Qccx}
\end{table}

\begin{table}[t!]
\centering
\renewcommand*{\arraystretch}{1.4}
\textbf{$\boldsymbol{pp}$ collision: associated electroweak-boson production}
\begin{tabular}[t]{lcccc}
\toprule
\multirow{2}{*}{\textbf{final state}}  & \multirow{2}{*}{$\left(\dfrac{m_c}{\rm GeV}, \dfrac{m_b}{\rm GeV}\right)$} & \multicolumn{3}{c}{$\boldsymbol{\sigma}~[\textbf{fb}]$} \\
& & \textbf{no resh.} & \textbf{FS resh.} & \textbf{IS resh.} \\
\midrule
\multirow{2}{*}{$ \psi(2\swave) + Z + X$} 
& $(1.55,4.7)$ & {$537.9(6)$} & {$493.8(5)$} & {$515.3(6)$} \\ 
& $(1.845,4.7)$ & {$509.7(5)$} & {$500.8(5)$} & {$505.1(5)$} \\ 
\hdashline
\multirow{2}{*}{$ \chi_{b0} + \gamma + X$} 
& $(1.55,4.7)$ & $742(1)$ & $633.4(9)$ & $662.2(9)$ \\ 
& $(1.55,4.93)$ & {$726.8(9)$} & {$693(1)$} & {$702(1)$} \\ 
\hdashline
\multirow{2}{*}{$\psi(2\swave) + W^+ + X$} 
& $(1.55,4.7)$  & $704.7(7)$ & $649.0(6) $ & $675.2(6)$ \\
& $(1.845,4.7)$ & {$618.6(6)$} & {$606.0(6)$} & {$611.5(6)$} \\ 
\bottomrule
\end{tabular}
\caption{\it\small Same as table~\ref{tab:sigma_pp2Qj}, but for the total (fiducial) cross sections of quarkonium production associated with electroweak bosons. Kinematic cuts are applied only for quarkonium production associated with photons: ${k_{T\gamma} > 5~{\rm GeV}}$ and ${|\eta_\gamma|< 5}$. For quark masses different from the default values, the LDMEs are rescaled by a mass-dependent factor (see text).}
\label{tab:sigma_pp2Qew}
\end{table}

As a first set of examples, we consider single-quarkonium production in proton-proton collisions at ${\sqrt{s} = 13\,{\rm TeV}}$, in association with a jet (table~\ref{tab:sigma_pp2Qj}), open heavy quarks (table~\ref{tab:sigma_pp2Qccx}), and electroweak gauge bosons (table~\ref{tab:sigma_pp2Qew}).
All numerical results include both CS and CO contributions up to order $v^7$ in the NRQCD velocity expansion (see table~\ref{tab:NRQCD_scaling}), and are evaluated at LO in $\alpha_s$.
For instance, the syntax to generate $p p \to h_{b} + j + X$ is:
\prompt{\tt import model sm\_onia}
\vskip -0.25truecm
\prompt{\tt generate p p > hb j; output; launch}

\noindent
For processes such as the example above, where no charm quark or charmonium is explicitly present in the final state, we adopt the fixed-flavour-number scheme with $n_f = 4$ light quark flavours. Otherwise, the {\tt c\_mass} restriction is applied, and $n_f = 3$ is used.
To avoid the soft and collinear divergences encountered in associated production with jets and photons, we impose kinematic cuts on the transverse momentum and rapidity of the massless final-state particles. For jets, we require that $k_{Tj} > 10\,{\rm GeV}$ and $|\eta_j| < 5$, while for photons we impose $k_{T\gamma} > 5\,{\rm GeV}$ and $|\eta_\gamma| < 5$.
All remaining processes discussed in this section are IR safe, and no cuts are applied.

The cross sections in table~\ref{tab:sigma_pp2Qj}, all evaluated in the same fiducial region, span the range $1$--$100\,{\rm nb}$ across the different quarkonia considered. Similar magnitudes are observed in table~\ref{tab:sigma_pp2Qccx} for associated open-heavy-quark production. This results from an accidental compensation among different effects, including the suppression from the additional power of $\alpha_s$, the presence of new types of diagrams, and the enhancement due to integration over a wider phase space.
As expected, the cross sections for associated weak-boson production, shown in table~\ref{tab:sigma_pp2Qew}, are significantly smaller, reflecting both the weaker electroweak coupling and the presence of heavier final-state particles.

Most of the processes included in table~\ref{tab:sigma_pp2Qj} exhibit similar features.
Momentum reshuffling induces deviations 
$\Delta \sigma^{\rm resh}$ at or below the percent level for all setups, with $\psi(2\swave) + j$ at $m_c = 1.55\,{\rm GeV}$ being the process least affected by momentum reshuffling. This is somewhat counter-intuitive, given that the alternative setup (for which $m_c = M_{\psi(2\swave)}/2$) does exhibit a shift in the fiducial cross section when momenta are reshuffled to accommodate the CO state masses. Moreover, the final-state reshuffling causes the cross section to shift in the opposite direction to that discussed in section~\ref{sec:momreshufftest}.
These illustrate that the impact of fiducial cuts on the cross section is generally difficult to anticipate. 
Overall, the uncertainties $\Delta\sigma^{\rm algo}$ associated with the treatment of the quarkonium masses are negligible. We attribute this mainly to the stringent cuts employed, which exclude the threshold regions and thereby reduce the impact of the reshuffling procedure, spoiling the expected ${\cal O}(\Delta M_{\cal Q}/M_{\cal Q})$ scaling.
However, since these behaviours may change significantly when different cuts are applied, we recommend using multiple reshuffling algorithms to reliably assess such uncertainties.

In table~\ref{tab:sigma_pp2Qccx}, momentum reshuffling affects the cross sections in different ways.
For $B_c^+ + b\bar{c}$, the mass-treatment deviation $\Delta \sigma^{\rm resh}$ remains at the percent level. For $\eta_b + c\bar{c}$, the deviation increases for the default setup, reaching $\Delta \sigma^{\rm resh}\sim5\%$, while the alternative setup yields cross sections that are compatible within MC uncertainties, independently of the reshuffling procedure employed.
The effect becomes more pronounced for $B_{c1}(L)^+ + b \bar{c}$ and $\psi(2\swave) + c \bar{c}$, with $\Delta \sigma^{\rm resh}$ reaching ${\sim}13\%$ for the former and ${\sim}20\%$ for the latter.
In addition, initial- and final-state reshuffling algorithms are found to differ at the percent level or below, with the largest deviation observed for $\psi(2\swave) + c \bar{c}$ ($\Delta\sigma^{\rm algo}\sim4\%$).
Based on the limited number of processes investigated here, the presence of two open heavy quarks appears to favour a reduction of the total production yield when physical masses are used.
Note that, in the case of associated open-quark production, the quark masses entering the open fermion line(s) should not differ significantly from their typical pole-mass values. Therefore, alternative setups that modify the mass parameters of the open quarks are discarded.

Moving to table~\ref{tab:sigma_pp2Qew}, all three processes considered exhibit sensitivity to both the heavy-quark and quarkonium masses. Among them, $\chi_{b0}+\gamma$ with the default setup shows the largest deviation associated with the mass treatment, with $\Delta \sigma^{\rm resh} \sim 13\%$, while the other two processes exhibit $\Delta \sigma^{\rm resh} \sim 6\%$.
The uncertainty associated with the reshuffling algorithm is instead $\Delta\sigma^{\rm algo} \sim 2\%$ for all processes. For the alternative setups, both $\Delta \sigma^{\rm resh}$ and $\Delta\sigma^{\rm algo}$ relative deviations decrease, as the heavier CO states become the only source of reshuffling uncertainty.
However, this does not imply that the alternative setups are necessarily preferable or that they are free from uncertainties. 
Moreover, we observe that, for $\chi_{b0} + \gamma$, changing the bottom-quark mass affects the reshuffled cross sections more than the non-reshuffled one. 
This is in contrast to our naive expectation that the momentum reshuffling prescription should reduce the sensitivity of the cross section to variations of the constituent-quark masses.  
As $\Delta \sigma^{\rm resh}$ is of the same order as the expected scaling $\Delta M_{\chi_{c2}}/M_{\chi_{c2}} \sim 5\%$, we infer that the neglected corrections are not excessively large in this case, and that the increased sensitivity to the constituent-quark mass is primarily driven by the naive LDME rescaling.

Before concluding the discussion of single-quarkonium hadroproduction, we elaborate further on $\chi_{b0}+\gamma$. Although both CS and CO $\chi_{b0}$ LDMEs have the same $v^2$ order (cf.\@ table~\ref{tab:NRQCD_scaling}), the former is forbidden at LO by charge-conjugation conservation. The \mgshort\ output confirms this, as the cross section is solely determined by the $\chi_{b0}[^3 \swave_1^{[8]}]$ state. 

Overall, the first examples reported in tables~\ref{tab:sigma_pp2Qj}, \ref{tab:sigma_pp2Qccx}, and \ref{tab:sigma_pp2Qew} illustrate different scenarios when a coherent mass treatment is employed. For the total cross sections, $\Delta \sigma^{\rm resh}$ follows the expected scaling of ${\cal O} \left( \Delta M_{\cal Q}/ M_{\cal Q}\right)$.
Once a fiducial region is imposed, however, this expectation no longer holds. Applying cuts that remove part of the threshold region should generally reduce the impact of reshuffling, but the size of this reduction remains difficult to anticipate, as it depends strongly on several factors, including the size of the accessible phase-space region and the specific structure of the matrix elements involved. 
In addition, our examples demonstrate that \madsons\ is capable of providing reliable cross sections that account for both \swave- and \pwavetxt\ contributions.

\subsection{Double-quarkonium production in proton-proton collisions}
\begin{table}[t!]
\centering
\renewcommand*{\arraystretch}{1.4}
\textbf{$\boldsymbol{pp}$ collision: double-quarkonium production}
\begin{tabular}[t]{lcccc}
\toprule
\multirow{2}{*}{\textbf{final state}}  & \multirow{2}{*}{$\left(\dfrac{m_c}{\rm GeV}, \dfrac{m_b}{\rm GeV}\right)$} & \multicolumn{3}{c}{$\boldsymbol{\sigma}~[\textbf{nb}]$} \\
& & \textbf{no resh.} & \textbf{FS resh.} & \textbf{IS resh.} \\
\midrule
$\psi(2\swave) + \psi(2\swave) + X$ & $(1.55,4.7)$ & $4.994(4)$ & $2.1123(8)$ & $2.9854(12)$ \\ \hdashline
$J/\psi + \psi(2\swave) + X$ & $(1.55,4.7)$ & $18.81(2)$ & $12.569(4)$ & $14.806(5)$ \\ \hdashline
$J/\psi + \eta_c + X$ & $(1.55,4.7)$ & $38.39(4)$ & $38.27(4)$ & $38.00(4)$ \\ \hdashline
$J/\psi + \Upsilon + X$ & $(1.55,4.7)$ & $0.15988(18)$ & $0.15475(18)$ & $0.15254(17)$ \\ \hdashline
$J/\psi + \eta_b + X$ & $(1.55,4.7)$ & $1.1832(5)$ & $1.173(4)$ & $1.1625(5)$ \\ \hdashline
$\Upsilon + \Upsilon + X$ & $(1.55,4.7)$ & $0.04703(3)$ & $0.04469(3)$ & $0.04559(3)$ \\
\bottomrule
\end{tabular}
\caption{\it\small Total cross sections for double-charmonium and double-bottomonium production in $p p$ collisions at $\sqrt{s} = 13\,{\rm TeV}$.}
\label{tab:sigma_pp2QQ}
\end{table}

The second set of examples concerns double-quarkonium production in proton-proton collisions at ${\sqrt{s}=13\,{\rm TeV}}$.
We consider the total cross sections for double-charmonium, double-bottomonium, and mixed charmonium-bottomonium production solely adopting the default setup. All these processes are IR finite for both CS and CO channels at LO, and no additional cuts are therefore applied.
They are generated in {\mgshort} following the same syntax as for single-quarkonium production. The numerical results are reported in table~\ref{tab:sigma_pp2QQ}. All CS and CO contributions up to order $v^7$ (see table~\ref{tab:NRQCD_scaling}) are included, without feed-down and DPS contributions.

The impact of reshuffling varies noticeably with the quarkonium states considered. For $J/\psi + \eta_c$, $J/\psi + \Upsilon$, and $\Upsilon + \Upsilon$, applying a momentum reshuffling procedure modifies the cross section at the percent level. In contrast, double-$\psi(2\swave)$ and $J/\psi + \psi(2\swave)$ production undergo a more significant reduction, with $\Delta\sigma^{\rm resh}\sim49\%$ and $\Delta\sigma^{\rm resh}\sim27\%$, respectively. For the same processes, large values of $\Delta\sigma^{\rm algo}$ are also observed, namely $\sim17\%$ for double-$\psi(2\swave)$ and $\sim8\%$ for $J/\psi + \psi(2\swave)$. The high sensitivity of these channels to the mass treatment has important phenomenological consequences for the description of the LHCb measurement of $J/\psi + \psi(2\swave)$~\cite{LHCb:2023wsl}.
Previous SPS predictions tended to fully cover the measured data, creating mild tension with interpretations involving sizeable DPS contributions. The reduction of both direct $J/\psi + \psi(2\swave)$ production and feed-down contributions from double-$\psi(2\swave)$ production is therefore important for improving our understanding of this process. 

An instructive comparison can be made between the analogous processes of double-$\psi(2\swave)$ and double-$\Upsilon$ production. For the non-reshuffled cross sections, we find 
$$\dfrac{\sigma\!\left(\psi(2\swave)+\psi(2\swave)\right)}{{\sigma\!\left(\Upsilon+\Upsilon\right)}} \sim 100\,,$$ which can be attributed to the different virtuality scales of the propagators, controlled by the charm-quark mass in one case and the bottom-quark mass in the other, as well as to the different LDMEs involved.
Moreover, the mass treatment generates corrections of different magnitudes for the two processes, with $\Delta\sigma^{\rm resh}$ decreasing by roughly a factor $10$ from double-$\psi(2\swave)$ to double-$\Upsilon$ production. This behaviour is consistent with the expected scaling of the reshuffling correction, namely $2\Delta M_{\psi(2\swave)}/M_{\psi(2\swave)} \sim 32\%$ for double-$\psi(2\swave)$ and $2\Delta M_{\Upsilon}/M_{\Upsilon}\sim 1\%$ for double-$\Upsilon$ production.
The remaining processes listed in table~\ref{tab:sigma_pp2QQ} follow the same pattern. For example, the expected scaling for $J/\psi + \psi(2\swave)$ is $\Delta M_{\psi(2\swave)}/M_{\psi(2\swave)} \sim 16\%$, in good agreement with the trend observed in the table.

To qualitatively understand the cross sections of other similar processes, we first observe that the $\mathcal{O}(\alpha_s^4)$ cross section of $J/\psi+\eta_b$ is almost an order of magnitude larger than that of $J/\psi+\Upsilon$ at the same order. Both processes are dominated by mixed CS-CO channels~\cite{Shao:2016wor}. However, the dominant contribution, $J/\psi[^3P_J^{[8]}]+\eta_b[^1S_0^{[1]}]$, in $J/\psi+\eta_b$ production (see table~\ref{tab:sigma_pp2JpsiEtab}) is further enhanced by $t$-channel gluon diagrams, similar to that shown in figure~1(e) of ref.~\cite{Shao:2016wor}. Such diagrams can only occur in CO-CO channels for $J/\psi+\Upsilon$ production.

Similar to the $J/\psi+J/\psi$ case, the total cross section of $J/\psi+\psi(2\swave)$ SPS production is dominated by the CS-CS channel due to the relative sizes of the LDMEs. However, owing to charge-conjugation conservation, the CS-CS contribution vanishes for $J/\psi+\eta_c$ at $\mathcal{O}(\alpha_s^4)$, making the LO contribution of this channel at $\mathcal{O}(\alpha_s^5)$ one order of magnitude smaller than the corresponding CS-CS contribution in $J/\psi+\psi(2\swave)$ production~\cite{Lansberg:2013qka}. This led to the expectation that the SPS contribution in $J/\psi+\eta_c$ hadroproduction is suppressed compared with the DPS contribution, making this process an ideal probe of the DPS mechanism. However, as shown by the results reported in table~\ref{tab:sigma_pp2QQ}, the total cross section of $J/\psi+\eta_c$ is even a factor of two larger than that of $J/\psi+\psi(2\swave)$. The dominant contributions in the former arise from the CS-CO channels $J/\psi[^3P_J^{[8]}]+\eta_c[^1S_0^{[1]}]$, analogous to the previously discussed $J/\psi+\eta_b$ case. These channels are enhanced by $t$-channel gluon diagrams, although they are suppressed by the smaller CO LDMEs. These diagrams lead to relatively flat di-quarkonium rapidity-gap $\Delta y$ distributions. Similar behaviour is also observed in the di-$J/\psi$ case, but only for CO-CO channels~\cite{Lansberg:2019fgm}. However, the double CO LDME suppression in the di-$J/\psi$ case makes this enhancement effect much less pronounced than in the $J/\psi+\eta_c$ case. Therefore, although sensitive to the CO LDMEs of $J/\psi$, the SPS contribution in $J/\psi+\eta_c$ hadroproduction remains sizeable, contrary to what has been assumed in the literature.
Note that, to observe this enhancement experimentally, large $\Delta y$ and low transverse momenta must be included, as the effect becomes negligible for $\Delta y \to 0$, where the expected LDME scaling becomes valid again.

\begin{table}[t!]
\centering
\renewcommand*{\arraystretch}{1.4}
\begin{tabular}[t]{lc||lc}
\toprule \multicolumn{4}{c}{\textbf{process: $\boldsymbol{p p \to J/\psi + \eta_b + X}$}}\\
\toprule
{\textbf{channel}} 
 & {$\boldsymbol{\sigma~[\textbf{pb}]}$} & {\textbf{channel}} 
 & {$\boldsymbol{\sigma~[\textbf{pb}]}$} \\ 
 \midrule
$J/\psi\big[{}^3{\swave}_1^{[1]}\big] + \eta_b\big[{}^1{\swave}_0^{[1]}\big]$ & {$0$} & $J/\psi\big[{}^3{\swave}_1^{[8]}\big] + \eta_b\big[{}^1{\swave}_0^{[1]}\big]$ & $203.73(9)$\\ \hdashline
$J/\psi\big[{}^3{\swave}_1^{[1]}\big] + \eta_b\big[{}^1{\swave}_0^{[8]}\big]$ & {$0$} & $J/\psi\big[{}^3{\swave}_1^{[8]}\big] + \eta_b\big[{}^1{\swave}_0^{[8]}\big]$ & $1.04(4)$ \\ \hdashline
$J/\psi\big[{}^3{\swave}_1^{[1]}\big] + \eta_b\big[{}^3{\swave}_1^{[8]}\big]$ & $0.5137(9)$
& $J/\psi\big[{}^3{\swave}_1^{[8]}\big] + \eta_b\big[{}^3{\swave}_1^{[8]}\big]$ & $1.34(2)$ \\ \hdashline
$J/\psi\big[{}^3{\swave}_1^{[1]}\big] + \eta_b\big[{}^1{\pwave}_1^{[8]}\big]$ & {$0$} & $J/\psi\big[{}^3{\swave}_1^{[8]}\big] + \eta_b\big[{}^1{\pwave}_1^{[8]}\big]$ & $6.44(9)$ \\ \hdashline
$J/\psi\big[{}^1{\swave}_0^{[8]}\big] + \eta_b\big[{}^1{\swave}_0^{[1]}\big]$ & $161.21(6)$ & $J/\psi\big[{}^3{\pwave}_J^{[8]}\big] + \eta_b\big[{}^1{\swave}_0^{[1]}\big]$ & $762(4)$ \\ \hdashline
$J/\psi\big[{}^1{\swave}_0^{[8]}\big] + \eta_b\big[{}^1{\swave}_0^{[8]}\big]$ & $0.81(4)$ & $J/\psi\big[{}^3{\pwave}_J^{[8]}\big] + \eta_b\big[{}^1{\swave}_0^{[8]}\big]$ & $4.1(3)$ \\ \hdashline
$J/\psi\big[{}^1{\swave}_0^{[8]}\big] + \eta_b\big[{}^3{\swave}_1^{[8]}\big]$ & $0.405(12)$ & $J/\psi\big[{}^3{\pwave}_J^{[8]}\big] + \eta_b\big[{}^3{\swave}_1^{[8]}\big]$ & $1.63(5)$ \\ \hdashline
$J/\psi\big[{}^1{\swave}_0^{[8]}\big] + \eta_b\big[{}^1{\pwave}_1^{[8]}\big]$ & $5.16(12)$ & $J/\psi\big[{}^3{\pwave}_J^{[8]}\big] + \eta_b\big[{}^1{\pwave}_1^{[8]}\big]$ & $25.29(9)$ \\ 
\bottomrule
\end{tabular}
\caption{\it\small Fock-state decomposition of the total cross section for associated $J/\psi+\eta_b$ production in $pp$ collisions at $\sqrt{s} = 13\,{\rm TeV}$.
Results are obtained using the final-state reshuffling algorithm.}
\label{tab:sigma_pp2JpsiEtab}
\end{table}

While we present a more detailed investigation of $J/\psi+\psi(2\swave)$ production in section~\ref{sec:jpsipsi2sresult}, we now further consider $pp \to J/\psi + \eta_b$ to highlight another feature of the \mgshort\ architecture: each  Fock-state contribution is stored independently, enabling a detailed decomposition of the process. 
The total cross section of this process is $\sigma_{J/\psi\textrm{--}\eta_b} = 1173(4)^{+258\%}_{-74.5\%}\,{\rm pb}$, while the breakdown into Fock-state combinations is given in table~\ref{tab:sigma_pp2JpsiEtab}.\footnote{Values of order $10^{-27}$ or smaller in the {\tt HTML} output are interpreted as numerical zeros, consistently with the symmetry arguments discussed in the text. Accordingly, they are reported as zero in the table.} 
As expected, the double-CS channel vanishes at LO due to charge-conjugation symmetry and colour conservation, but becomes non-zero at $\mathcal{O}(\alpha_s^5)$ through the emission of an additional gluon~\cite{Lansberg:2013qka}. The dominant contributions at $\mathcal{O}(\alpha_s^4)$ therefore arise from the mixed-colour channels. 
Among them, most CS $J/\psi$ channels also vanish, and the only non-zero contribution is about one order of magnitude smaller than the CS $\eta_b$ channels. This can again be traced back to charge-conjugation symmetry: three gluons must couple to the $c \bar{c}$ line to satisfy this symmetry, which at LO is only possible when the $\eta_b$ is in a $^3 \swave_1^{[8]}$ configuration. However, as mentioned before, this channel involves propagators with larger virtualities than the other CS-CO mixed channels, resulting in a suppression of $m_b^4/m_c^4 \sim 85$. 
Interestingly, while the same argument does not apply to the double-$^3 \swave_1^{[8]}$ channel, for which charge parity is not a good quantum number, this contribution is still suppressed by the additional CO LDME. Since the two suppression mechanisms are of comparable size, the two channels, $J/\psi[^3\swave_1^{[1]}]+\eta_b[^3\swave_1^{[8]}]$ and $J/\psi[^3\swave_1^{[8]}]+\eta_b[^3\swave_1^{[8]}]$, are found to have similar magnitudes.
Finally, we observe that the largest contribution to the cross section comes from $J/\psi[^3\pwave_J^{[8]}]+\eta_b[^1\swave_0^{[1]}]$.
Although this channel is, in principle, a potential probe of CO $\psi(N\swave)$ LDMEs, the $\eta_b$ meson is challenging to detect, and no experimental measurement has yet demonstrated the feasibility of such a study.

\subsection{Charmonium production in electron-positron collisions}
\begin{table}[t!]
\centering
\renewcommand*{\arraystretch}{1.4}
\textbf{$\boldsymbol{e^+e^-}$ collision: inclusive and exclusive charmonium production}
\begin{tabular}[t]{lcccc}
\toprule
\multirow{2}{*}{\textbf{final state}}  & \multirow{2}{*}{$\left(\dfrac{m_c}{\rm GeV}, \dfrac{m_b}{\rm GeV}\right)$}& \multicolumn{3}{c}{$\boldsymbol{\sigma}~[\textbf{fb}]$} \\
& & \textbf{no resh.} & \textbf{FS resh.} & \textbf{IS resh.} \\
\midrule
\multirow{3}{*}{$\psi(2\swave) + g + X$} 
& $(1.55,4.7)$ & $202.82 (3)$ & $189.10(3)$ & $211.78(3)$ \\ 
& $(1.845,4.7)$ & {$299.20(4)$} & {$293.58(4)$} & {$303.21(4)$} \\  
& $(1.945,4.7)$ & $338.490(4)$ & $338.490(4)$ & $338.490(4)$ \\ 
\hdashline
\multirow{1}{*}{$\psi(2\swave)\big[{}^3\swave_1^{[1]}\big] + gg$} 
& $(1.55,4.7)$ & $127.27(11)$ & $90.86(8)$ & $116.25(10)$ \\ 
{\qquad\qquad\quad$+\ X$}
& $(1.845,4.7)$ & {$130.12(11)$} & $130.12(11)$ & $130.12(11)$ \\ 
\hdashline
\multirow{1}{*}{$\psi(2\swave) + c \bar{c} + X$}
& $(1.55,4.7)$ & $63.08 (4)$ & $46.27 (3)$ & $50.54 (3)$ \\ 
\hdashline
\multirow{1}{*}{$h_c + c \bar{c} + X$} 
& $(1.55,4.7)$ & $7.924(6)$ & $6.551(5)$ & $6.599(5)$ \\ 
\hdashline
\multirow{2}{*}{$J/\psi + \chi_{c0}$} 
& $(1.55,4.7)$ & $4.6551(4)$ & $4.4892(3)$ & $4.9535(4)$ \\ 
& $(1.705,4.7)$ & $7.2112(7)$ & $7.4915(7)$ & $6.8110(7)$ \\ 
\hdashline
\multirow{2}{*}{$J/\psi + \chi_{c1}$} 
& $(1.55,4.7)$ & $0.76830(9)$ & $0.73091(8)$ & $0.81499(9)$ \\ 
& $(1.755,4.7)$ & $1.16544(16)$ & $1.22910(17)$ & $1.11965(16)$ \\ 
\hdashline
\multirow{2}{*}{$J/\psi + \chi_{c2}$} 
& $(1.55,4.7)$ & $1.04835(4)$ & $0.99371(3)$ & $1.09331(4)$ \\ 
& $(1.78,4.7)$ & $1.43786(7)$ & $1.52470(8)$ & $1.41769(7)$ \\ 
\hdashline
\multirow{3}{*}{$\chi_{c0} + \chi_{c2}$}
& $(1.55,4.7)$ & $1.3914(3)\cdot10^{-5}$ & $1.2619(2)\cdot10^{-5}$ & $1.2242(2)\cdot10^{-5}$ \\ 
& $(1.705,4.7)$ & $1.7487(3)\cdot10^{-5}$ & $1.7131(3)\cdot10^{-5}$ & $1.6918(3)\cdot10^{-5}$ \\ 
& $(1.78,4.7)$ & $1.8697(3)\cdot10^{-5}$ & $1.9098(4)\cdot10^{-5}$ & $1.9407(4)\cdot10^{-5}$ \\  
\hdashline
\multirow{3}{*}{$\chi_{c1} + h_{c}$}
& $(1.55,4.7)$ & $0.63456(7)$ & $0.57200(8)$ & $0.69509(7)$ \\ 
& $(1.755,4.7)$ & $0.88768(9)$ & $0.88523(9)$ & $0.88871(9)$ \\ 
& $(1.765,4.7)$ & $0.89843(9)$ & $0.9009(1)$ & $0.89745(9)$ \\ 
\bottomrule
\end{tabular}
\caption{\it\small Same as table~\ref{tab:sigma_pp2Qj}, but for charmonium production in $e^+ e^-$ collisions at $\sqrt{s} = 10.58\,{\rm GeV}$, with $E_{e^-} = 7\,{\rm GeV}$ and $E_{e^+} = 4\,{\rm GeV}$. No kinematic cuts are applied. For quark masses different from the default values, the LDMEs are rescaled by a mass-dependent factor (see text).}
\label{tab:sigma_e+e-}
\end{table}

The last class of processes we consider are inclusive and exclusive charmonium production processes in $e^+ e^-$ collisions at $\sqrt{s} = 10.58\,{\rm GeV}$. All processes are IR finite at LO, and no kinematic cuts are applied. The \mgshort\ results are reported in table~\ref{tab:sigma_e+e-}.

For some of the processes presented here, rescaling the quark mass parameter can lead to substantially larger cross sections than those obtained in the default setup.
In addition, unlike the hadronic collisions discussed previously, initial- and final-state reshuffling affect some of the cross sections in opposite directions.
Taken together, these effects lead to the relatively larger deviations observed in the table when different mass parameters and reshuffling procedures are employed.
To understand this behaviour, it is again useful to compare the observed variation with the expected mass scaling, $\sum_{\cal Q}\Delta M_{\cal Q}/M_{\cal Q}$.
The $\Delta \sigma^{\rm resh}$ associated with the $2\to3$ production processes $\psi(2\swave) + c\bar{c}$ and $h_c + c\bar{c}$ satisfy the expected scaling.
The same conclusion holds for the CS contribution to inclusive $\psi(2\swave)$ production, $\psi(2\swave) + gg$, for which we find $\Delta \sigma^{\rm resh} \approx 19\%$, while $\Delta M_{\psi(2\swave)}/M_{\psi(2\swave)} \approx 16\%$. Hence, in this case, we consider the cross section obtained within the reshuffling procedure to provide a more realistic estimate than that obtained from the charm-mass rescaling proposed here.
By contrast, $\Delta \sigma^{\rm resh}\approx1\%$ for the CO channel, $\psi(2\swave) + g$, does not reproduce the expected scaling, $\Delta M_{\psi(2\swave)}^{[8]}/M_{\psi(2\swave)}^{[8]} \approx 18\%$. In this case, adopting a heavier charm-mass value appears to be the more reliable approach. 
For the exclusive double-quarkonia processes, the situation is more nuanced, even among processes that might appear similar a priori.
We observe that for $J/\psi + \chi_{cJ}$ and $\chi_{c1} + h_c$, where initial- and final-state reshuffling shift the cross section in opposite directions, we find $\Delta \sigma^{\rm resh} < \sum_{\cal Q}\Delta M_{\cal Q}/M_{\cal Q}$ across the different charm masses considered. This suggests that, for these processes, our approximation does not capture the dominant physical mass effects and that the contribution from relativistic corrections is underestimated. In these cases, rescaling the quark-mass parameter again appears to provide a more realistic result. By contrast, for $\chi_{c0} + \chi_{c2}$, we find that $\Delta \sigma^{\rm resh}$ and the charm-mass variations are of comparable size, both being consistent with the expected scaling. However, the two methods modify the cross section in opposite directions, leading to a less clear-cut situation.

A few comments on notable physical features of our results are in order.
All exclusive double-charmonium processes have cross sections of ${\cal O}(1\,{\rm fb})$, except for $\chi_{c0} + \chi_{c2}$, whose cross section is significantly smaller. At $\mathcal{O}(\alpha_s^2\alpha^2)$, these processes are produced via $s$-channel diagrams mediated by a photon or a $Z$ boson. For $\chi_{c0} + \chi_{c2}$ production, single-photon exchange violates charge-conjugation symmetry, and the process therefore proceeds only via a highly off-shell $Z$ boson, resulting in a much smaller cross section. However, the $\mathcal{O}(\alpha^4)$ contribution to $\chi_{c0} + \chi_{c2}$ production via two virtual $s$-channel photons is non-zero, and exceeds the cross section presented at $\mathcal{O}(\alpha_s^2\alpha^2)$ reported in table~\ref{tab:sigma_e+e-} by roughly one order of magnitude. 
The other exclusive processes discussed in table~\ref{tab:sigma_e+e-} exhibit a distinct hierarchy. The most instructive example is $J/\psi + \chi_{c1}$, whose cross section is significantly smaller than that of the $J/\psi + \chi_{c0}$ processes and comparable in magnitude to $h_c + \chi_{c1}$, although the $h_c$ LDME is much smaller than that of $J/\psi$. The associated production of $J/\psi + \chi_{c1}$ is subjected to helicity and parity suppressions~\cite{Dong:2011fb}. 
Each helicity configuration exhibits an asymptotic behaviour governed by a power of ${m_c^2}/{s}$, as dictated by the helicity selection rule~\cite{Brodsky:1981kj}.
By contrast, the double-\pwavetxt\ exclusive process $h_c + \chi_{c1}$ is not affected by these suppressions. At the energy considered here, the smaller $h_c$ LDME in the $h_c+\chi_{c1}$ process and the suppressions in the $J/\psi+\chi_{c1}$ are of comparable size, yielding similar cross sections.
These provide two new examples of the point highlighted in ref.~\cite{ColpaniSerri:2025vdz}: the impact of subleading contributions is often difficult to predict using simple counting arguments based solely on the hierarchy of couplings and velocity-scaling rules.

Another interesting process is inclusive $\psi(2\swave) + g$ production, which at $\mathcal{O}(\alpha_s\alpha^2)$ is entirely dominated by CO contributions, making it a benchmark channel for CO LDME extraction, similarly to inclusive $J/\psi$~\cite{Braaten:1995ez,Zhang:2009ym}. 
Although this channel is LO in the perturbative expansion in $\alpha_s$, as in the $J/\psi$ case, a significant contribution to inclusive $\psi(2\swave)$ production at $B$-factory energies arises from the CS channel $e^+e^-\to\gamma^*\to \psi(2\swave)+gg+X$, whose leading contribution is at $\mathcal{O}(\alpha_s^2\alpha^2)$, as shown in table \ref{tab:sigma_e+e-}. The suppression by one additional power of $\alpha_s$ is partially compensated by the larger CS LDME.  
We also note that the hierarchy observed in table~\ref{tab:sigma_e+e-}, namely $\sigma(e^+e^-\to \psi(2\swave)+g+X) > \sigma(e^+e^-\to \psi(2\swave)\big[{}^3\swave_1^{[1]}\big]+gg+X)$, may change when different sets of CO LDMEs are used. In addition, the impact of the aforementioned relativistic corrections~\cite{He:2009uf,Jia:2009np,Li:2026zsu,Jiang:2026ieb}, QCD radiative corrections~\cite{Zhang:2009ym,Ma:2008gq,Gong:2009kp}, and initial-state photon radiation~\cite{Shao:2014rwa,Gong:2019rpd}, which are not (fully) included here, should not be underestimated.

Investigating this class of processes at $e^+ e^-$ colliders shows a different behaviour of momentum reshuffling compared to the hadronic case discussed previously. In particular, the majority of the processes considered here exhibit $\Delta \sigma^{\rm resh} < \Delta \sigma^{\rm algo}$, while for the exclusive double-charmonium processes this effect is compounded by the two reshuffling procedures shifting the cross section in opposite directions. We again emphasise that such behaviour is not easily predictable, as it depends on the interplay of multiple factors. As a result, it may differ significantly even between processes that, a priori, one might expect to behave similarly, such as $\chi_{c0} + \chi_{c2}$ and $\chi_{c1} + h_c$.
The variety of situation presented here further highlight that a proper estimate of mass-treatment uncertainties requires a systematic comparison of multiple reshuffling algorithms for multiple constituent-mass choices, rather than reliance on a single prescription.

\subsection[\texorpdfstring{$J/\psi + \psi(2\swave)$}{J/psi + psi(2S)} hadroproduction: a case study]{\boldmath $J/\psi + \psi(2\swave)$ hadroproduction: a case study}\label{sec:jpsipsi2sresult}

\begin{figure}[htb!]
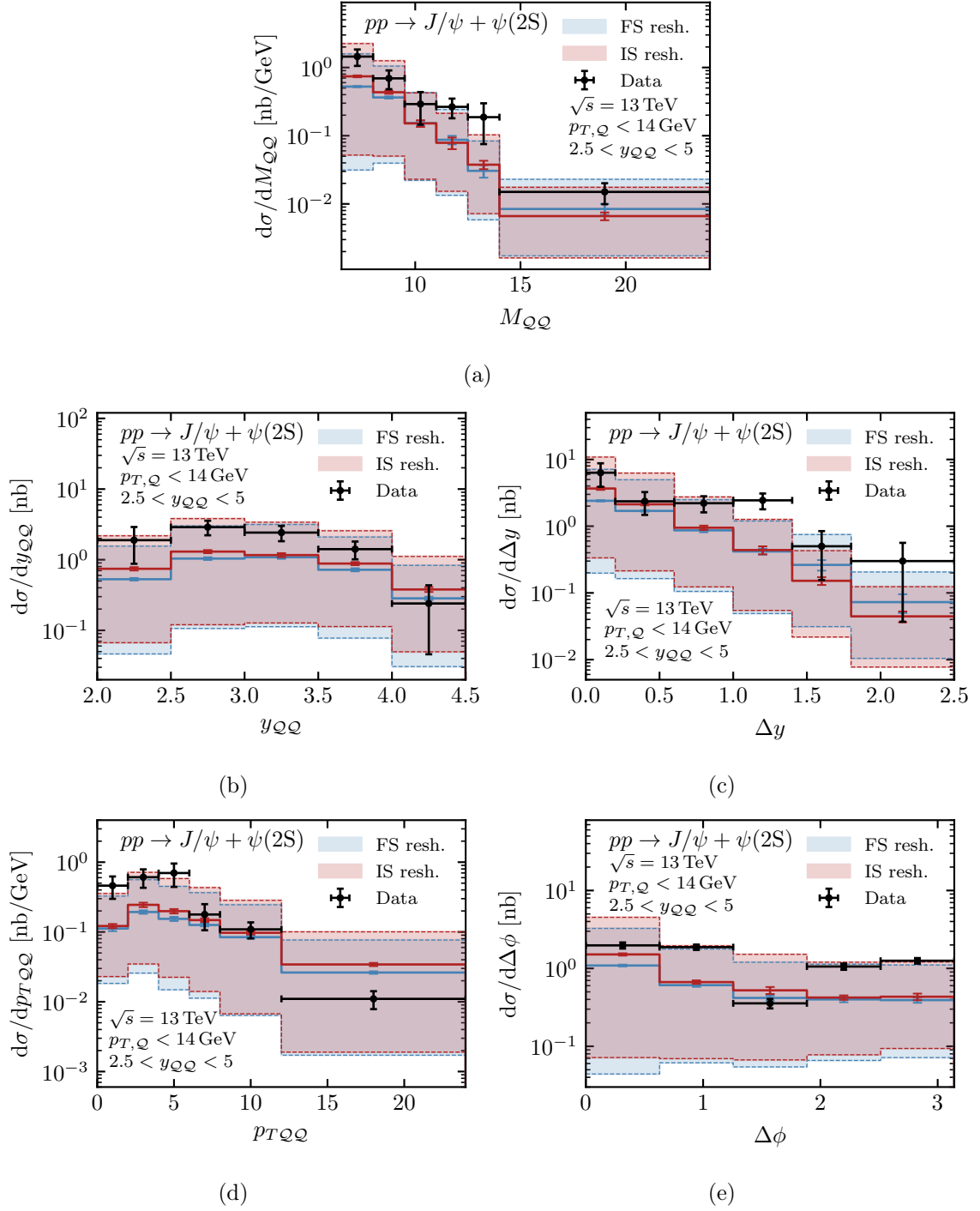

  \centering    
  \begin{subfigure}[t]{0.49\textwidth}
      \input{figures/mQQ.pgf}
      \caption{}
      \label{subfig:mqq}
  \end{subfigure}\\
  \begin{subfigure}[t]{0.49\textwidth}
      \input{figures/rapQQ.pgf}
      \caption{}
      \label{subfig:yqq}
  \end{subfigure}
  \begin{subfigure}[t]{0.49\textwidth}
      \input{figures/rap_diff.pgf}
      \caption{}
      \label{subfig:Dy}
  \end{subfigure}
  \begin{subfigure}[t]{0.49\textwidth}
      \input{figures/ptQQ.pgf}
      \caption{}
      \label{subfig:ptqq}
  \end{subfigure}
  \begin{subfigure}[t]{0.49\textwidth}
      \input{figures/ang_diff.pgf}
      \caption{}
      \label{subfig:Dphi}
  \end{subfigure}
  \caption{\it Differential cross sections for $J/\psi + \psi(2\swave)$ production at the LHC as functions of (a) the pair invariant mass, (b) the pair rapidity, (c) the absolute rapidity difference, (d) the pair transverse momentum, and (e) the azimuthal angle difference. Fiducial cuts are indicated in the figures. Theoretical predictions are obtained by combining {\tt \mgshort} with {\tt \pythia8}. The LHCb data are taken from ref.~\cite{LHCb:2023wsl}.\vspace*{-0.2\baselineskip}
  }
\label{fig:xsec-jpsi-psi2S}
\end{figure}

As our last example, we present a realistic phenomenological application of our framework to the associated production of $J/\psi+\psi(2\swave)$ at the LHC.
This process has been measured by the LHCb collaboration~\cite{LHCb:2023wsl} at $\sqrt{s}=13$ TeV. The motivation for studying this process is multifaceted. It provides an ideal probe of quarkonium production mechanisms~\cite{Qiao:2002rh,Qiao:2009kg,Lansberg:2013qka,Sun:2014gca,Baranov:2015cle,Lansberg:2015lva,He:2019qqr,Lansberg:2019fgm,Lansberg:2020rft,Sun:2023exa,He:2024ugx,He:2025kkw}, DPS~\cite{Kom:2011bd,Lansberg:2014swa,Borschensky:2016nkv,Prokhorov:2020owf}, fully charmed tetraquark states~\cite{ATLAS:2023bft,CMS:2025vnq}, and transverse-momentum-dependent gluon distributions in protons~\cite{Lansberg:2017dzg,Scarpa:2019fol}. We consider only the SPS contribution at $\mathcal{O}(\alpha_s^4)$, including all CS and CO Fock states up to order $v^7$ (see table \ref{tab:NRQCD_scaling}) and the relevant feed-down contributions. 
Our theory-data comparison is shown in figure~\ref{fig:xsec-jpsi-psi2S}. The bands represent the $7$-point uncertainty obtained from independent variations of the factorisation and renormalisation scales by factors of two, while the statistical MC uncertainties are shown as error bars. 
Following the experimental analysis, we require each quarkonium to satisfy $p_{T{\cal Q}} < 14\,{\rm GeV}$ and $2.5 < y_{\cal Q} < 5$. In addition, we employ our default setup (see appendix~\ref{app:setup}) to generate the parton-level events.

This process provides an ideal benchmark to demonstrate the capabilities of \madsons. 
First, the mass difference between the two produced quarkonia, the $J/\psi$ and $\psi(2\swave)$, is $590~{\rm MeV}$, hence a sensitivity to the mass treatment is expected. More specifically, as we take $m_c = 1.55~{\rm GeV}$, the order of the deviation of the direct channel is ${\Delta M_{\psi(2\swave)}/M_{\psi(2\swave)} \sim 16\%}$; similarly for the feed-down contribution from $\psi(2\swave) + \psi(2\swave)$ we have $2 \Delta M_{\psi(2\swave)}/M_{\psi(2\swave)} \sim 32\%$, while $\Delta M_{\psi(2\swave)}/M_{\psi(2\swave)} + \Delta M_{\chi_{cJ}}/M_{\chi_{cJ}} \sim 25\textrm{--}28\%$ for $\chi_{cJ} + \psi(2\swave)$, depending on $J=0,1,2$.
For all these channels, momentum reshuffling is thus required for a consistent treatment.
We consider initial- and final-state reshuffling separately to explicitly illustrate the differences between the two algorithms. Although not performed in this work, the two results can be combined in a further step into a single band that includes the additional theoretical uncertainty associated with the choice of momentum reshuffling algorithm.
Moreover, {\pwavetxt} contributions in both direct and feed-down channels are not negligible.
While the process is dominated by the CS-CS channel, it also receives contributions from CO states, including single and double {\pwavetxt} channels which collectively account for $\sim17\%$ of the total fiducial cross section, and $J/\psi$ feed-down from $\psi(2\swave)$ and $\chi_{cJ}$ states, which contribute roughly $25\%$ of the total fiducial cross section. 

To account for quarkonium decays, we interface the {\mgshort} LO events with the parton-shower, hadronisation, and decay modules of \pythia~{\tt v8.315}~\cite{Bierlich:2022pfr}. We enable multi-parton interactions, intrinsic-$k_T$ smearing, and initial- and final-state radiation. The $J/\psi$ is kept stable, while all other quarkonia are allowed to decay. After the full event generation with \pythia{\tt 8}, double-$J/\psi$ events are removed from both direct and feed-down channels by requiring exactly one $J/\psi$ in the final state. The resulting prediction is then reweighted by the branching ratio ${\rm Br}\big(\psi(2\swave)\to {\rm non-}J/\psi\big)\approx 46\%$ to match the unfolded experimental cross section.

The measurement reports a fiducial cross section of
\begin{equation}
\sigma^{\rm LHCb}_{J/\psi+\psi(2\swave)} = 4.5\pm{0.7}({\rm stat.})\pm{0.3} ({\rm syst.})\,{\rm nb}\,,
\end{equation}
which should be compared with our theoretical predictions,
\begin{align}
    \sigma^{\rm IS~resh.}_{J/\psi+\psi(2\swave)} & = 2.23^{+192\%}_{-89\%}\,{\rm nb}\,,\\ 
    \sigma^{\rm FS~resh.}_{J/\psi+\psi(2\swave)} & = 1.83^{+191\%}_{-90\%}\,{\rm nb}\,,
\end{align}
corresponding to initial- and final-state reshuffling, respectively. The quoted theoretical uncertainties are estimated using the standard $7$-point scale variation prescription. The difference between the two central values corresponds to $\Delta \sigma^{\rm algo} \approx 10\%$. Although both central values lie below the measured cross section, the large scale uncertainties affecting our LO prediction render them compatible with the data.
The comparison between our theoretical predictions and the experimental data for five different distributions is shown in figure~\ref{fig:xsec-jpsi-psi2S}. 
Overall, we find satisfactory agreement, with the predictions being either comparable to or below the data. 
Although leading-logarithmic QCD radiation is included through the parton shower, it should be noted that our study does not include full NLO QCD corrections, complete relativistic corrections, or DPS contributions.
Their inclusion could compensate for the mismatch with the data observed in some bins, leading to an increase of the integrated cross section as well. We further comment on this possibility in the discussion of each distribution below.

We first consider figure~\ref{subfig:mqq}, which shows the differential cross section as a function of the quarkonium-pair invariant mass $M_{\cal QQ}$. For $M_{\cal QQ} \lesssim 10\,{\rm GeV}$, where SPS is expected to dominate, both reshuffling algorithms provide predictions in satisfactory agreement with the data.
Nonetheless, the large scale uncertainties prevent us from reaching firmer conclusions. A priori, including higher-order terms in the $\alpha_s$ expansion should reduce these uncertainties, provided that the perturbative expansion remains stable.\footnote{Possible remedies for perturbative instabilities in quarkonium processes are discussed in refs.~\cite{Shao:2018adj,Maxia:2026fbz}.}
For $M_{\cal QQ} > 10\,{\rm GeV}$, the DPS contribution is expected to become ever more relevant, as its cross section should decrease more slowly with $M_{\cal QQ}$ than that of SPS. Consistently, our predictions tend to undershoot the data in this region. Hence, compared with the predictions presented in ref.~\cite{LHCb:2023wsl}, our result improves the description by leaving more room for DPS. The reduction of the cross section is primarily driven by momentum reshuffling, as also shown in table~\ref{tab:sigma_pp2QQ}.
Moreover, the sizeable $\Delta \sigma^{\rm algo}$ correction mentioned above  is reflected in the distribution of figure~\ref{subfig:mqq}. This effect is driven by the first bins, which contribute most to the cross section and whose central values differ significantly depending on the algorithm employed. By contrast, the separation observed in the last two bins is largely due to the increased statistical MC uncertainty at large $M_{\cal QQ}$ and should not be interpreted as a genuine deviation between the two algorithms.

Figures~\ref{subfig:yqq} and~\ref{subfig:Dy} show the cross-section dependence on the rapidity of the quarkonium pair and the absolute difference between the quarkonium rapidities, respectively. As before, our theoretical predictions improve upon those shown in ref.~\cite{LHCb:2023wsl}, since they maintain good agreement with the data while leaving room for DPS contributions. For instance, figure~\ref{subfig:Dy} presents features analogous to those observed in figure~\ref{subfig:mqq}, where the SPS prediction falls faster than the measurement. This behaviour is expected from studies of $J/\psi$-pair production~\cite{Kom:2011bd,Lansberg:2014swa}, as the large $\Delta y$ region is where DPS, which produces a broader spectrum, becomes dominant. Again, while the discrepancy between the two reshuffling procedures in the first bin represents a genuine difference, the separation observed in the last bins can be traced back to the increased statistical MC uncertainty.

Figures~\ref{subfig:ptqq} and~\ref{subfig:Dphi} show the distributions in the transverse plane. It is important to note that, at LO, these distributions for $p_{T\mathcal{QQ}}>0$ and $\Delta \phi<\pi$ are entirely determined by the perturbative parton-shower and non-perturbative intrinsic-$k_T$ smearing models implemented in \pythia{\tt 8}, and are consequently sensitive to their modelling choices. For this reason, together with the large scale uncertainties affecting the transverse-momentum spectrum of the quarkonium pair, we find the description of the $p_{T{\cal QQ}}$ distribution satisfactory. Similar conclusions apply to the azimuthal angle difference, where we note that the last bin is particularly sensitive to the $k_T$-smearing procedure. A similar sensitivity of the azimuthal difference to $k_T$ effects in double-$J/\psi$ production has already been pointed out in the literature~\cite{Kom:2011bd}.

Overall, figure~\ref{fig:xsec-jpsi-psi2S} illustrates the  capabilities of \madsons. The inclusion of \pwavetxt\ states in the framework enables the systematic treatment of all phenomenologically relevant CO channels and feed-down contributions.
Moreover, the improved mass treatment reduces the overall $J/\psi + \psi(2\swave)$ cross section, leading to an improved description of the experimental data compared with previous predictions.

\section{Leptonium production}
\label{sec:leptionium}

The QED analogues of quarkonia are bound states composed of a charged lepton and an antilepton. The six possible flavour combinations of purely leptonic atoms, collectively referred to as leptonia, are positronium ($e^-e^+$), muonium ($e^{-}\mu^{+}$), tauonium ($e^{-}\tau^{+}$), dimuonium ($\mu^-\mu^+$), mu--tauonium ($\mu^{-}\tau^{+}$), and ditauonium ($\tau^-\tau^+$), together with their charge-conjugated counterparts where applicable. Unlike NRQCD, the LDMEs governing bound-state formation in NRQED can be computed perturbatively. The explicit expressions for the \swave- and {\pwavetxt} leptonium LDMEs, derived from the Coulomb potential, are given in eqs.~\eqref{eq:ldme_leptonia_swave} and \eqref{eq:ldme_leptonia_pwave} in appendix~\ref{app:setup}, respectively.

\begin{figure}[t]
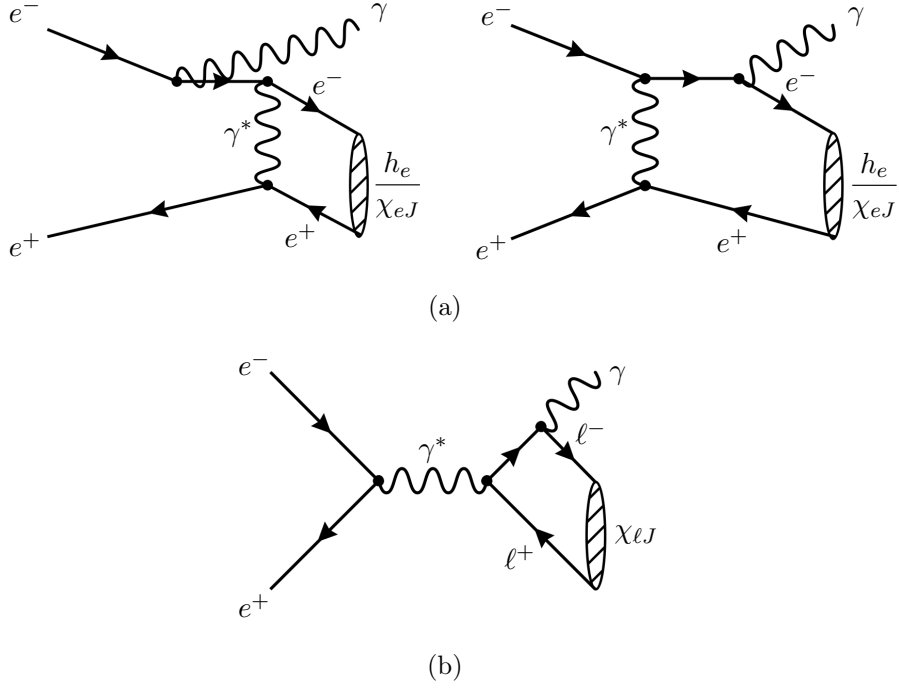

    \centering
    \begin{subfigure}[b]{\textwidth}
        \centering
        \def\svgwidth{0.4\textwidth}
        \input{feynman/lep_1.tex}
        \def\svgwidth{0.4\textwidth}
        \input{feynman/lep_2.tex}
        \caption{}\label{fig:diag_leptonium_a}
    \end{subfigure}
    \begin{subfigure}[b]{0.4\textwidth}
        \centering
        \def\svgwidth{\textwidth}
        \input{feynman/lep_4.tex}
        \caption{}\label{fig:diag_leptonium_b}
    \end{subfigure}
    \caption{\it\small {Representative Feynman diagrams for leptonium production at $e^+e^-$ colliders: (a) para- or ortho-positronium production via $t$-channel photon exchange; (b) ortho-leptonium ($\ell=e,\mu,\tau$) production in association with a photon from final-state radiation. The diagrams were generated using {\tt FeynGame}~\cite{feyngame,feyngame2,feyngame3}.}}
    \label{fig:diag_leptonium}
\end{figure}

With the upgraded implementation in {\madsons}, it is now possible to compute cross-section predictions not only for \swavetxt{} leptonia, as already demonstrated in ref.~\cite{ColpaniSerri:2025vdz}, but also for \pwavetxt{} configurations. To validate the NRQED implementation, we compare cross sections obtained with \mgshort{} against analytic calculations. In particular, we consider the processes $e^+e^- \to {\cal L}(2\pwave)+\gamma$ in the limit of infinitely large $Z$- and Higgs-boson masses, complementing the study of \swavetxt{} para- and ortho-positronium presented in figure~6 of ref.~\cite{ColpaniSerri:2025vdz}. The final-state leptonia ${\cal L}(2\pwave)$ considered here are para-positronium $h_e(2\pwave)$, corresponding to the ${}^1\pwave_1$ Fock state, as well as ortho-positronium $\chi_{eJ}(2\pwave)$ and ortho-ditauonium $\chi_{\tau J}(2\pwave)$, corresponding to the ${}^3\pwave_J$ Fock states. Para-ditauonium production, $e^+e^-\to h_\tau(2\pwave)+\gamma$, is instead forbidden by charge-conjugation symmetry in QED.
The representative Feynman diagrams contributing to these processes are shown in figure~\ref{fig:diag_leptonium}. All diagrams contribute to positronium production, while only the diagram in figure~\ref{fig:diag_leptonium_b} contributes to ditauonium production.

Before presenting the benchmark results, we comment on the different conventions for the principal quantum number used in quarkonium and leptonium spectroscopy. In quarkonium physics, the principal quantum number is conventionally defined as $N=n_r+1$, where $n_r$ denotes the radial quantum number, counting the number of radial nodes and vanishing for the ground state. In hydrogen-like Coulombic systems, however, the principal quantum number is defined such that the lowest \pwavetxt{} state corresponds to $N=2$. We adopt this latter convention for leptonia, as it naturally reflects the degeneracy of the Coulomb potential among states with the same principal quantum number, such as $N\swave$ and $N\pwave$ levels.


Since the leptons forming the bound state must be kept massive, the \ufo{} model has to be loaded with the \texttt{lepton\_masses} restriction. In addition, the standard \mgshort{} syntax can be used to exclude diagrams involving $Z$ or Higgs bosons, thereby reproducing the limit in which electroweak bosons are infinitely heavy. For example, the para-positronium production process is generated via the commands
\prompt{import model sm\_onia-lepton\_masses}
\vskip -0.25truecm
\prompt{generate e+ e- > Ps(2|1P1) a / z h; output; launch}
\noindent
To account for the low energy scale relevant for leptonium production, we employ the $\alpha(0)$ scheme rather than the $G_\mu$ scheme, which is the default setup of the \texttt{sm\_onia} model. Since we only consider contributions involving internal quasi-real photons, this can be achieved by setting the first entry of \texttt{BLOCK SMINPUTS} in the \texttt{param\_card.dat} file to $137.036$, corresponding to the inverse electromagnetic coupling in the Thomson limit,
\begin{equation}
\alpha(0)=\dfrac{1}{137.036}\,.
\end{equation}
In all calculations, the leptonium mass is chosen as the sum of the constituent lepton masses,
\begin{equation}
M_{\mathcal L}=m_{\ell^-}+m_{\ell^{\prime+}}\,,
\end{equation}
\vspace{\fill}
\begin{figure}[h]
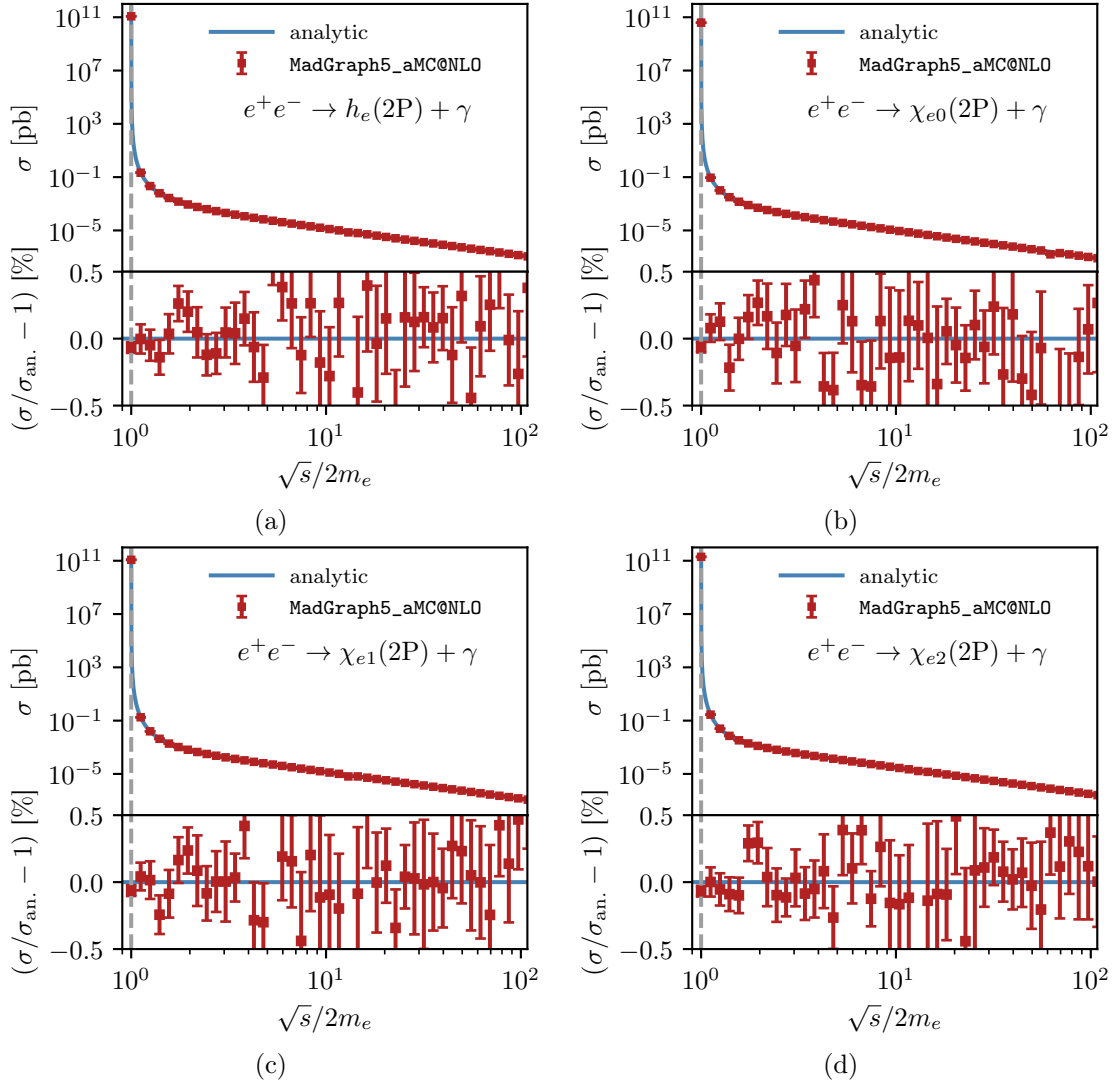

    \centering
    \begin{subfigure}[b]{0.49\textwidth}
        \input{figures/xsec_paraEE.pgf}
        \caption{}
        \label{fig:xsec_EE_a}
    \end{subfigure}
    \begin{subfigure}[b]{0.49\textwidth}
        \input{figures/xsec_orthoEE0.pgf}
        \caption{}
        \label{fig:xsec_EE_b}
    \end{subfigure}
    \begin{subfigure}[b]{0.49\textwidth}
        \input{figures/xsec_orthoEE1.pgf}
        \caption{}
        \label{fig:xsec_EE_c}
    \end{subfigure}
    \begin{subfigure}[b]{0.49\textwidth}
        \input{figures/xsec_orthoEE2.pgf}
        \caption{}
        \label{fig:xsec_EE_d}
    \end{subfigure}
    \caption{\small\it Cross sections as functions of the c.m.\@ energy for the production in $e^+ e^-$ collisions of a photon in association with a \pwavetxt\ (a) para- and (b-d) ortho-positronium. In each figure, the upper panels show the total cross section obtained with {\tt\mgshort} (red dots) and from analytical expressions (blue line), whereas the lower panels display the ratio, normalised to the analytical reference values. Error bars on the {\tt \mgshort} points indicate statistical MC uncertainties.}
    \label{fig:xsec_EE}
\end{figure}
\vspace{\fill}
\newpage\noindent
such that no momentum reshuffling is required. This choice also considerably simplifies the analytic evaluation of the cross section, allowing for a direct comparison with the \mgshort{} predictions. 

For $e^+e^-\to h_e(N\pwave)+\gamma$, the analytic cross section reads
\begin{equation}
\begin{aligned}
    & \lim_{m_Z,m_H\to\infty}\sigma\left(e^+e^-\to h_e(N\pwave)+\gamma\right)\\
    & \qquad = \dfrac{\pi\alpha^8}{24 s}\dfrac{N^2-1}{N^5}\dfrac{1}{\left(1 - \xi_{e}\right)^5}\, 
    \left\lbrace\sqrt{1 - \xi_{e}}\,\bigg\lbrack 36 - 110 \xi_{e} + 82 \xi_{e}^{2} + 18 \xi_{e}^{3} - 38 \xi_{e}^{4} \bigg\rbrack\right.\\
    &\qquad\quad \left. - \xi_{e}\log\!\left(\dfrac{1 - \sqrt{1 -\xi_{e}}}{1 + \sqrt{1 - \xi_{e}}}\right) \bigg\lbrack 16 - 5 \xi_{e} - 29 \xi_{e}^{2} + 31 \xi_{e}^{3} - 7 \xi_{e}^{4} \bigg\rbrack \right\rbrace\\
    & \quad 
    \overset{s\gg m_e^2}{\longrightarrow} \dfrac{3\pi\alpha^8}{2s}\dfrac{N^2-1}{N^5}\,,
\end{aligned}
\end{equation}
where $\xi_e=4m_e^2/s$.
The comparison of the total exclusive cross section is shown in figure~\ref{fig:xsec_EE_a} as a function of the ratio of the c.m.\@ energy to the positronium mass, $\sqrt{s}/2m_e$. The agreement between the analytic calculation and the independent \mgshort{} prediction is excellent over the entire energy range considered, from the production threshold ($\sqrt{s}\gtrsim 2m_e$) to the asymptotic high-energy regime ($\sqrt{s}\gg 2m_e$), where the electron mass becomes negligible. The relative differences remain well within the numerical MC uncertainties and are typically at the few-permille level.
Similar conclusions hold for the spin-triplet ortho-positronium production processes, $e^+e^-\to\chi_{eJ}+\gamma$, shown for $J=0,\ 1,\ \text{and}\ 2$ in figures~\ref{fig:xsec_EE_b}, \ref{fig:xsec_EE_c}, and \ref{fig:xsec_EE_d}, respectively. The analytic expressions used for comparison are given by
\begin{align}
    & \lim_{m_Z,m_H\to\infty}\sigma\left(e^+e^-\to\chi_{e 0}(N\pwave)+\gamma\right)\nonumber \\
    &\qquad = \dfrac{\pi\alpha^8}{216s}\dfrac{N^2-1}{N^5}\dfrac{1}{\left(1-\xi_e\right)^5}\left\lbrace\sqrt{1-\xi_e}\,\bigg\lbrack 236  - 550 \xi_{e} + 348\xi_{e}^{\!2} + 176 \xi_{e}^{3} - 406 \xi_{e}^{4} + 148 \xi_{e}^{5}\right.\nonumber \\
    & \qquad\quad\left. + 30 \xi_{e}^{6} - 18 \xi_{e}^{7} \bigg\rbrack +3\xi_{e}\log\!\left(\dfrac{1-\sqrt{1-\xi_e}}{1+\sqrt{1-\xi_e}}\right)\bigg\lbrack 4 - 23 \xi_{e} + 13 \xi_{e}^{2}+ 27 \xi_{e}^{3} - 45 \xi_{e}^{4} + 18 \xi_{e}^{5} \bigg\rbrack\right\rbrace\nonumber\\
    &\quad\overset{s\gg m_e^2}{\longrightarrow} \dfrac{59\pi\alpha^8}{54s}\dfrac{N^2-1}{N^5}\,,
\end{align}
\begin{align}
    & \lim_{m_Z,m_H\to\infty}\sigma\left(e^+e^-\to\chi_{e 1}(N\pwave)+\gamma\right)\nonumber\\
    &\qquad = \dfrac{\pi\alpha^8}{72s}\dfrac{N^2-1}{N^5}\dfrac{1}{\left(1-\xi_e\right)^5}\left\lbrace\sqrt{1-\xi_e}\,\bigg\lbrack 112 - 264 \xi_{e} + 304\xi_{e}^{2} - 206 \xi_{e}^{3} + 70 \xi_{e}^{4} \right.\nonumber\\
    &\qquad\quad\left. - 48 \xi_{e}^{5} - 4 \xi_{e}^{6} \bigg\rbrack + 3\xi_{e}\log\!\left(\dfrac{1-\sqrt{1-\xi_e}}{1+\sqrt{1-\xi_e}}\right)\bigg\lbrack 8 - 24 \xi_{e} + 36 \xi_{e}^{2} - 33 \xi_{e}^{3} + 7 \xi_{e}^{4} \bigg\rbrack\right\rbrace\nonumber\\
    &\qquad\overset{s\gg m_e^2}{\longrightarrow} \dfrac{14\pi\alpha^8}{9s}\dfrac{N^2-1}{N^5}\,,
\end{align}
\begin{align}
    &\lim_{m_Z,m_H\to\infty}\sigma\left(e^+e^-\to\chi_{e 2}(N\pwave)+\gamma\right)\nonumber\\
    &\qquad = \dfrac{\pi\alpha^8}{216s}\dfrac{N^2-1}{N^5}\dfrac{1}{\left(1-\xi_e\right)^5}\left\lbrace\sqrt{1-\xi_e}\,\bigg\lbrack 736 - 2528 \xi_{e} + 3912\xi_{e}^{2} - 1166 \xi_{e}^{3} - 1634 \xi_{e}^{4}\right.\nonumber\\
    &\qquad\quad + 512 \xi_{e}^{5} + 12 \xi_{e}^{6} - 24 \xi_{e}^{7}  \bigg\rbrack+3\xi_{e}\log\!\left(\dfrac{1-\sqrt{1-\xi_e}}{1+\sqrt{1-\xi_e}}\right)\bigg\lbrack 8 - 124 \xi_{e} + 608 \xi_{e}^{2} - 717 \xi_{e}^{3}\nonumber \\
    &\qquad\quad\left. + 195 \xi_{e}^{4} \bigg\rbrack\right\rbrace \nonumber\\
    &\quad\overset{s\gg m_e^2}{\longrightarrow} \dfrac{92\pi\alpha^8}{27s}\dfrac{N^2-1}{N^5}\,.
\end{align}
In all cases, the \mgshort{} predictions agree with the analytic results within statistical MC uncertainties over the full energy range considered. This provides a strong validation of the implementation of the \pwavetxt{} projectors and LDMEs for NRQED bound states.

\begin{figure}[t]
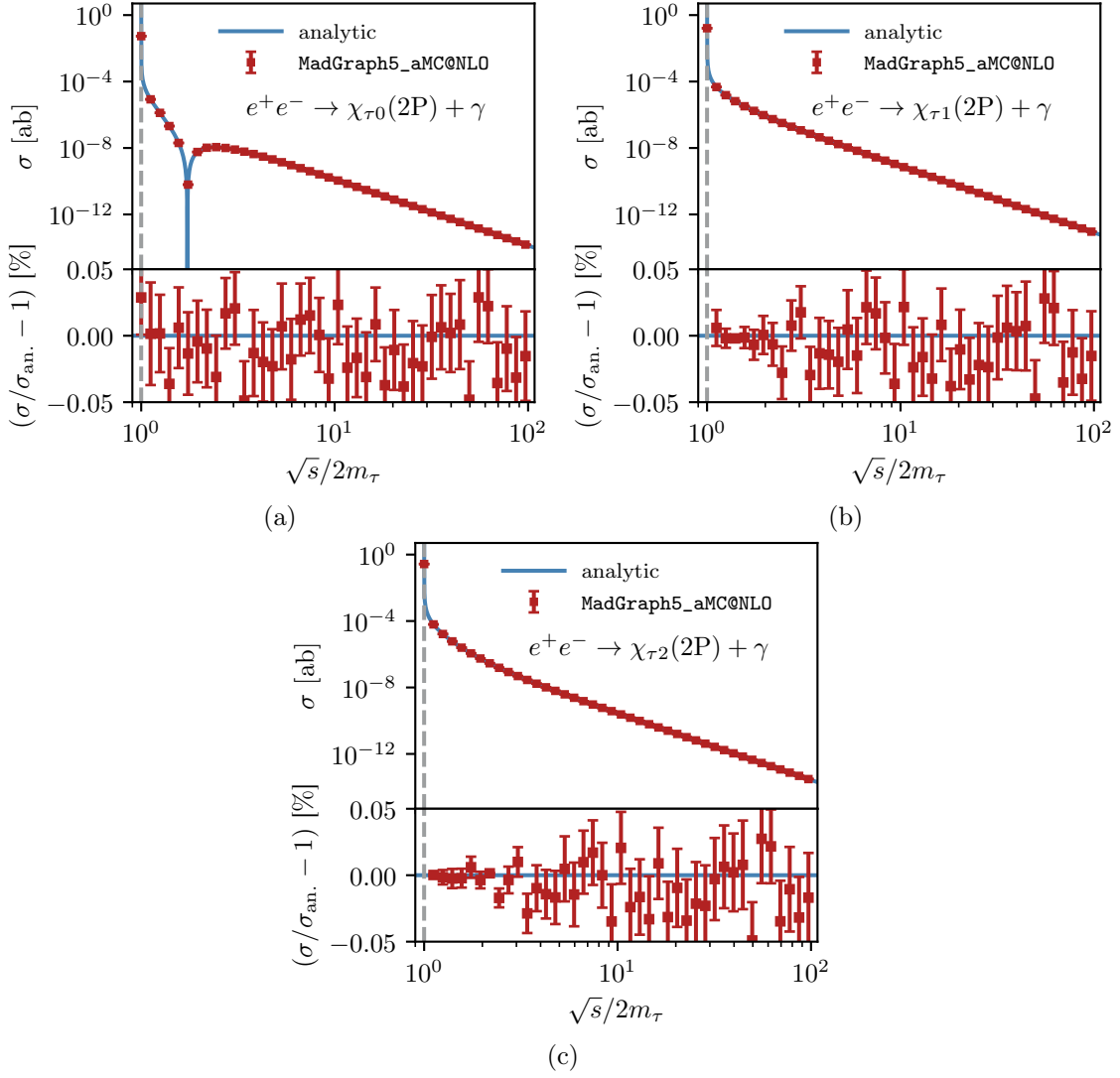

    \centering
    \begin{subfigure}[b]{0.49\textwidth}
        \input{figures/xsec_orthoTT0.pgf}
        \caption{}
        \label{fig:xsec_TT_a}
    \end{subfigure}
    \begin{subfigure}[b]{0.49\textwidth}
        \input{figures/xsec_orthoTT1.pgf}
        \caption{}
        \label{fig:xsec_TT_b}
    \end{subfigure}
    \begin{subfigure}[b]{0.49\textwidth}
        \input{figures/xsec_orthoTT2.pgf}
        \caption{}
        \label{fig:xsec_TT_c}
    \end{subfigure}
    \caption{\small\it Same as figure~\ref{fig:xsec_EE}, but for $\chi_{\tau J}$ associated production. The state $h_{\tau}(2\mathrm{P})$ is not considered, as the process $e^+e^-\to h_{\tau}(2\mathrm{P})+\gamma$ is forbidden in QED by charge-conjugation symmetry.}
    \label{fig:xsec_TT}
\end{figure}

Owing to the absence of the $t$-channel diagrams shown in figure~\ref{fig:diag_leptonium_a}, the analytic expressions for the ortho-ditauonium production processes $e^+e^-\to\chi_{\tau J}(N\pwave)+\gamma$ take a particularly compact form,
\begin{equation}
    \lim_{m_Z,m_H\to\infty}\sigma\left(e^+e^-\to\chi_{\tau 0}(N\pwave)+\gamma\right) = \dfrac{\pi\alpha^8}{54s}\dfrac{N^2-1}{N^5}\dfrac{\xi_\tau\left(1-3\xi_\tau\right)^2}{(1-\xi_\tau)}\,,
\end{equation}
\begin{equation}
    \lim_{m_Z,m_H\to\infty}\sigma\left(e^+e^-\to\chi_{\tau 1}(N\pwave)+\gamma\right) = \dfrac{\pi\alpha^8}{9s}\dfrac{N^2-1}{N^5}\dfrac{\xi_\tau\left(1+\xi_\tau\right)}{(1-\xi_\tau)}\,,
\end{equation}
\begin{equation}
    \lim_{m_Z,m_H\to\infty}\sigma\left(e^+e^-\to\chi_{\tau 2}(N\pwave)+\gamma\right) = \dfrac{\pi\alpha^8}{27s}\dfrac{N^2-1}{N^5}\dfrac{\xi_\tau\left(1+3 \xi_\tau+6\xi_\tau^2\right)}{(1-\xi_\tau)}\,,
\end{equation}
where $\xi_\tau = 4m_\tau^2/s$. The comparison between the analytic results and the \mgshort{} predictions for the three total angular momentum states $J=0,\ 1,\ \text{and}\ 2$ is shown in figures~\ref{fig:xsec_TT_a}, \ref{fig:xsec_TT_b}, and \ref{fig:xsec_TT_c}, respectively.
Over the entire kinematic range, the analytic predictions are reproduced by \mgshort{} at the sub-permille level, fully consistent with the quoted MC uncertainties. This provides further strong evidence for the validity of the implementation.

Before concluding this section, we elaborate further on the $\chi_{\tau J}$ results.
While the cross sections for $\chi_{\tau 1}(2\pwave)$ and $\chi_{\tau 2}(2\pwave)$ exhibit a smooth behaviour over the full energy range, the $\chi_{\tau 0}(2\pwave)$ production cross section vanishes at $\sqrt{s}=\sqrt{3}\,M_{\chi_{\tau 0}(2\pwave)}$. This feature is related to the composite nature of the bound state and is also present in the analogous charmonium and bottomonium production processes~\cite{Chung:2008km,Li:2009ki,Sang:2009jc,Xu:2014zra,Brambilla:2017kgw}. At LO, the observed cancellation arises from a dynamical mechanism in which different terms cancel at the amplitude level. Higher-order corrections in both the strong/electromagnetic coupling and velocity expansions are not expected to preserve this exact cancellation, so improved predictions would no longer yield a vanishing cross section at the same c.m.\@ energy. Nevertheless, since these are only subleading corrections, a pronounced dip is expected to survive, providing a characteristic signature that could be studied to test \pwavetxt\ leptonium and charmonium production experimentally, for example at future Super Tau-Charm Facilities~\cite{Achasov:2023gey,SCTF_CDR_PhysicsDetector_2018}. By contrast, this cancellation is not a generic feature of scalar particle production in association with a photon. For instance, no analogous dip occurs in the process $e^+e^-\to H+\gamma$ in an effective field theory containing an effective $\gamma\gamma H$ interaction. 

\section{Conclusions}\label{sec:conclusions}

We have presented \madsons, a new \mgshort\ module that automates the simulation of
arbitrary tree-level processes involving quarkonium and leptonium production.
Building upon the previous implementation of \swavetxt\ states~\cite{ColpaniSerri:2025vdz}, the present work completes the automation of the most phenomenologically relevant LO bound-state production processes by incorporating the orbital- and total-angular-momentum projectors required for \pwavetxt\ states.
The derivatives entering the orbital angular-momentum projector are evaluated using multivariate dual numbers, yielding exact and numerically stable results without relying on symbolic differentiation. 
The implementation has been extensively validated through comparisons with \helaconia. The squared helicity amplitudes agree for individual phase-space points at the level expected from double-precision arithmetic, while integrated cross sections are found to agree within the numerical MC uncertainties. 

We have further introduced a momentum-reshuffling procedure that reconciles the physical bound-state masses with the on-shell constituent masses required to evaluate the matrix elements. This treatment correctly reproduces the physical production thresholds and removes a limitation of conventional NRQCD calculations, particularly in processes involving several bound states composed of identical constituents but possessing different physical masses. Multiple reshuffling prescriptions, based on initial- and final-state recoil schemes, allow the residual ambiguity of the procedure to be quantified, providing a new source of theoretical uncertainty which is, in general, not negligible and should therefore be included in phenomenological studies.
The overall effect of including physical masses in the cross-section calculation is satisfactory. For most of the processes investigated in this work, our mass treatment reproduces the expected cross-section scaling, thereby improving the theoretical predictions. However, in some cases residual tensions remain, which may indicate that our implementation captures only part of the higher-order corrections in the $v$ expansion.

We illustrated the capabilities of our framework through single- and double-quarkonium production in proton-proton collisions at $\sqrt{s} = 13\,{\rm TeV}$ and charmonium production in $e^+ e^-$ collisions at $B$-factory energies. We also investigated the impact of using the measured quarkonium masses and quantified the associated reshuffling uncertainty. As a dedicated phenomenological application, we revisited $J/\psi + \psi(2\swave)$ production at the LHC, considering prompt $J/\psi$ production, including feed-down contributions from $\psi(2\swave)$ and $\chi_{cJ}$ states. The decay of the heavier states, parton showering, and hadronisation were simulated by interfacing the parton-level \mgshort\ events with \pythia{\tt 8}. Employing the physical quarkonium masses lowers the predicted cross section relative to the conventional treatment, leading to an improved description of the LHCb data while still leaving room for sizeable DPS contributions.

The framework also enables the study of leptonium production processes, providing a flexible tool for investigations of exotic atoms. In this context, we derived closed-form NRQED predictions for the considered cross sections and showed that \mgshort\ reproduces the analytic results for \pwavetxt\ para- and ortho-positronium production at the few-permille level, and for ortho-ditauonium production at the sub-permille level, across the full range of collision energies. This provides a fully independent validation of the \pwavetxt\ implementation. 

The \madsons\ module seamlessly integrates into the \mgshort\ framework, benefiting from its numerous existing features. Together with the previous \swavetxt\ implementation, this work enables the fully automated event generation of processes involving arbitrary numbers of quarkonia and leptonia within the established workflows. These developments provide a solid foundation for future extensions of \madsons, most notably towards the automation of NLO predictions within the same framework.

\begin{acknowledgments}
We thank Valentin Hirschi for suggesting the use of dual numbers for automatic differentiation. This work is supported by the ERC (grant 101041109 ``BOSON'') and the French ANR (grant
ANR-20-CE31-0015, ``PrecisOnium''). Views and opinions expressed are however those of the authors only and do not necessarily reflect those of the European Union or the European Research Council Executive Agency. Neither the European Union nor the granting authority can be held responsible for them.
\end{acknowledgments}

\appendix

\section{Computational setup}
\label{app:setup}

In this appendix, we report our computational setup, including the default parameters of the {\tt sm\_onia} {\ufo} model. The setup given here is the general one used for the validations and applications presented in the paper; deviations from it are indicated on a case-by-case basis where applicable.

\begin{table}[H]
\centering
\renewcommand*{\arraystretch}{1.4}
\begin{tabular}[t]{cc|cc}
\toprule
\textbf{parameter}& \multicolumn{1}{c}{value} & \textbf{parameter}& value \\
\midrule
$m_H$ & $125\,\mathrm{GeV}$ & $\Gamma_H$ & $6.3823393\,{\rm MeV}$ \\
$m_Z$ & $91.188\,\mathrm{GeV}$ & $\Gamma_Z$ & $2.0476\,{\rm GeV}$ \\
$m_t$\phantom{\textsuperscript{2}}  & $173\,\mathrm{GeV}$ & $\Gamma_t$ & $1.4915\,{\rm GeV}$ \\
$m_b$\phantom{\textsuperscript{2}} & $4.7\,\mathrm{GeV}$ & $\Gamma_W$ & $2.441404\,{\rm GeV}$ \\
$m_c$\footnotemark & $1.55\,\mathrm{GeV}$  & $G_\mathrm{F}$ & $1.166390\cdot10^{-5}\,\mathrm{GeV}^{-2}$ \\
$m_\tau$\textsuperscript{\ref{fnt:lepton_masses_restriction}} & $1.777\,\mathrm{GeV}$ & $\alpha_{{\scriptscriptstyle G}_\mu}$ & $1/132.507$ \\
$m_\mu$\textsuperscript{\ref{fnt:lepton_masses_restriction}} & $105.66\,\mathrm{MeV}$ &  & \\ 
$m_e$\textsuperscript{\ref{fnt:lepton_masses_restriction}} & $511\,\mathrm{keV}$ & & \\
\bottomrule
\end{tabular}
\caption{\it\small Summary of the global SM parameter settings used as defaults in the \texttt{sm\_onia} \ufo\ model and throughout all analyses presented in this article.}
\label{tab:sm_input}
\end{table}
\footnotetext{Takes a non-zero value only when the \texttt{c\_mass} restriction is enabled.\label{fnt:c_mass_restriction}}
\addtocounter{footnote}{1}
\footnotetext{Takes a non-zero value only when the \texttt{lepton\_masses} restriction is enabled.\label{fnt:lepton_masses_restriction}}

The SM input parameters (elementary particle masses, decay widths, the Fermi constant, and the electromagnetic coupling) are given in table~\ref{tab:sm_input}. These values are chosen to agree with the default settings of the {\tt sm\_onia} model, with the optional application of the {\tt c\_mass} restriction for a non-zero charm mass ($m_c$) or the {\tt lepton\_masses} restriction for non-zero lepton masses ($m_\tau$, $m_\mu$, and $m_\e$).
When the charm quark is treated as massive through the application of the \texttt{c\_mass} restriction, we adopt the three-flavour number scheme (3FS). In the default setup, where the charm quark is massless, we instead employ the four-flavour number scheme (4FS). We note that the \texttt{lepton\_masses} restriction does not modify the charm-quark mass, which therefore remains zero. Consequently, this restriction also uses the 4FS by default. Accordingly, the 3FS is used when discussing the production of charmonia or $B_c^{(*)}$ mesons, whereas the 4FS is used for processes containing only bottomonia.
In the electroweak sector, the $Z$-boson mass $m_Z$, the fine-structure constant $\alpha_{G_\mu}$ in the $G_\mu$ scheme, and the Fermi constant $G_F$ from muon decay are taken as independent inputs. Therefore, the $W$-boson mass is not a free parameter, but is instead uniquely determined by the others as
\begin{equation}
    m_W = 80.419\big[002445756163\big]\,{\rm GeV}\, ,
\end{equation}
where the digits in the squared brackets are included to reach double-precision accuracy. They are relevant for performing high-precision benchmark comparisons.
Additionally, we take the CKM matrix to be the identity matrix.

The default masses assigned to the CS states of quarkonia are listed in table~\ref{tab:onia_masses}. The corresponding CO states are taken to be heavier by $200\,\mathrm{MeV}$, reflecting the typical energy scale used in \pythia{\tt 8} for the hadronisation of CO states into physical CS mesons via the emission of a soft gluon.
On the other hand, the default leptonium masses are given by the sum of the constituent lepton masses forming the bound state.

The default LDMEs assigned to charmonium, bottomonium, and $B_c$ states are listed in tables~\ref{tab:ldme_charmonia}, \ref{tab:ldme_bottomonia}, and \ref{tab:ldme_Bc}, respectively.
The values of CS states are obtained using the relation between LDMEs and wavefunctions at the origin,
\begin{align}
    \langle \mathcal O_{^{2S+1}\swave_J^{[1]}}^{\mathcal Q} \rangle & = (2J+1)N_{[1]}\frac{\left|R_\mathcal{Q}(0)\right|^2}{4\pi}\ ,\\
    \langle \mathcal O_{^{2S+1}\pwave_J^{[1]}}^{\mathcal Q} \rangle & = (2J+1)N_{[1]}\frac{3\left|R^{\prime}_\mathcal{Q}(0)\right|^2}{4\pi}\ ,
\label{eq:ldme_cs}
\end{align}
with $N_{[1]} = 2N_c = 6$. The values of $\vert R_\mathcal{Q}(0)\vert^2$ and $\vert R_\mathcal{Q}^\prime(0)\vert^2$ are taken from ref.~\cite{Eichten:1995ch}, based on calculations using the Buchmüller-Tye QCD potential~\cite{Buchmuller:1980su}.
For the CO states of charmonia and bottomonia, we consider existing fits: ref.~\cite{Han:2014jya} for the former and ref.~\cite{Han:2014kxa} for the latter.\footnote{For those sets where only combinations of \pwave- and \swavetxt\ LDMEs are provided, we adopt values satisfying the expected scaling rule (see table~\ref{tab:NRQCD_scaling}).}
For the $B_c^{(*)}$ and $B_{cJ}$ mesons, we are not aware of any fit of their CO LDMEs in the literature. In this case, we use the expected $v^2$ scaling to estimate the CO {\swavetxt} LDME for $B_c^{(*)\pm}$ by dividing the corresponding spin-singlet CS LDME by a factor of $100$. For the remaining cases, we assume the following scaling relations for the spin-singlet states:
\begin{align}
    \langle\mathcal{O}_{^{1}\pwave_1^{[8]}}^{B_c^\pm}\rangle & = \dfrac{\langle\mathcal{O}_{^{1}\swave_0^{[1]}}^{B_c^\pm}\rangle}{100}\dfrac{(m_c+m_b)^2}{4}\ ,\\
    \langle\mathcal{O}_{^{1}\swave_0^{[8]}}^{B_{c1}(L)}\rangle & = \langle\mathcal{O}_{^{1}\pwave_1^{[1]}}^{B_{c1}(L)}\rangle/\dfrac{(m_c+m_b)^2}{4}\ ,
\label{eq:ldme_co_cb}
\end{align}
with $m_c = 1.5\,{\rm GeV}$ and $m_b = 4.75\,{\rm GeV}$. The spin-triplet mesons are then obtained via heavy-quark spin symmetry.
In the case of leptonia, the LDMEs can be obtained directly
\begin{table}[H]
\centering
\renewcommand*{\arraystretch}{1.4}
\begin{tabular}[t]{cc|cc}
\toprule
\textbf{meson} & \multicolumn{1}{c}{mass} & \textbf{meson} & mass \\
\midrule
$\eta_c$ & $2.98\,\mathrm{GeV}$ & $J/\psi$ & $3.10\,\mathrm{GeV}$ \\
& & $\chi_{c0}$ & $3.41\,\mathrm{GeV}$ \\
$h_c$ & $3.53\,\mathrm{GeV}$ & $\chi_{c1}$ & $3.51\,\mathrm{GeV}$ \\
& & $\chi_{c2}$ & $3.56\,\mathrm{GeV}$ \\
$\eta_c(2\swave)$ & $3.64\,\mathrm{GeV}$ & $\psi(2\swave)$ & $3.69\,\mathrm{GeV}$ \\
\midrule
$\eta_b$ & $9.40\,\mathrm{GeV}$ & $\Upsilon$ & $9.46\,\mathrm{GeV}$ \\
& & $\chi_{b0}$ & $9.86\,\mathrm{GeV}$ \\
$h_b$ & $9.90\,\mathrm{GeV}$ & $\chi_{b1}$ & $9.89\,\mathrm{GeV}$ \\
& & $\chi_{b2}$ & $9.91\,\mathrm{GeV}$ \\
$\eta_b(2\swave)$ & {$10.00\,\mathrm{GeV}$} & $\Upsilon(2\swave)$ & $10.02\,\mathrm{GeV}$ \\
& & $\chi_{b0}(2\pwave)$ & $10.23\,\mathrm{GeV}$ \\
$h_b(2\pwave)$ & $10.26\,\mathrm{GeV}$ & $\chi_{b1}(2\pwave)$ & $10.26\,\mathrm{GeV}$ \\
& & $\chi_{b2}(2\pwave)$ & $10.27\,\mathrm{GeV}$ \\
$\eta_b(3\swave)\footnotemark$ & {$10.34\,\mathrm{GeV}$} & $\Upsilon(3\swave)$ & $10.36\,\mathrm{GeV}$ \\
& & $\chi_{b0}(3\pwave)^{\ref{fnt:mass_filled}}$ & {$10.49\,\mathrm{GeV}$} \\
$h_b(3\pwave)^{\ref{fnt:mass_filled}}$ & {$10.51\,\mathrm{GeV}$} & $\chi_{b1}(3\pwave)$ & $10.51\,\mathrm{GeV}$ \\
& & $\chi_{b2}(3\pwave)$ & $10.52\,\mathrm{GeV}$ \\
\midrule
${B_c^\pm}$ & $6.27\,\mathrm{GeV}$ & ${B_c^{\ast \pm}}^{\ref{fnt:mass_filled_bc}}$ & {$6.33\,\mathrm{GeV}$} \\
& & ${B_{c0}^{\ast \pm}}^{\ref{fnt:mass_filled}}$ & {$6.70\,\mathrm{GeV}$}\\
${B_{c1}(L)^\pm}^{\ref{fnt:mass_filled}}$ &{$6.77\,\mathrm{GeV}$} & ${B_{c1}(H)^{\pm}}^{\ref{fnt:mass_filled}}$ & {$6.76\,\mathrm{GeV}$} \\
& & ${B_{c2}^{\ast \pm}}^{\ref{fnt:mass_filled}}$ & {$6.79\,\mathrm{GeV}$}\\
$B_c^\pm(2\swave)$ & $6.87\,\mathrm{GeV}$ & $B_c^{\ast \pm}(2\swave)^{\ref{fnt:mass_filled_bc}}$ & {$6.90\,\mathrm{GeV}$} \\
\bottomrule
\end{tabular}
\caption{\it\small Mass values of the quarkonia included in the {\tt sm\_onia} model. Most entries correspond to the PDG values available at the time of this work, except for $B_c^*(1\swave)$ and $B_c^*(2\swave)$ which are instead inferred from ref.~\cite{CMS:2019uhm}. Unknown masses are estimated based on symmetries with analogous observed states.
The values reported in this table are directly used for CS states, whereas the corresponding CO states are shifted upward by $200\,\mathrm{MeV}$.}
\label{tab:onia_masses}
\end{table}
\footnotetext{Reconstructed.
\label{fnt:mass_filled}}
\addtocounter{footnote}{1}\footnotetext{Based on ref.~\cite{CMS:2019uhm}.
\label{fnt:mass_filled_bc}}
\noindent
from the perturbative bound-state wavefunctions at the origin in a Coulomb potential. For \swave- and \pwavetxt\ states, they are respectively given by
\begin{align}
    \langle \mathcal O_{^{2S+1}\swave_S}^{{\mathcal L}(N\swave)} \rangle &= (2S+1)\dfrac{\alpha^3}{\pi}\dfrac{1}{N^3}\left(\dfrac{m_{\ell^-}m_{\ell^{\prime +}}}{m_{\ell^-}+m_{\ell^{\prime +}}}\right)^{\!3},\label{eq:ldme_leptonia_swave}\\
    \langle \mathcal O_{^{2S+1}\pwave_J}^{{\mathcal L}(N\pwave)} \rangle &= (2J+1)\dfrac{\alpha^5}{\pi}\dfrac{N^2-1}{N^5}\left(\dfrac{m_{\ell^-}m_{\ell^{\prime +}}}{m_{\ell^-}+m_{\ell^{\prime +}}}\right)^{\!5}.\label{eq:ldme_leptonia_pwave}
\end{align}

The renormalisation and factorisation scales used in this work are chosen as half the sum of the final-state transverse energies,
\begin{equation}\label{eq:setup_mu}
    \mu_R = \mu_F \equiv \frac{H_T}{2} = \frac{1}{2} \sum_{i=3}^{{\rm N}+2} E_{i,{\scriptscriptstyle T} }\,,
\end{equation}
where the sum runs over all final-state particles in a $2 \to {\rm N}$ process, and $E_{i,{\scriptscriptstyle T}} = \sqrt{m_i^2 + k_{i,{\scriptscriptstyle T}}^2}$ is the transverse energy of the $i^{\mathrm{th}}$ particle.
In addition, the {\tt NNPDF40\_nlo\_nf\_4\_pdfas} ({\tt LHAPDF6} index: {\tt 335700}) PDF set~\cite{NNPDF:2021njg} is used throughout this work.

\newpage
\vspace*{\fill}
\renewcommand*{\arraystretch}{1.4}
\begin{longtable}[t]{cc|cc}
\toprule
${\mathcal Q}[n]$ & $\langle \mathcal O_{n}^{\mathcal Q} \rangle$ 
&
${\mathcal Q}[n]$ & $\langle \mathcal O_{n}^{\mathcal Q} \rangle$ \\
\midrule\endhead
$\eta_c\big[{}^1\swave_0^{[1]}\big]$ & $0.3867  \,\mathrm{GeV}^{3}$ & $J/\psi\big[{}^3\swave_1^{[1]}\big]$ & $1.16    \,\mathrm{GeV}^{3}$ \\
$\eta_c\big[{}^3\swave_1^{[8]}\big]$ & $0.0146  \,\mathrm{GeV}^{3}$ & $J/\psi\big[{}^1\swave_0^{[8]}\big]$ & $0.0146  \,\mathrm{GeV}^{3}$ \\
$\eta_c\big[{}^1\swave_0^{[8]}\big]$ & $0.00301 \,\mathrm{GeV}^{3}$ & $J/\psi\big[{}^3\swave_1^{[8]}\big]$ & $0.00903 \,\mathrm{GeV}^{3}$ \\
                                     &                              &  
$J/\psi\big[{}^3\pwave_0^{[8]}\big]$ & $0.0343  \,\mathrm{GeV}^{5}$ \\
$\eta_c\big[{}^1\pwave_1^{[8]}\big]$ & $0.1029  \,\mathrm{GeV}^{5}$ & $J/\psi\big[{}^3\pwave_1^{[8]}\big]$ & $0.1029  \,\mathrm{GeV}^{5}$ \\
                                     &                              &  
$J/\psi\big[{}^3\pwave_2^{[8]}\big]$ & $0.1715  \,\mathrm{GeV}^{5}$ \\
\midrule
                                        &                              &  
$\chi_{c0}\big[{}^3\pwave_0^{[1]}\big]$ & $0.1074  \,\mathrm{GeV}^{5}$ \\
                                        &                              &  
$\chi_{c0}\big[{}^3\swave_1^{[8]}\big]$ & $0.00215 \,\mathrm{GeV}^{3}$ \\
$h_c\big[{}^1\pwave_1^{[1]}\big]$       & $0.3223  \,\mathrm{GeV}^{5}$ & $\chi_{c1}\big[{}^3\pwave_1^{[1]}\big]$ & $0.3223  \,\mathrm{GeV}^{5}$ \\
$h_c\big[{}^1\swave_0^{[8]}\big]$       & $0.00645 \,\mathrm{GeV}^{3}$ & $\chi_{c1}\big[{}^3\swave_1^{[8]}\big]$ & $0.00645 \,\mathrm{GeV}^{3}$ \\
                                        &                              &  
$\chi_{c2}\big[{}^3\pwave_2^{[1]}\big]$ & $0.5371  \,\mathrm{GeV}^{5}$ \\
                                        &                              &  
$\chi_{c2}\big[{}^3\swave_1^{[8]}\big]$ & $0.0107  \,\mathrm{GeV}^{3}$ \\
\midrule
$\eta_c(2\swave)\big[{}^1\swave_0^{[1]}\big]$ & $0.2526     \,\mathrm{GeV}^{3}$ & $\psi(2\swave)\big[{}^3\swave_1^{[1]}\big]$   & $0.7577     \,\mathrm{GeV}^{3}$ \\
$\eta_c(2\swave)\big[{}^3\swave_1^{[8]}\big]$ & $0.00396029 \,\mathrm{GeV}^{3}$ & $\psi(2\swave)\big[{}^1\swave_0^{[8]}\big]$   & $0.00396029 \,\mathrm{GeV}^{3}$ \\
$\eta_c(2\swave)\big[{}^1\swave_0^{[8]}\big]$ & $0.00349746 \,\mathrm{GeV}^{3}$ & $\psi(2\swave)\big[{}^3\swave_1^{[8]}\big]$   & $0.00349746 \,\mathrm{GeV}^{3}$ \\
                                              &                                 & 
$\psi(2\swave)\big[{}^3\pwave_0^{[8]}\big]$   & $0.0088889  \,\mathrm{GeV}^{5}$ \\
$\eta_c(2\swave)\big[{}^1\pwave_1^{[8]}\big]$ & $0.026667   \,\mathrm{GeV}^{5}$ & $\psi(2\swave)\big[{}^3\pwave_1^{[8]}\big]$   & $0.026667   \,\mathrm{GeV}^{5}$ \\
                                              &                                 & 
$\psi(2\swave)\big[{}^3\pwave_2^{[8]}\big]$   &  $0.044444  \,\mathrm{GeV}^{5}$ \\
\bottomrule
\caption{\it\small Default LDME values in {\tt \mgshort} for charmonium mesons. CS values are computed according to eq.~\eqref{eq:ldme_cs} using the wavefunctions at the origin from ref.~\cite{Eichten:1995ch} ($m_c=1.48\,{\rm GeV}$). CO values are taken from ref.~\cite{Han:2014jya} ($m_c=1.5\,{\rm GeV}$).
}
\label{tab:ldme_charmonia}
\end{longtable}
\vspace*{\fill}

\newpage
\vspace*{\fill}
\renewcommand*{\arraystretch}{1.4}
\begin{longtable}[t]{cc|cc}
\toprule
${\mathcal Q}[n]$ & $\langle \mathcal O_{n}^{\mathcal Q} \rangle$ 
&
${\mathcal Q}[n]$ & $\langle \mathcal O_{n}^{\mathcal Q} \rangle$ \\
\midrule\endhead

$\eta_b\big[{}^1\swave_0^{[1]}\big]$   & $3.093   \,\mathrm{GeV}^{3}$ & $\Upsilon\big[{}^3\swave_1^{[1]}\big]$ & $9.28    \,\mathrm{GeV}^{3}$ \\
$\eta_b\big[{}^3\swave_1^{[8]}\big]$   & $0.0208  \,\mathrm{GeV}^{3}$ & $\Upsilon\big[{}^1\swave_0^{[8]}\big]$ & $0.0208  \,\mathrm{GeV}^{3}$ \\
$\eta_b\big[{}^1\swave_0^{[8]}\big]$   & $0.00920 \,\mathrm{GeV}^{3}$ & $\Upsilon\big[{}^3\swave_1^{[8]}\big]$ & $0.0276  \,\mathrm{GeV}^{3}$ \\
                                       &                              &  
$\Upsilon\big[{}^3\pwave_0^{[8]}\big]$ & $0.69    \,\mathrm{GeV}^{5}$ \\
$\eta_b\big[{}^1\pwave_1^{[8]}\big]$   & $2.07    \,\mathrm{GeV}^{5}$ & $\Upsilon\big[{}^3\pwave_1^{[8]}\big]$ & $2.07    \,\mathrm{GeV}^{5}$ \\
                                       &                              & 
$\Upsilon\big[{}^3\pwave_2^{[8]}\big]$ & $3.45    \,\mathrm{GeV}^{5}$ \\
\midrule
                                        &                             &  
$\chi_{b0}\big[{}^3\pwave_0^{[1]}\big]$ & $2.03   \,\mathrm{GeV}^{5}$ \\
                                        &                             &  
$\chi_{b0}\big[{}^3\swave_1^{[8]}\big]$ & $0.006  \,\mathrm{GeV}^{3}$ \\
$h_b\big[{}^1\pwave_1^{[1]}\big]$       & $6.089  \,\mathrm{GeV}^{5}$ & $\chi_{b1}\big[{}^3\pwave_1^{[1]}\big]$ & $6.089  \,\mathrm{GeV}^{5}$ \\
$h_b\big[{}^1\swave_0^{[8]}\big]$       & $0.019  \,\mathrm{GeV}^{3}$ & $\chi_{b1}\big[{}^3\swave_1^{[8]}\big]$ & $0.019  \,\mathrm{GeV}^{3}$ \\
                                        &                             &  
$\chi_{b2}\big[{}^3\pwave_2^{[1]}\big]$ & $10.15  \,\mathrm{GeV}^{5}$ \\
                                        &                             &  
$\chi_{b2}\big[{}^3\swave_1^{[8]}\big]$ & $0.0315 \,\mathrm{GeV}^{3}$ \\
\midrule 
$\eta_b(2\swave)\big[{}^1\swave_0^{[1]}\big]$   & $1.544  \,\mathrm{GeV}^{3}$ & $\Upsilon(2\swave)\big[{}^3\swave_1^{[1]}\big]$ & $4.63   \,\mathrm{GeV}^{3}$ \\
$\eta_b(2\swave)\big[{}^3\swave_1^{[8]}\big]$   & $0.0102 \,\mathrm{GeV}^{3}$ & $\Upsilon(2\swave)\big[{}^1\swave_0^{[8]}\big]$ & $0.0102 \,\mathrm{GeV}^{3}$ \\
$\eta_b(2\swave)\big[{}^1\swave_0^{[8]}\big]$   & $0.0059 \,\mathrm{GeV}^{3}$ & $\Upsilon(2\swave)\big[{}^3\swave_1^{[8]}\big]$ & $0.0177 \,\mathrm{GeV}^{3}$ \\
                                                &                             &  
$\Upsilon(2\swave)\big[{}^3\pwave_0^{[8]}\big]$ & $0.3    \,\mathrm{GeV}^{5}$ \\
$\eta_b(2\swave)\big[{}^1\pwave_1^{[8]}\big]$   & $0.9    \,\mathrm{GeV}^{5}$ & $\Upsilon(2\swave)\big[{}^3\pwave_1^{[8]}\big]$ & $0.9    \,\mathrm{GeV}^{5}$ \\
                                                &                             &  
$\Upsilon(2\swave)\big[{}^3\pwave_2^{[8]}\big]$ & $1.5    \,\mathrm{GeV}^{5}$ \\
\midrule
                                                 &                            &  
$\chi_{b0}(2\pwave)\big[{}^3\pwave_0^{[1]}\big]$ & $2.368 \,\mathrm{GeV}^{5}$ \\
                                                 &                            &  
$\chi_{b0}(2\pwave)\big[{}^3\swave_1^{[8]}\big]$ & $0.011 \,\mathrm{GeV}^{3}$ \\
$h_b(2\pwave)\big[{}^1\pwave_1^{[1]}\big]$       & $7.103 \,\mathrm{GeV}^{5}$ & 
$\chi_{b1}(2\pwave)\big[{}^3\pwave_1^{[1]}\big]$ & $7.103 \,\mathrm{GeV}^{5}$ \\
$h_b(2\pwave)\big[{}^1\swave_0^{[8]}\big]$       & $0.028 \,\mathrm{GeV}^{3}$ & 
$\chi_{b1}(2\pwave)\big[{}^3\swave_1^{[8]}\big]$ & $0.028 \,\mathrm{GeV}^{3}$ \\
                                                 &                            &  
$\chi_{b2}(2\pwave)\big[{}^3\pwave_2^{[1]}\big]$ & $11.84 \,\mathrm{GeV}^{5}$ \\
                                                 &                            &  
$\chi_{b2}(2\pwave)\big[{}^3\swave_1^{[8]}\big]$ & $0.046 \,\mathrm{GeV}^{3}$ \\
\midrule
$\eta_b(3\swave)\big[{}^1\swave_0^{[1]}\big]$       & $1.18    \,\mathrm{GeV}^{3}$ & $\Upsilon(3{\swave})\big[{}^3{\swave}_1^{[1]}\big]$ & $3.54    \,\mathrm{GeV}^{3}$ \\
$\eta_b(3\swave)\big[{}^3\swave_1^{[8]}\big]$       & $0.00725 \,\mathrm{GeV}^{3}$ & $\Upsilon(3\swave)\big[{}^1\swave_0^{[8]}\big]$     & $0.00725 \,\mathrm{GeV}^{3}$ \\
$\eta_b(3\swave)\big[{}^1\swave_0^{[8]}\big]$       & $0.00373 \,\mathrm{GeV}^{3}$ & $\Upsilon(3\swave)\big[{}^3\swave_1^{[8]}\big]$     & $0.0112  \,\mathrm{GeV}^{3}$ \\
                                                    &                              &  
$\Upsilon(3\swave)\big[{}^3\pwave_0^{[8]}\big]$     & $0.125   \,\mathrm{GeV}^{5}$ \\
$\eta_b(3\swave)\big[{}^1\pwave_1^{[8]}\big]$       & $0.375   \,\mathrm{GeV}^{5}$ & $\Upsilon(3\swave)\big[{}^3\pwave_1^{[8]}\big]$     & $0.375   \,\mathrm{GeV}^{5}$ \\
                                                    &                              & 
$\Upsilon(3\swave)\big[{}^3\pwave_2^{[8]}\big]$     & $0.625   \,\mathrm{GeV}^{5}$ \\
\midrule
                                                     &                           & 
$\chi_{b0}(3{\pwave})\big[{}^3{\pwave}_0^{[1]}\big]$ & $2.57  \,\mathrm{GeV}^{5}$ \\
                                                     &                            & 
$\chi_{b0}(3\pwave)\big[{}^3\swave_1^{[8]}\big]$     & $0.016 \,\mathrm{GeV}^{3}$ \\
$h_b(3\pwave)\big[{}^1\pwave_1^{[1]}\big]$           & $7.709 \,\mathrm{GeV}^{5}$ & 
$\chi_{b1}(3\pwave)\big[{}^3\pwave_1^{[1]}\big]$     & $7.709 \,\mathrm{GeV}^{5}$ \\
$h_b(3\pwave)\big[{}^1\swave_0^{[8]}\big]$           & $0.047 \,\mathrm{GeV}^{3}$ & 
$\chi_{b1}(3\pwave)\big[{}^3\swave_1^{[8]}\big]$     & $0.047 \,\mathrm{GeV}^{3}$ \\
                                                     &                            & 
$\chi_{b2}(3\pwave)\big[{}^3\pwave_2^{[1]}\big]$     & $12.85 \,\mathrm{GeV}^{5}$ \\
                                                     &                            & 
$\chi_{b2}(3\pwave)\big[{}^3\swave_1^{[8]}\big]$     & $0.079 \,\mathrm{GeV}^{3}$ \\
\bottomrule
\caption{\it\small Default LDME values in {\tt \mgshort} for bottomonium mesons. CS values are computed according to eq.~\eqref{eq:ldme_cs} using wavefunctions at the origin from ref.~\cite{Eichten:1995ch} ($m_b=4.88\,{\rm GeV}$). CO values are taken from ref.~\cite{Han:2014kxa} ($m_b = 4.75\,{\rm GeV}$).
}
\label{tab:ldme_bottomonia}
\end{longtable}

\newpage
\vspace*{\fill}
\renewcommand*{\arraystretch}{1.4}
\begin{longtable}[t]{cc|cc}
\toprule
${\mathcal Q}[n]$ & $\langle \mathcal O_{n}^{\mathcal Q} \rangle$ 
&
${\mathcal Q}[n]$ & $\langle \mathcal O_{n}^{\mathcal Q} \rangle$ \\
\midrule\endhead
$B_c^\pm\big[{}^1\swave_0^{[1]}\big]$        & $0.784   \,\mathrm{GeV}^{3}$ & 
$B_c^{\ast \pm}\big[{}^3\swave_1^{[1]}\big]$ & $2.352   \,\mathrm{GeV}^{3}$ \\
$B_c^\pm\big[{}^3\swave_1^{[8]}\big]$        & $0.00784 \,\mathrm{GeV}^{3}$ & 
$B_c^{\ast \pm}\big[{}^1\swave_0^{[8]}\big]$ & $0.00784 \,\mathrm{GeV}^{3}$ \\
$B_c^\pm\big[{}^1\swave_0^{[8]}\big]$        & $0.00784 \,\mathrm{GeV}^{3}$ & 
$B_c^{\ast \pm}\big[{}^3\swave_1^{[8]}\big]$ & $0.02352 \,\mathrm{GeV}^{3}$ \\
                                             &                              & 
$B_c^{\ast \pm}\big[{}^3\pwave_0^{[8]}\big]$ & $0.02552 \,\mathrm{GeV}^{5}$ \\
$B_c^\pm\big[{}^1\pwave_1^{[8]}\big]$        & $0.07656 \,\mathrm{GeV}^{5}$ & 
$B_c^{\ast \pm}\big[{}^3\pwave_1^{[8]}\big]$ & $0.07656 \,\mathrm{GeV}^{5}$ \\
                                             &                              & 
$B_c^{\ast \pm}\big[{}^3\pwave_2^{[8]}\big]$ & $0.1276  \,\mathrm{GeV}^{5}$ \\
\midrule
                                                &                             & 
$B_{c0}^{\ast \pm}\big[{}^3\pwave_0^{[1]}\big]$ & $0.288  \,\mathrm{GeV}^{5}$ \\
$B_{c0}^{\ast \pm}\big[{}^3\swave_1^{[8]}\big]$ & $0.0295 \,\mathrm{GeV}^{3}$ \\
$B_{c1}(L)^\pm\big[{}^1\pwave_1^{[1]}\big]$     & $0.864  \,\mathrm{GeV}^{5}$ & 
$B_{c1}(H)^{\pm}\big[{}^3\pwave_1^{[1]}\big]$   & $0.864  \,\mathrm{GeV}^{5}$ \\
$B_{c1}(L)^\pm\big[{}^1\swave_0^{[8]}\big]$     & $0.0884 \,\mathrm{GeV}^{3}$ & 
$B_{c1}(H)^{\pm}\big[{}^3\swave_1^{[8]}\big]$   & $0.0884 \,\mathrm{GeV}^{3}$ \\
                                                &                             & 
$B_{c2}^{\ast \pm}\big[{}^3\pwave_2^{[1]}\big]$ & $1.44   \,\mathrm{GeV}^{5}$ \\
                                                &                             & 
$B_{c2}^{\ast \pm}\big[{}^3\swave_1^{[8]}\big]$ & $0.147  \,\mathrm{GeV}^{3}$ \\
\midrule
$B_c^\pm(2\swave)\big[{}^1\swave_0^{[1]}\big]$        & $0.469   \,\mathrm{GeV}^{3}$ & 
$B_c^{\ast \pm}(2\swave)\big[{}^3\swave_1^{[1]}\big]$ & $1.41    \,\mathrm{GeV}^{3}$ \\
$B_c^\pm(2\swave)\big[{}^3\swave_1^{[8]}\big]$        & $0.00469 \,\mathrm{GeV}^{3}$ & 
$B_c^{\ast \pm}(2\swave)\big[{}^1\swave_0^{[8]}\big]$ & $0.00469 \,\mathrm{GeV}^{3}$ \\
$B_c^\pm(2\swave)\big[{}^1\swave_0^{[8]}\big]$        & $0.00469 \,\mathrm{GeV}^{3}$ & 
$B_c^{\ast \pm}(2\swave)\big[{}^3\swave_1^{[8]}\big]$ & $0.0141  \,\mathrm{GeV}^{3}$ \\
                                                      &                              & 
$B_c^{\ast \pm}(2\swave)\big[{}^3\pwave_0^{[8]}\big]$ & $0.0153  \,\mathrm{GeV}^{5}$ \\
$B_c^\pm(2\swave)\big[{}^1\pwave_1^{[8]}\big]$        & $0.0458  \,\mathrm{GeV}^{5}$ & 
$B_c^{\ast \pm}(2\swave)\big[{}^3\pwave_1^{[8]}\big]$ & $0.0458  \,\mathrm{GeV}^{5}$ \\
                                                      &                              & 
$B_c^{\ast \pm}(2\swave)\big[{}^3\pwave_2^{[8]}\big]$ & $0.0764  \,\mathrm{GeV}^{5}$ \\
\bottomrule
\caption{\it\small Default LDME values in {\tt \mgshort} for $B_c$ mesons. CS values are computed according to eq.~\eqref{eq:ldme_cs} using wavefunctions at the origin from ref.~\cite{Eichten:1995ch} ($m_c=1.48\,{\rm GeV}$ and $m_b=4.88\,{\rm GeV}$). CO values are obtained assuming velocity scaling, heavy-quark spin symmetries, and dimensional analysis ($m_c=1.5\,{\rm GeV}$ and $m_b=4.75\,{\rm GeV}$).}
\label{tab:ldme_Bc}
\end{longtable}
\vspace*{\fill}

\bibliographystyle{JHEP} 
\bibliography{references}

\end{document}